%% file: DESKiDS_cosmic_shear.tex
\documentclass[a4paper,twocolumn, numberedappendix, twocolappendix,revtex4,iop]{openjournal} 
\usepackage[T1]{fontenc}
\usepackage[utf8]{inputenc}
\usepackage{natbib}
\usepackage{multirow}
\usepackage{comment}
\usepackage[a4paper,marginpar=10pt,headheight=20pt, right=1cm, top=1.5in, left=1in, bottom=0pt, footskip=0pt, footnotesep=10pt]{geometry}

\DeclareRobustCommand{\VAN}[3]{#2}
\let\VANthebibliography\thebibliography
\def\thebibliography{\DeclareRobustCommand{\VAN}[3]{##3}\VANthebibliography}

\usepackage{graphicx}	
\usepackage{amsmath}	
\usepackage{amssymb}	
\usepackage{xspace}        
\usepackage{xcolor}          
\usepackage{newtxtext,newtxmath}
\usepackage{booktabs}
\usepackage{fancyhdr}
\usepackage{times}
\newcommand{\decimalarcmin}{\(\stackrel{\:'}{\textstyle.}\)}

\usepackage[colorlinks,linkcolor=blue,citecolor=blue,urlcolor=blue ]{hyperref}
\usepackage{xurl} 

\definecolor{purple}{RGB}{156,81,182}
\definecolor{hotpink}{RGB}{255, 0,255}
\definecolor{applegreen}{rgb}{0.55, 0.71, 0.0}
\definecolor{dodgerblue}{RGB}{30,144,255}
\definecolor{lavenderpink}{rgb}{0.98, 0.68, 0.82}
\definecolor{orange}{RGB}{255,132,0}
\definecolor{ferngreen}{RGB}{73, 119, 73}

\newcommand{\Mpc}{\,h^{-1}\mathrm{Mpc}}

\newcommand{\Om}{\Omega_\mathrm{m}}
\newcommand{\Ob}{\Omega_\mathrm{b}}

\newcommand{\ob}{\omega_\mathrm{b}}
\newcommand{\ns}{n_\mathrm{s}}
\newcommand{\As}{A_\mathrm{s}}
\newcommand{\lcdm}{$\Lambda$CDM }
\newcommand{\bb}[1]{\left[ #1 \right]}
\newcommand{\vek}[1]{\mbox{\boldmath $#1$}}
\newcommand{\SeK}{S_8^{\rm KiDS}}
\newcommand{\SeD}{S_8^{\rm DES}}
\newcommand{\SeJ}{S_8^{\rm Joint}}
\newcommand{\SeJS}{S_8^{\rm J1IA}}
\def\C{{C_{\epsilon \epsilon}}}
\def\d{{\rm d}}

\renewcommand{\arraystretch}{1.2}

\newcommand{\be}{\begin{equation}}  \newcommand{\ee}{\end{equation}}

\begin{document}
\vspace*{-\headsep}\vspace*{\headheight}
{\footnotesize \hfill FERMILAB-PUB-23-267-PPD}\\
\vspace*{-\headsep}\vspace*{\headheight}
{\footnotesize \hfill DES-2023-0769}

\count\footins = 1000

\title{DES Y3 + KiDS-1000: Consistent Cosmology combining cosmic shear surveys}
\author{\vspace{-1.5cm}
Dark Energy Survey and Kilo-Degree Survey Collaboration:  \\
T.~M.~C.~Abbott,$^{1}$
M.~Aguena,$^{2}$
A.~Alarcon,$^{3}$
O.~Alves,$^{4}$
A.~Amon,$^{5,6}$
F.~Andrade-Oliveira,$^{4}$
M.~Asgari,$^{7}$
S.~Avila,$^{8}$
D.~Bacon,$^{9}$
K.~Bechtol,$^{10}$
% ...
\\
{\it (Affiliations can be found after the references)}
}
\begin{abstract}
We present a joint cosmic shear analysis of the Dark Energy Survey (DES Y3) and the Kilo-Degree Survey (KiDS-1000) in a collaborative effort between the two survey teams. We find consistent cosmological parameter constraints between DES Y3 and KiDS-1000 which, when combined in a joint-survey analysis, constrain the parameter $S_8 = \sigma_8 \sqrt{\Omega_{\rm m}/0.3}$ with a mean value of $0.790^{+0.018}_{-0.014}$.   The mean marginal is lower than the maximum a posteriori estimate, $S_8=0.801$, owing to skewness in the marginal distribution and projection effects in the multi-dimensional parameter space.  Our results are consistent with $S_8$ constraints from observations of the cosmic microwave background by {\it Planck}, with agreement at the $1.7\sigma$ level.  We use a Hybrid analysis pipeline, defined from a mock survey study quantifying the impact of the different analysis choices originally adopted by each survey team. We review intrinsic alignment models, baryon feedback mitigation strategies, priors, samplers and models of the non-linear matter power spectrum. 
\end{abstract}
\vspace{20pt}
\keywords{Cosmology, Weak Gravitational Lensing}
\maketitle
{\small \tableofcontents}

\section{Introduction}
\label{sec:intro}
A cosmic shear analysis exploits the weak gravitational lensing of background galaxy images by foreground large-scale structures in order to probe the growth of structures and the expansion of the Universe.    Since the first detection \citep{bacon/etal:2000, kaiser/etal:2000, vanwaerbeke/etal:2000,wittman/etal:2000} significant developments in instrumentation, data, theory, simulation and statistical analysis have led to an established and robust cosmological probe.  Cosmic shear is one of the primary science drivers for the next generation of cosmological surveys imaged with {\it Euclid}\footnote{{\it Euclid}: \url{https://sci.esa.int/web/euclid}}, the Nancy Grace Roman Space Telescope\footnote{Hereafter {\it Roman}: \url{https://roman.gsfc.nasa.gov/}} and the Vera C. Rubin Observatory\footnote{Hereafter {\it Rubin}: \url{https://www.lsst.org/}}. These surveys are designed to attain sub-percent level precision on joint measurements of the dark energy equation of state parameter, $w_0$, and the scale factor evolution parameter, $w_{\rm a}$. Their goal is to understand the mechanism that drives cosmic acceleration.  Large-scale pathfinders include the Deep Lens Survey \citep[DLS,][]{wittman/etal:2002,jee/etal:2016}, the Canada France Hawaii Telescope Lensing Survey \citep[CFHTLenS,][]{heymans/etal:2012,joudaki/etal:2017}, the Dark Energy Survey (DES, \citealt{DESDR1}; \citealt{Sevilla-Noarbe/etal:2021}; \citealt{amon/etal:2022}; \citealt*{secco/etal:2022}), 
the Hyper Suprime Camera Survey \citep[HSC,][]{aihara/etal:2018,dalal/etal:2023,li/zhang/etal:2023} and the Kilo Degree Survey \citep[KiDS,][]{kuijken/etal:2015,kuijken/etal:2019,asgari/etal:2021}.   All have set tight and consistent constraints on the parameter\footnote{There is no consensus in the literature on the name to best describe the $S_8$ parameter.  It has been referred to as the clustering amplitude parameter, the structure growth parameter and the lensing amplitude parameter.} $S_8 = \sigma_8 \sqrt{\Om/0.3}$, which accounts for the inherent cosmic shear degeneracy between $\Om$, the matter density parameter, and $\sigma_8$, the linear-theory standard deviation of matter density fluctuations in spheres of radius $8 \Mpc$. These constraints all show a preference for $S_8$ to be lower than the value derived from observations of the Cosmic Microwave Background when adopting the flat-\lcdm model \citep[CMB;][]{planck/etal:2020}. Any significant offset between direct measurements of $S_8$ and the CMB-informed \lcdm prediction would, in principle, indicate issues with the \lcdm model.
The community has not reached a consensus on whether the `tension' seen between these early and late Universe observations arises from sampling variance, systematic errors in the theoretical modelling and/or data analysis, or whether the existing results are already an indication of beyond-$\Lambda$CDM physics. 

The weak lensing community has a long history of collaborative initiatives to improve the robustness of cosmic shear cosmology with shear and redshift measurement challenges analysing image simulations and mock catalogues \citep{heymans/etal:2006, hildebrandt/etal:2010, kitching/etal:2012,mandelbaum/etal:2015,euclid-desprez/etal:2020,schmidt/etal:2020}.  The strong track record in releasing all relevant data products has also allowed for examination and verification by independent groups (\citealt{maccrann/etal:2015, efstathiou/lemos:2018, troxel/etal:2018,asgari/etal:2019}; \citealt*{asgari/etal:2020}; \citealt{joudaki/etal:2020,chang/etal:2019,garcia-garcia/etal:2021,amon/efstathiou:2022,longley/etal:2022}).    Comparison studies find consistent weak lensing measurements between different surveys, as quantified through measurements of the projected surface mass density around luminous red galaxies \citep*{amon/etal:2018,leauthaud/etal:2022,amon/etal:2023}.  Unified analyses of cosmic shear surveys also find consistency between the surveys tested, but highlight the impact of different analysis choices on the recovered cosmological parameter constraints (\citealt{benjamin/etal:2007,chang/etal:2019}; \citealt*{asgari/etal:2020}; \citealt{joudaki/etal:2020}). Most recently, \citet{longley/etal:2022} combined the first year data from DES and HSC with the fourth data release from KiDS, reporting the tightest Stage-III cosmic shear constraints to date.  The 1.6-1.9\% constraints on $S_8$ range from $S_8 = 0.777^{+0.016}_{-0.017}$ to $S_8 = 0.791^{+0.013}_{-0.013}$, depending on the different modelling approaches and methods used to account for the cross-survey covariance between the overlapping KiDS and HSC surveys.

This paper presents a joint collaboration analysis of the latest data releases from the Dark Energy Survey (hereafter DES Y3), and the Kilo-Degree Survey (hereafter KiDS-1000)\footnote{We have chosen to limit this joint analysis to DES and KiDS. Accurately modelling the cross-survey covariance with the addition of HSC is non-trivial: roughly half of HSC overlaps the Northern stripe of KiDS with the other half overlapping with the DES equatorial stripe (see Appendix~\ref{app:DESminusKiDS}).}. Table~\ref{tab:params} summarises the survey specifications for these two complementary data sets.  The DES footprint spanning a factor of five times the area of the KiDS footprint, can be contrasted with KiDS utilising three times as many filters as DES in the wide field cosmic shear measurement, with matched-depth photometry in both the optical and near infrared.  In the DES Deep Field calibration analysis, DES extend both the depth and wavelength range of the survey into the near infrared \citep*{hartley/etal:2021}. To advance photometric redshift calibration, both surveys incorporate significant imaging of calibration fields that have been targeted by a range of different deep spectroscopic redshift campaigns \citep*{hildebrandt/etal:2021,myles/etal:2021}. Using self-organising maps (SOM), both teams mitigate incompleteness in these spectroscopic training samples (\citealt{buch/etal:2019,wright/etal:2020}; \citealt*{myles/etal:2021}), incorporating near-infrared information (\citealt{wright/etal:2019}; \citealt*{hartley/etal:2021,everett/etal:2022}), with validation using cross-correlation clustering measurements (\citealt{vandenbusch/etal:2020}; \citealt*{hildebrandt/etal:2021}; \citealt*{gatti/etal:2022}). DES and KiDS both adopt blinding procedures to avoid the inclusion of confirmation bias in their findings \citep{kuijken/etal:2015, muir/etal:2020}. The cosmic shear constraints from the two surveys are consistent, with similar prescriptions chosen to derive analytical covariance matrices \citep{schneider/etal:2002,joachimi/etal:2021,friedrich/etal:2021}. \citet{asgari/etal:2021} present the KiDS-1000 cosmic shear analysis finding $S_8= 0.759_{-0.021}^{+0.024}$.   \citet{amon/etal:2022}; \citet*{secco/etal:2022} present the DES Y3 joint cosmic shear and shear ratio analysis finding $S_8= 0.772^{+0.018}_{-0.017}$ in their $\Lambda$CDM-optimised analysis.   

Whilst there are many similarities between the DES and KiDS data processing techniques, catalogue production and calibration methods, it is worth noting that the two teams take rather different approaches to the shear measurement \citep{miller/etal:2013,huff/mandelbaum:2017,sheldon/huff:2017}.  This has been motivated in part by the different tiling strategies and PSF properties of each survey. The DES analysis operates on astrometrically sheared and stacked images with the KiDS analysis operating on individual unsheared exposures using analytic astrometric corrections.  These core differences in data handling make it highly non-trivial to conduct a KiDS-like shear measurement of DES and vice versa.  Many in the lensing community argue that the significant resources required to make a pixel-level comparison of shear measurement methods using real data would be unwarranted, given the significant investment already made in the production of realistic image simulations \citep{kannawadi/etal:2019, maccrann/etal:2021, li/etal:2023}.  This argument is strengthened by the finding that the simulation-informed calibration correction is fairly insensitive to modifications of the simulation's input parameters, well within the uncertainty adopted by each team \citep[see for example appendix C of][]{li/etal:2023b}.  We are therefore confident in the compatibility and robustness of the DES and KiDS shear catalogues for this joint-survey analysis.

In this analysis we quantify how the cosmological parameter constraints for DES and KiDS are impacted by differences in the model and analysis framework chosen by each survey.  Table~\ref{tab:hybrid} highlights these methodological differences, which, in each case, have been chosen based on well-reasoned and sound scientific arguments detailed in \citet{joachimi/etal:2021} and \citet{krause/etal:2021}. We present constraints from a Hybrid pipeline where we adopt a unified set of choices designed to provide the most robust joint-survey analysis.  We adopt the findings of each team for their overall covariance matrix and their quantification of redshift and shear calibration uncertainty, following any survey-specific data-related systematic limitations and mitigation strategies.  

This paper is organised as follows.  In Section~\ref{sec:analysis} we review the differences in methodology and the analysis framework between the two survey teams, along with our choices and rationale behind the Hybrid pipeline.  Cosmological parameter constraints are presented in Section~\ref{sec:results} with our conclusions presented in Section~\ref{sec:conclusions}.   Detailed information about the analysis can be found in the appendices.  Appendix~\ref{app:DESminusKiDS} presents the DES data vector used in the joint-survey analysis, re-measured from the DES Y3 catalogues with the overlapping region of the KiDS footprint excised to mitigate cross-survey covariance.  Appendix~\ref{app:scalecuts} defines the required scale cuts for the DES and KiDS data vectors when using the DES baryon feedback mitigation strategy.   Appendix~\ref{app:mocks} presents a study of the impact of the different analysis choices within a controlled mock data environment.  We compare constraints when the astrophysical systematics are matched to each survey's framework and consider potential systematic biases in a joint-survey analysis when the underlying astrophysical systematic models are no-longer matched to the models adopted by each survey.  Appendix~\ref{app:samplers} compares the differences between the recovered cosmological constraints from a range of samplers and in Appendix~\ref{app:hybrid} we verify the accuracy of our Hybrid pipeline using mock data based on the {\sc EuclidEmulatorv2} \citep{EuclidEmv2/etal:2021}.  Appendix~\ref{app:KiDSCOSEBIs} quantifies the probability of finding an offset in $S_8$ when comparing noisy cosmic shear constraints with and without scale cuts. Finally in Appendix~\ref{app:extras} we present additional tables and figures to complement the primary analysis in Section~\ref{sec:results}.

\begin{table}
\centering                                      
\begin{tabular}{llll}          
\toprule
&  DES Y3 &  KiDS-1000 & HSC Year 3\\    
\midrule
\multicolumn{3}{l}{\bf Cosmic shear catalogue:}\\   [+0.1cm] 
Area ${\rm [deg^2]}$ & 4143 & 777 & 416 \\ [+0.1cm]  
Wavebands & {\it riz} (Wide) + & ${\it ugriZYJHK_s}$ & {\it grizy}\\ [+0.1cm]  
 & ${\it grizJHK_s}$ (Deep)  & & \\ [+0.1cm]  
$n_{\rm eff}$ & 5.59 &  6.22 & 14.96\\ [+0.1cm]   
$z_{\rm median}$ & 0.63 & 0.67 &  0.80 \\ [+0.1cm]  
2-pt statistic & $\xi_\pm(\theta)$ &  COSEBIs & $\xi_\pm(\theta)\C(\ell)$\\ [+0.1cm]  
\midrule
& DES Y3 & \multicolumn{2}{c}{KiDS-1000}\\   
\midrule
\multicolumn{3}{l}{\bf Data calibration uncertainty:}\\    [+0.1cm]                              
Shear  & $m_1: {\cal G}(-0.006;0.009)$ & \multicolumn{2}{l}{$\vek{\Delta m}:{\cal N}(\vek{\mu}_m;\vek{C}_ m)$}\\     [+0.1cm]                       
	 & $m_2: {\cal G}(-0.020;0.008)$ & \multicolumn{2}{l}{$\mu_m^1=-0.009 \,\,\,\,\sigma_m^1= 0.019$}\\       [+0.1cm]                                  
	 & $m_3: {\cal G}(-0.024;0.008)$ & \multicolumn{2}{l}{$\mu_m^2=-0.011 \,\,\,\,\sigma_m^2= 0.020$ }\\        [+0.1cm]                     
	 & $m_4: {\cal G}(-0.037;0.008)$ & \multicolumn{2}{l}{$\mu_m^3=-0.015 \,\,\,\,\sigma_m^3= 0.017$}\\  [+0.1cm]                         
	 &     & \multicolumn{2}{l}{$\mu_m^4=\,\,\,\,0.002 \,\,\,\,\sigma_m^4= 0.012$}\\ [+0.1cm] 
	 &     & \multicolumn{2}{l}{$\mu_m^5=\,\,\,\,0.007 \,\,\,\,\sigma_m^5= 0.010$ }\\ [+0.1cm] 
Redshift & $\Delta z_1: {\cal G}(0;0.018)$ & \multicolumn{2}{l}{$\vek{\Delta z}:{\cal N}(\vek{\mu}_z;\vek{C}_ z)$}\\    [+0.1cm]    
& $\Delta z_2: {\cal G}(0;0.015)$ &  \multicolumn{2}{l}{$\mu_z^1=\,\,\,\,0.000 \,\,\,\,\sigma_z^1= 0.011$}\\        [+0.1cm]        
& $\Delta z_3: {\cal G}(0;0.011)$ & \multicolumn{2}{l}{$\mu_z^2=\,\,\,\,0.002 \,\,\,\,\sigma_z^2= 0.011$}\\     [+0.1cm]         
& $\Delta z_4: {\cal G}(0;0.017)$ & \multicolumn{2}{l}{$\mu_z^3=\,\,\,\,0.013 \,\,\,\,\sigma_z^3= 0.012$}\\	 [+0.1cm]   
&     &  \multicolumn{2}{l}{$\mu_z^4=\,\,\,\,0.011 \,\,\,\,\sigma_z^4= 0.009$}\\ [+0.1cm] 
&     & \multicolumn{2}{l}{$\mu_z^5=-0.006 \,\,\,\,\sigma_z^5= 0.010$}\\ [+0.1cm] 
\bottomrule
\end{tabular}
\caption{The properties of the DES Y3, KiDS-1000 and HSC Year 3 cosmic shear catalogue (upper section). The area listed corresponds to the total area of the survey footprint after masking.   The wavebands utilised for the analysis are listed, with DES analysing the  {\it riz} filters in the Wide Survey area, using extra depth and wavelength coverage in the Deep Fields for calibration \citep*{hartley/etal:2021}. The effective number density of sources, $n_{\rm eff}$, is given in units of galaxies per square arcmin \citep{heymans/etal:2012}. For DES Y3 and KiDS-1000 we list the observational systematic nuisance parameters used in this analysis (lower section). Uncertainty on the shear, $m_i$, and redshift calibration corrections, $\Delta z_i = \langle n_i^{\rm estimate}(z) \rangle - \langle n_i^{\rm true}(z) \rangle$, per tomographic bin $i$, are modelled as uncorrelated between the four DES tomographic bins.  DES uses independent Gaussian priors ${\cal G}(\mu;\sigma)$, with mean $\mu$ and standard deviation $\sigma$.  Data calibration uncertainty is modelled as correlated between the five KiDS tomographic bins, using a five dimensional multivariate Gaussian prior ${\cal N}(\vek{\mu};\vek{C})$ with mean, $\vek{\mu}$, and covariance, $\vek{C}$.  The amplitude of the diagonal of the covariance is listed as $\sigma^i = \sqrt{C^{ii}}$.}   
\label{tab:params}           
\end{table}

\section{Cosmic Shear Analysis}
\label{sec:analysis}
In this paper, the term `cosmic shear' refers to the two-point statistical analysis of the observed correlations between galaxy ellipticities, $\epsilon^{\rm obs}$, which are related to the weak gravitational lensing shear\footnote{We adopt complex notation for the two ellipticity and shear components where, for example, $\epsilon = \epsilon_1 + {\rm i} \epsilon_2$ and $\gamma = \gamma_1 + {\rm i} \gamma_2$.} $\gamma$ as
\be
\epsilon^{\rm obs} = (1+m)\left[\epsilon^{\rm int} + \eta + \alpha\epsilon^* + \beta\delta\epsilon^* + c + \gamma \right],
\label{eqn:eobs}
\ee
where $\epsilon^{\rm int}$ is the intrinsic galaxy ellipticity, $\eta$ is the random noise on the measurement, $\epsilon^*$ is the model ellipticity of the image point spread function (PSF), $\delta\epsilon^*$ is the error on the PSF ellipticity model,  $c$ is an additive bias that is uncorrelated with the PSF and $\alpha$, $\beta$ and $m$ are scalars that may vary for different galaxy samples \citep{heymans/etal:2006,paulin-henriksson/etal:2008,jarvis/etal:2016}.

For a perfect shape measurement method, $m=0$, $c=0$, $\alpha=0$ and $\delta\epsilon^*=0$, for all galaxies.  Both DES and KiDS use image simulations to quantify the amplitude and uncertainty of the multiplicative calibration correction $m$ \citep{kannawadi/etal:2019,maccrann/etal:2021}\footnote{Note that in both cases, image simulations are used to quantify \emph{residual} biases, after an initial calibration step (see Section~\ref{sec:datacal} for details).}. They also both quantify the additive calibration correction terms empirically with a significant detection of mean ellipticity in both DES, $\langle \epsilon^{\rm obs}_1 \rangle = (3 \pm 1)\times 10^{-4}$ \citep{gatti/etal:2021}, and KiDS\footnote{In \citet{li/etal:2023b} an anisotropic error in the original likelihood sampler of the \emph{lens}fit shear measurement software was corrected which resulted in a reduction of the KiDS-1000 additive bias to $\langle \epsilon^{\rm obs}_2 \rangle = (3 \pm 1)\times 10^{-4}$.  In the same KiDS-1000 cosmic shear re-analysis, the PSF correction scheme, shear and redshift calibration were also improved following \citet{li/etal:2023, vandenbusch/etal:2022} but the cosmological constraints on $S_8$ were essentially unchanged.  Compared to the analysis of \citet{asgari/etal:2021}, the goodness of fit of the best-fit model was, however, significantly improved with the $\chi^2$ reducing from $\chi^2=88.3$ with $p(\chi^2 > \chi^2_{\rm min} | \nu)=0.05$, to $\chi^2=62.7$ with $p(\chi^2 > \chi^2_{\rm min} | \nu)=0.66$.  As these enhancements to the KiDS-1000 catalogue and calibration were only finalised towards the end this project, we have used the original \citet{giblin/etal:2021} shear catalogue and \citet{hildebrandt/etal:2021} redshift distributions in this analysis.}, $\langle \epsilon^{\rm obs}_2 \rangle = (6 \pm 1)\times 10^{-4}$ \citep{giblin/etal:2021}.  This additive bias is corrected by subtracting the average ellipticity, per tomographic bin, to ensure that the mean shear is zero by definition. Both teams then verify that the mild levels of PSF contamination detected are sufficiently low\footnote{\cite{giblin/etal:2021} find the KiDS-1000 PSF contribution to the observed two point correlation function to be either less than $\sim 2\%$ of the expected cosmic shear signal, and/or within $10\%$ of the expected noise on the measurement at all scales.  \citet{amon/etal:2022} find the DES Y3 PSF contribution to be at most 30\% of the expected cosmic shear signal and limited to small scales for a few tomographic bin pairs.  Both teams analyse a PSF-contaminated mock cosmic shear data vector finding the residual PSF in each shear catalogue induces a $<0.1\sigma$ change in $S_8$ \citep{giblin/etal:2021, amon/etal:2022}.} so as not to bias cosmological parameter constraints (\citealt{giblin/etal:2021};
\citealt{jarvis/etal:2021}; \citealt{gatti/etal:2021}; 
\citealt{amon/etal:2022}). As such, we only correct for the multiplicative and additive shear calibration terms, $m$ and $c$, in Equation~\ref{eqn:eobs}. The resulting corrected shear angular power spectrum $\C^{\rm corr}(\ell)$ then contains signal from both gravitational lensing (G) and the intrinsic (I) alignment of galaxies, 
\be
\label{eqn:c_ee}
\C^{\rm corr}(\ell) = C_{\rm GG}(\ell) + C_{\rm GI}(\ell) + C_{\rm II}(\ell)\;.
\ee
In this section, we summarise the approach taken in the DES (\citealt{amon/etal:2022}; \citealt*{secco/etal:2022}) and KiDS \citep{asgari/etal:2021} cosmic shear analyses to model each component in equation~\ref{eqn:c_ee} and constrain cosmological parameters.  The differing methodology has been rigorously tested by each collaboration to the accuracy required by each survey \citep{joachimi/etal:2021,krause/etal:2021}.   Given the additional constraining power of a joint-survey analysis, we review each survey's analysis choices in Appendices~\ref{app:mocks} and~\ref{app:hybrid} utilising information from recent N-body and hydrodynamical simulations \citep{EuclidEmv2/etal:2021,vandaalen/etal:2020}.  Motivated by our findings, we define a set of unified analysis choices which we adopt in our fiducial Hybrid analysis in Section~\ref{sec:results}.  Here our aim has been to design the most robust pipeline possible for a joint-survey analysis. In the case of some of our Hybrid analysis choices, the decisions are either subjective by nature (for example the choice of priors and parameters), or else there is simply not enough information at present to make an informed decision (for example the complexity of the IA model). For these elements, we have made decisions based on input from the two collaborations, adopting the best available options suited for a joint-survey analysis at this point in time.

\begin{table*}
\centering                                      
\begin{tabular}{llll}          
\toprule
 &  DES Y3 &  KiDS-1000 & Hybrid \\    
\midrule
\multicolumn{3}{l}{\bf Cosmological parameter priors:}\\  
Amplitude & $\As: \bb{0.5,\,5.0}$ & $S_8: \bb{0.1,1.3}$  & $S_8: \bb{0.1,1.3}$\\
Hubble constant & $h: \bb{0.55,0.91}$ & $h: \bb{0.64,0.82}$ & $h: \bb{0.64,0.82}$\\
Matter density& $\Om: \bb{0.1,0.9}$ & $\omega_{\rm c}: \bb{0.051,0.255}$ & $\omega_{\rm c}: \bb{0.051,0.255}$\\
Baryon density& $\Ob: \bb{0.03,0.07}$ & $\ob: \bb{0.019,0.026}$ & $\ob: \bb{0.019,0.026}$\\
Spectral index & $\ns: \bb{0.87,\,1.07}$ & $\ns: \bb{0.84,\,1.1}$ & $\ns: \bb{0.84,\,1.1}$\\
Neutrinos & $1000\,\Omega_\nu h^2: \bb{0.6,6.44}$ & $\Sigma m_\nu = 0.06 {\rm eV}$  & $\Sigma m_\nu =  \bb{0.055,0.6} {\rm eV}$\\
\midrule      
\multicolumn{3}{l}{\bf Astrophysical systematic models and priors:}\\                             
Intrinsic Alignments & TATT:  $b_{\rm TA}: \bb{0,2}$; $a_1, a_2, \eta_1, \eta_2: \bb{-5,5}$  &  NLA: $A_{\rm IA}: \bb{-6,6}$ &   NLA-z: $A_{\rm IA}, \eta_{\rm IA}: \bb{-5,5}$ \\
Non-linear Model & {\sc Halofit} & {\sc HMCode2016} & {\sc HMCode2020} \\
Baryon Feedback &  {\rm Scale cuts} & $A_{\rm bary}:\bb{2,3.13}$  &  {\rm Scale cuts} \& $\log_{10}(T_{\rm AGN}/{\rm K}):\bb{7.3,8.0}$ \\
Neutrino Model & \citet{bird/etal:2012} & {\sc HMCode2016} &  {\sc HMCode2020} \\
\midrule
\multicolumn{3}{l}{\bf Sampling Algorithm:}\\                                
 &{\sc PolyChord} & {\sc MultiNest}  &{\sc PolyChord}\\
\bottomrule
\end{tabular}
\caption{Comparison of the modelling choices in the DES-like, KiDS-like and Hybrid cosmic shear analyses:  cosmological parameters, astrophysical systematic models and the chosen sampling algorithm. The values in square brackets are the limits of the adopted top-hat priors.  The listed cosmological parameters are: the amplitude of the primordial power spectrum of scalar density fluctuations, $A_{\rm s}$;  the Hubble constant, $h=H_0/(100\,{\rm km}\,{\rm s}^{-1}\,{\rm Mpc}^{-1})$; the matter density, $\Omega_{\rm m}$; the baryon density, $\Omega_{\rm b}$;  the scalar spectral index, $n_{\rm s}$; the neutrino mass density parameter, $\Omega_\nu$; $S_8=\sigma_8 \sqrt{\Om/0.3}$, where $\sigma_8$ is the linear-theory standard deviation of matter density fluctuations in spheres of radius $8 \Mpc$; $\omega_{\rm c} = \Omega_{\rm c}h^2$, where $\Omega_{\rm c}$ is the cold dark matter density; $\omega_{\rm b} = \Omega_{\rm b}h^2$; and the sum of the neutrino masses, $\Sigma m_\nu$. } 
\label{tab:hybrid}           
\end{table*}

\subsection{Non-linear matter power spectrum}
The DES and KiDS teams both choose to relate the convergence power spectrum, $C_{\rm GG}(\ell)$, to the non-linear matter power spectrum $P_{\rm \delta}(k,z)$, using a modified flat sky Limber approximation \citep{loverde/afshordi:2008,kilbinger/etal:2017}
\begin{align}
\label{eq:limber}
 C_{\rm GG}^{(ij)}(\ell) = \int_0^{\chi_\mathrm{hor}}\d\chi\:\frac{W^{(i)}(\chi)W^{(j)}(\chi)}{f_{\rm K}^2(\chi)} 
   P_{\rm \delta} \left(\frac{\ell+1/2}{f_\mathrm{K}(\chi)},\chi\right)\;.
\end{align} 
Here $f_{\rm K}(\chi)$ is the comoving angular diameter distance which simplifies to $\chi$, the radial comoving distance, for a spatially flat Universe, and the integral is taken to the horizon, $\chi_{\rm hor}$. The kernel, $W$, depends on the redshift distribution of the correlated tomographic populations, $i$ or $j$, with the mathematical form found in equations 6.19 and 6.22 of \citet{bartelmann/schneider:2001}.  

Each team calculates the linear matter power spectrum using {\sc CAMB} \citep{CAMB}, and uses a halo model approach, calibrated with numerical simulations, to determine the non-linear correction.   The details of this correction, however, differ.  The DES team adopts the \citet{smith/etal:2003} {\sc Halofit} fitting formula, re-calibrated by \citet{takahashi/etal:2012}, where the impact of a non-zero neutrino mass is modelled using the {\sc CAMB}-version of the \citet{bird/etal:2012} fitting formula\footnote{We refer the reader to appendix B of \citet{mead/etal:2021} for a discussion of the differences between the \citet{bird/etal:2012}, {\sc HMCode2016} and {\sc HMCode2020} non-linear models which diverge for neutrino masses $\Sigma m_{\nu} \gtrsim 0.3$eV and scales $k \gtrsim 0.1 h {\rm Mpc}^{-1}$.  In this `heavy' neutrino regime, the different techniques used to include massive neutrinos in numerical simulations lead to significant differences in the high-$k$ scales of the non-linear matter power spectrum. Both \citet{bird/etal:2012} and {\sc HMCode2016} are calibrated using simulations created with neutrinos as a separate particle species.  The two sets of simulations differ, however, in their approach to the injection of these particles.  {\sc HMCode2020} is calibrated using {\sc MiraTitan} \citep{heitmann/etal:2016}, where `linear neutrinos' are included in the evolution of the background and the gravitational potential, but there is no back-reaction on the neutrinos from the simulation particles.  There is no community consensus on the simulation technique that provides the most accurate estimate of the non-linear matter power spectrum.  Most recently the linear approach has been adopted for the {\sc EuclidEmulatorv2} simulations, and a particle-based approach has been adopted for the {\sc MillenniumTNG} \citep{hernandez-aguayo/etal:2022} and {\sc Aemulus} $\nu$ simulations \citep{derose/etal:2023}.  This issue is largely academic, however, as the impacted $k$-scales are the same scales where there is significant uncertainty on the non-linear suppression of power caused baryon feedback \citep[see for example][]{harnois-deraps/etal:2015}.}. The KiDS team adopts the {\sc HMCode2016} model from \citet{mead/etal:2015}, with the appropriate extensions for non-zero neutrino mass \citep{mead/etal:2016}.  In our Hybrid analysis we adopt {\sc HMCode2020}, an updated version of this model, which delivers improved accuracy at the level of $<2.5\%$ to $k  < 10 \, h\, \mathrm{Mpc}^{-1}$ \citep[see][and Appendix~\ref{app:NLmodels}]{mead/etal:2021}.

\subsection{Intrinsic galaxy alignments}
\label{sec:IA}
The intrinsic alignment (IA) of galaxies with their local environment can mimic weak lensing and therefore this phenomenon is included as an astrophysical systematic uncertainty in all cosmic shear analyses, with a range of different analytical models proposed to mitigate this effect \citep[see][and references therein]{troxel/ishak:2015,joachimi/etal:2015,kiessling/etal:2015}.   

The KiDS team adopts the non-linear linear-alignment model (NLA) which describes the linear tidal alignment of galaxies with the density field \citep{hirata/etal:2004}.  This model also includes an ad hoc non-linear correction to the linear matter power spectrum \citep{bridle/king:2007}, as motivated by intrinsic alignment measurements in numerical simulations and the Sloan Digital Sky Survey \citep{heymans/white/etal:2006,hirata/etal:2007}.  The fiducial KiDS analysis allows for only one free nuisance parameter\footnote{We hereafter refer to the KiDS fiducial NLA parametrisation, with one free $A_{\rm IA}$ parameter, as NLA (no-z).}, $A_{\rm IA}$, modulating the amplitude of the intrinsic alignment model (see equations 3-5 in \citealt{bridle/king:2007} for the NLA intrinsic alignment power spectra, $C_{\rm GI}$ and $C_{\rm II}$). The NLA model can also include flexibility where the amplitude of the IA signal depends on the average luminosity of each tomographic sample \citep{joachimi/etal:2011}.  More commonly used, however, is an NLA model amplitude that evolves with redshift, using a power-law with $[(1+z)/(1+z_{\rm pivot})]^{\eta_{\rm IA}}$, hereafter referred to as the NLA-z model\footnote{In this analysis we use $z_{\rm pivot}=0.62$.}.

The DES team adopts the Tidal Alignment and Tidal Torquing model \citep[TATT,][]{blazek/etal:2019}, which extends the linear alignment model with the inclusion of a tidal torquing alignment mechanism.   The fiducial DES analysis allows for five free nuisance parameters: a tidal alignment amplitude, $a_1$, with redshift evolution, $\eta_1$; a tidal torquing amplitude, $a_2$, with redshift evolution, $\eta_2$; a linear bias amplitude, $b_{\rm TA}$ \citep[see equations 37-39 in][for the TATT intrinsic alignment power spectra, $C_{\rm GI}$ and $C_{\rm II}$, highlighting the expectation of a non-zero B-mode from the `II' term]{blazek/etal:2019}. In the limit $a_2$, $\eta_2$, $b_{\rm TA} \rightarrow 0$, the TATT model reduces to the NLA-z model with $a_1 = A_{\rm IA}$ and $\eta_1 = \eta_{\rm IA}$.

Both the NLA and TATT models are found to provide a sufficiently good fit to numerical simulations (\citealt*{secco/etal:2022}; \citealt{hoffmann/etal:2022}).  Currently there is no observational evidence to support the use of one model over the other in two-point cosmic shear analyses.   For our Hybrid analysis we choose to adopt the NLA-z model. In terms of complexity, this is between the original choices of the two survey teams.  In our fiducial Hybrid analysis we adopt independent IA parameters for each survey to reflect the different survey depths, selection functions, shape measurement methods and photometric redshift errors which can be absorbed by the IA model (for further discussion, see Appendix~\ref{sec:J2IA}).

\subsection{Baryon feedback}
\label{ssec:BF}
Studies of hydrodynamical simulations find significant differences in the small-scale, $k>0.1 h {\rm Mpc}^{-1}$, total matter distribution relative to dark matter-only simulations, as a result of baryon cooling, star formation and active galactic nuclei (AGN) feedback \citep[see][and references therein]{chisari/etal:2019}.   Theoretical models of the non-linear matter power spectrum, $P_{\delta}(k,z)$, that have been calibrated using dark matter-only simulations are therefore inaccurate at high-$k$ \citep{white/etal:2004,semboloni/etal:2011}.  There is, however, a large degree of uncertainty on the scale, amplitude and redshift dependence of baryon feedback.

To account for this uncertainty the KiDS team adopts the {\sc HMCode} non-linear matter power spectrum $P_{\rm \delta}(k,z)$ model, which incorporates uncertainty from baryon feedback through a single free parameter.  This parameter scales the halo concentration and the stellar and gas content, leading to a modification in the overall amplitude and shape of the `one-halo' term in the halo model.   \cite{asgari/etal:2021} utilise the \citet{mead/etal:2016} version of {\sc HMCode}, calibrated with the {\sc OWLS} hydrodynamical simulations \citep{vandaalen/etal:2011}.  \citet{troester/etal:2021} present a KiDS-1000 cosmic shear band power spectrum re-analysis using {\sc HMCode2020}, calibrated with the updated\footnote{In Appendix~\ref{app:KiDSCOSEBIs} we find a $0.19\sigma$ change in $S_8$ when changing from {\sc HMCode2016} to {\sc HMCode2020} in a KiDS-like reanalysis of the KiDS-1000 COSEBIs data vector.} {\sc BAHAMAS} hydrodynamical simulations \citep{mccarthy/etal:2017,vandaalen/etal:2020}.

DES take a different approach to mitigating baryon feedback uncertainty in their analysis, eliminating data on scales which are expected to be impacted significantly.   Taking the `worst-case scenario' for the extent of the impact as the AGN model from the suite of {\sc OWLS} hydrodynamical simulations \citep{vandaalen/etal:2011}, \citet{krause/etal:2021} contaminate a mock \lcdm cosmic shear data vector.   Small-scale information is progressively removed until a dark matter-only cosmological analysis biases the recovered cosmology from the {\sc OWLS} mock with a maximum bias of $0.3\sigma_{\rm 2D}$ in the $\Om-S_8$ 2D plane. For the DES Y3 analysis, the `fiducial' scale cuts are defined for a $w$CDM analysis of the joint cosmic shear, galaxy-galaxy lensing and galaxy clustering likelihood \citep{3x2ptDES/etal:2021}. This probe combination is hereafter referred to as $3\times2$pt. In this analysis we adopt the DES Y3 alternative `$\Lambda$CDM-optimised' scale cuts, which allow for the inclusion of smaller scale cosmic shear information while satisfying that the predicted baryon feedback bias is $<0.14\sigma_{\rm 2D}$ (\citealt{amon/etal:2022}; \citealt*{secco/etal:2022}). These scale cuts were shown to be robust at the level of a few percent against a range of hydrodynamic simulations (see figure 5 and section G2 in \citealt*{secco/etal:2022}). 

In our Hybrid analysis we combine the two survey strategies, adopting both the DES-Y3 `$\Lambda$CDM-optimised' and equivalently defined scale cuts for KiDS (see Appendix~\ref{app:scalecuts}), along with the marginalisation over a free baryon feedback parameter in the cosmological analysis.  For our Hybrid non-linear model choice, {\sc HMCode2020}, the free parameter $T_{\rm AGN}$ maps to the {\sc BAHAMAS}-defined heating temperature of the AGN\footnote{Note that $T_{\rm AGN}$ is not a physical parameter.  The hydrodynamical {\sc BAHAMAS} simulations inject black hole seed particles into halos which then grow via gas accretion and mergers.  The mass energy of the accreted gas heats neighbouring gas particles by $\Delta T_{\rm heat}$.  Out of the many free parameters for this AGN feedback recipe, changes in $\Delta T_{\rm heat}$ were found to introduce the greatest impact to the resulting simulations \citep{lebrun/etal:2014}.  The single-parameter baryon feedback model implemented within {\sc HMCode2020} is calibrated using three {\sc BAHAMAS} simulations with $\Delta T_{\rm heat} = 10^{7.6}$K, $10^{7.8}$K and $10^{8.0}$K.  Here a linear relationship is fit between $\log_{10}(\Delta T_{\rm heat}/{\rm K})$ and each of the six halo-model parameters that are sensitive to feedback.  Variation in the strength of baryon feedback is then modulated within {\sc HMCode2020} using one parameter, denoted $T_{\rm AGN}$, which modifies all six halo parameters in line with the changes seen in {\sc BAHAMAS} \citep[see section 6.2 and 6.3 of][for details]{mead/etal:2021}.} which modulates the strength of the baryon feedback suppression of the matter power spectrum\footnote{During the course of this work we isolated unusual behaviour at low-$k$ for {\sc HMCode2020} predictions of feedback models outwith the {\sc BAHAMAS} range where {\sc HMCode2020} was calibrated; $\log_{10}(T_{\rm AGN}/{\rm K})= \bb{7.6,8.0}$.  This arose from the unphysical behaviour of the one-halo term on these scales which is magnified when the feedback parameter is significantly raised.  The {\sc HMCode2020} software has since been updated within \href{https://github.com/cmbant/CAMB}{\sc CAMB} v1.4.0 and \href{https://github.com/alexander-mead/HMcode-python}{\sc HMCode-python} manually suppressing this unwanted effect, following \citet{mead/etal:2015}.   This update was not available when we were conducting this analysis, but we have since verified that it does not introduce any significant change in our results.  Specifically $S_8$ changes by only $0.02\sigma$.}. We adopt a tophat prior on $\log_{10}(T_{\rm AGN}/{\rm K})$ with the range $\bb{7.3,8.0}$.  This prior range is allowed by a range of observational constraints on baryon feedback suppression: small-scale cosmic shear analyses \citep{harnois-deraps/etal:2015,yoon/jee:2021,chen/etal:2023,arico/etal:2023}; joint analyses of KiDS-1000 cosmic shear with shear-thermal Sunyaez Zel'dovich (SZ) cross-correlation measurements \citep{troester/etal:2022} or with X-ray cluster gas fractions and kinematic SZ gas profiles \citep{schneider/etal:2022}; DES Y3 shear-thermal SZ cross-correlation measurements \citep{pandey/etal:2023}.  Whilst stronger feedback models are permitted by these observations, we choose to set the upper limit\footnote{We refer the reader to \citet{yoon/jee:2021,amon/efstathiou:2022,preston/etal:2023} for DES and KiDS cosmic shear studies where more extreme AGN feedback models are explored.} for our baryon feedback prior informed by the maximum feedback strength with which the {\sc BAHAMAS} simulations produce a baryon fraction at group scale that is consistent with observations \citep{vandaalen/etal:2020}. The lower limit is set to reproduce cosmic shear predictions that are equivalent to a dark matter only model (see Appendix~\ref{app:hybrid}).  Whilst this low level of feedback cannot reproduce the observed group-scale baryon fraction in {\sc BAHAMAS}, we choose to retain a dark matter only power spectrum in our model space for consistency with the {\sc HMCode2016} prior range used in \citet{asgari/etal:2021} and the dark matter power spectrum used in \citet{amon/etal:2022}; \citet*{secco/etal:2022}.  We verify the accuracy and discuss the benefits of this combined-approach in Appendix~\ref{app:hybrid}.

\subsection{Cosmic shear statistic}
\label{sec:2ptstats}
DES present cosmological constraints from measurements of the real-space two-point shear correlation functions $\xi_\pm(\theta)$, utilising the full spherical sky expression\footnote{This full sky expression is more accurate for large area surveys compared to the flat-sky Hankel transform approximation used in previous studies.} to relate these statistics to the cosmic shear power spectrum, $\C(\ell)$,
\be
\xi_\pm (\theta) = \frac{1}{4\pi} \sum_{\ell=2}^{\infty} (2\ell+1) \, d^{\ell}_{2,\pm2}(\theta) \left[C^{\rm EE}_{\epsilon \epsilon}(\ell) \pm C^{\rm BB}_{\epsilon \epsilon}(\ell) \right]  \, .
\ee
Here $d^{\ell}_{m,n}$ are the reduced Wigner D-matrices \citep{chon/etal:2004,kilbinger/etal:2017}, and the cosmic shear power spectrum has been separated into its E- and B-modes.  Any significant B-mode component would originate from data-related systematics and/or the intrinsic alignment signal, $C_{\rm II}$.   The $\xi_\pm(\theta)$ estimator, given in equation 10 of \citet{amon/etal:2022}, is used to measure the auto and cross-correlation between 4 tomographic bins using twenty angular logarithmic bins spanning $2.5 < \theta \leq 250.0$ arcmin, although not all $\theta$-bins are used in the analysis. Tomographic bins are defined using the {\sc SOMPZ} method (\citealt{buch/etal:2019}; \citealt*{myles/etal:2021}). 

KiDS present cosmological constraints from measurements of complete orthogonal sets of E/B-integrals (COSEBIs), which cleanly separate the E- and B-mode signals.  The COSEBIs, $E_n$ and $B_n$, are discrete values which can be estimated by integrating over finely binned $\xi_\pm$ measurements \citep[see equation 7 of][]{asgari/etal:2021}.  They are related to the cosmic shear power spectrum as
\be
\label{eqn:cosebis}
E_n = \int_0^{\infty} \frac{\d\ell\,\ell}{2\pi}C^{\rm EE}_{\epsilon \epsilon}(\ell)\,W_n(\ell)\;,
\ee
with $B_n$ following the same form.  Here the weight function, $W_n$, depends on the angular range that can be accessed from the data, which KiDS define as $0.5 < \theta \leq 300.0$ arcmin, and serves to limit the effective $\ell$-range entering the cosmic shear analysis \citep[see section 2.2 of][for details]{asgari/etal:2021}.   In Appendix~\ref{app:scalecuts}, we restrict this angular range for the DES-like and Hybrid joint-survey analyses, following the baryon feedback mitigation strategy of \citet{krause/etal:2021}.  KiDS define five tomographic bins using {\sc BPZ} photometric redshifts between $0.1<z_{\rm phot} \leq 1.2$. \citep*{benitez:2000,hildebrandt/etal:2021}

The fiducial DES cosmic shear analyses combine $\xi_\pm(\theta)$ with additional data from the shear ratio (SR) statistic (\citealt*{sanchez/etal:2022}). \citet{amon/etal:2022} demonstrate the importance of including this extra observable to better constrain the parameters of the TATT model.  With an NLA-z analysis, however, the $S_8$ constraining power is unchanged by the inclusion of the SR data.  Given our Hybrid analysis choice for the NLA-z  IA model, we choose not to include the additional SR data in this joint-survey analysis for pragmatic reasons: a new shear ratio measurement would have otherwise been required for KiDS; we wished to retain the cosmic shear-only nature of the survey comparison.

In addition to the fiducial statistics analysed by each collaboration, there are a range of alternative two-point cosmic shear statistics that have been studied in both real and Fourier space \citep[see, for example][]{asgari/etal:2021,doux/etal:2022,loureiro/etal:2021,schneider/asgari/etal:2022}.   Through a study of noisy mock surveys, \citet{asgari/etal:2021,hamana/etal:2022} find that $\sim 15\%$ of the time, shot noise can lead to $>1\sigma$ offsets in $S_8$ when comparing constraints from a range of two-point statistical analyses of the same data set.   Throughout this analysis we have therefore chosen to retain each survey's fiducial statistic of choice with a joint-survey data vector composed of the $\xi_\pm(\theta)$ tomographic measurements from DES combined with the $E_n$ tomographic measurements from KiDS.  By fixing each statistic in this way, we can directly compare how a change in the analysis framework influences each survey's published $S_8$ constraint without having to also quantify the impact of noise on an alternative two-point statistic (although see Appendix~\ref{app:KiDSCOSEBIs} where this complication nevertheless arises as a result of modifying the angular range for KiDS).

\subsection{Data calibration uncertainty}
\label{sec:datacal}
The shear estimation methods for KiDS and DES differ. The DES team adopts a {\sc Metacalibration} Gaussian model fit \citep{huff/mandelbaum:2017,sheldon/huff:2017} which involves artificially shearing and remeasuring galaxies in the real survey images to calibrate the response of each galaxy to shear.  This approach mitigates both noise bias, the dominant source of bias in shear measurement \citep{melchior/viola:2012, refregier/etal:2012}, and model bias \citep{voigt/bridle:2010}. The KiDS team adopts a `self-calibrating' bulge-disk model fit \citep[\emph{lens}fit:][]{miller/etal:2013}, whereby the initial best-fit galaxy model is simulated with the appropriate noise level,  and re-run through the measurement pipeline. The per-galaxy noise bias calibration correction is then given by the difference between the input and measured ellipticity \citep{fenechconti/etal:2017}. 

In order to reach the required percent-level accuracy, both survey teams quantify residual biases using image simulations built from HST-COSMOS observations \citep{kannawadi/etal:2019,maccrann/etal:2021, li/etal:2023}. These studies highlight the importance of including realistic fractions of blended objects in image simulations for accurate shear calibration. The DES Y3 calibration included the first multi-band image simulation study for lensing.   This approach allows for the replication of the redshift calibration process, thereby also mitigating the impact of blending on redshift calibration. The coupled calibration errors identified between the {\sc Metacalibration} shear estimates and phenotype redshift distributions are mitigated in the DES Y3 analysis with an image-simulation calibrated effective redshift distribution for the tomographic source samples. 

DES account for the uncertainty on their shear calibration by marginalising over four independent free parameters, $m_{i=1..4}$, adopting Gaussian priors (see Table~\ref{tab:params}). KiDS differs\footnote{We note that both teams explore how their method to account for shear calibration errors impacts their results. \citet{asgari/etal:2021} find a $0.2\sigma$ increase in $S_8$ when adopting uncorrelated free $m$-parameters with Gaussian priors of width $\sigma_m^{i}$.  The `full blending treatment' analysis in \citet{amon/etal:2022} includes the correlation between shear calibration parameters as measured with the \citet{maccrann/etal:2021} multi-wavelength image simulations.  For this analysis they find their constraints are indistinguishable from the fiducial result.}, assuming 100\% correlation, between the tomographic bins, for the uncertainty in the calibration correction, $\sigma_m^{i=1..5}$.  The correlated uncertainty is then subsumed into the analytical covariance for the cosmic shear data vector \citep[see equation 37 in][]{joachimi/etal:2021}.

The primary redshift estimation methods for KiDS and DES both utilise a SOM approach. KiDS make exclusive use of spectroscopic training sets, creating a `gold' sample of photometric galaxies that are sufficiently well represented in the training sample \citep{wright/etal:2020}.  DES supplements their spectroscopic training sample with very high-quality photometric redshifts\footnote{See also \cite{vandenbusch/etal:2022} who take this approach with an updated analysis of KiDS-1000.} and include magnitude limits to ensure sufficient coverage between the training samples and the Deep Field data \citep*{myles/etal:2021}. The DES approach also relies on an estimation of the survey transfer function using {\sc BALROG} \citep{suchyta/etal:2016}, which injects simulated galaxies with photometry drawn from the Deep Field into real Wide Field images \citep*{everett/etal:2022}.

KiDS utilise a multi-band mock galaxy survey to estimate the uncertainty on the mean redshift of each tomographic bin \citep{vandenbusch/etal:2020}, verified through a cross-correlation analysis \citep*{hildebrandt/etal:2021}.   The correlated uncertainty between the bins is accounted for in the cosmological analysis using five free parameters drawn from a five dimensional multivariate Gaussian prior ${\cal N}(\vek{\mu}_z;\vek{C}_ z)$ \citep[for details see section 3.3 of][]{joachimi/etal:2021}.  

The fiducial DES cosmic shear analysis accounts for uncertainty on the mean redshift in a similar way, using four independent free parameters, $\Delta z_{i=1..4}$ with Gaussian priors\footnote{See also alternative analyses in \citet{amon/etal:2022,stolzner/etal:2021} which account for uncertainty in the full shape of the tomographic redshift distributions. Here the DES team use the {\sc Hyperrank} method \citep{cordero/etal:2021}, sampling over multiple realisations of the SOM-calibrated distributions, and the KiDS team use a flexible Gaussian mixture model.}.  The DES prior is set by combining information from: sample variance in the SOM training data; the Deep Field  photometric calibration error \citep*{hartley/etal:2021}; the systematic error related to the choice of training sample and the uncertainty estimated from the \citet{maccrann/etal:2021} multi-band image simulation analysis; a cross-correlation analysis \citep*{gatti/etal:2022}.  For details see section 5.6 of \citet*{myles/etal:2021}. We caution against direct comparisons of the prior widths for the calibration corrections listed in Table~\ref{tab:params} as this neglects the nuances in the different approaches.  In DES, the adopted formalism to determine the calibration of redshift-dependent biases including the impact of blending, also absorbs additional shear calibration uncertainty.  In KiDS, the uncertainty on both the shear and redshift calibration is included as a correlated error across tomographic bins. 

Throughout this analysis, we retain the survey-specific methodology to account for data calibration uncertainty in each section of the joint-survey data vector.  Revisiting the studies that establish the robustness of these choices is beyond the scope of this analysis. The DES pipeline was modified in order to analyse KiDS data using a five-dimensional multivariate Gaussian $\Delta_z$ prior (hereafter a DES-like analysis), and the KiDS pipeline was modified in order to analyse DES data using four free shear calibration parameters, $m_i$, with Gaussian priors (hereafter a KiDS-like analysis). The Hybrid pipeline also permits survey-specific data calibration nuisance parameters and priors.  

\vspace{1cm}
\subsection{Cosmological parameter inference}
\label{sec:inference}

Sampling of the posterior is carried out using {\sc Polychord}\footnote{{\sc\,Polychord}: \url{https://github.com/PolyChord/}} \citep{handley/etal:2015} for DES Y3 and {\sc Multinest}\footnote{{\sc\,Multinest}:\,\url{https://github.com/farhanferoz/MultiNest}} \citep{feroz/etal:2009} for KiDS-1000.  \citet*{lemos/etal:2022} demonstrate that for the DES first year (hereafter Y1) $3\times2$pt analysis, {\sc Multinest} systematically underestimates the 68\% credible intervals for $S_8$, at the level of $\sim 10\%$.  We revisit this study using the DES Y3 cosmic shear likelihood and the KiDS-chosen {\sc Multinest} settings in Appendix~\ref{app:samplers}, defining the true posterior as that estimated using two Markov Chain Monte Carlo (MCMC) algorithms.  For the {\sc Polychord} sampler we find an accurate recovery of both the 68\% and 95\% credible intervals.  With {\sc Multinest}, however, we find a $\sim 12\% (15\%)$ underestimate of the 68\% (95\%) credible intervals.  These findings are independent of the choice of IA model.   Given the enhanced performance of {\sc Polychord} over {\sc Multinest}, we adopt this sampler for our Hybrid analysis.  This choice is also beneficial in terms of estimating the Bayesian evidence, where \textsc{MultiNest} estimates have been shown to be biased in all but the strictest of settings \citep*{lemos/etal:2022}.   We note that there is a significant extra cost in terms of computational time when using {\sc Polychord}, supporting the development and future use of time-saving measures such as analytical marginalisation over nuisance parameters \citep{ruiz-zapatero/etal:2023,hadzhiyska/etal:2023}, likelihood emulators \citep{spurio-mancini/etal:2022} and neural network assisted sampling techniques\footnote{See for example {\sc Nautilus}: \url{https://github.com/johannesulf/nautilus}}.

Table~\ref{tab:hybrid} lists the cosmological parameter priors chosen by each survey for the flat-\lcdm analysis. In contrast to the DES analysis, the KiDS team fixes the neutrino mass, $\Sigma m_\nu$, at the minimum mass allowed by oscillation experiments and chooses more informative priors on the Hubble constant, $h$. The survey teams also differ over their choice of parameter to marginalise over the amplitude of the matter power spectrum. The DES team chooses to sample using the amplitude of the primordial power spectrum of scalar density fluctuations $\As$, and the KiDS team chooses to sample in $S_8$ (see figures 15 and 16 in \citealt{joachimi/etal:2021} and figure 17 in \citealt*{secco/etal:2022} to visualise how these different parameter and prior choices inform the multi-dimensional parameter space\footnote{\citet{sugiyama/etal:2020} provide a novel solution to the question of optimal parameter choice.  They derive a correction weight which converts samples from a chain using flat priors in $(10^{10}A_{\rm s}, \Omega_{\rm m})$ such that the weighted chain has, to first order, the desired flat $(S_8, \Omega_{\rm m})$ priors that are automatically delivered when adopting $S_8$ sampling.  Using this approach \citet{li/zhang/etal:2023} reweigh {\sc Polychord} chains to recover constraints in the $(S_8, \Omega_{\rm m})$ plane that are identical, irrespective of the chosen amplitude sampling parameter: $A_{\rm s}$, $\ln A_{\rm s}$ or $S_8$.  \citet{dalal/etal:2023} argue that with the use of correction weights, adopting $A_{\rm s}$ priors allows for the most efficient sampling of the posterior.}).

When reporting the headline cosmological parameter constraints, \citet{amon/etal:2022}; \citet*{secco/etal:2022} report the mean of the 1D marginal distribution, along with a credible interval that encompasses 68\% of the marginal highest posterior density. \citet{asgari/etal:2021} report the maximum a posteriori (MAP) and an associated 68\% credible region given by the projected joint highest posterior density region \citep[PJ-HPD, see section 6.4 of][for details]{joachimi/etal:2021}.   
In addition to these headline results, \citet{amon/etal:2022}; \citet*{secco/etal:2022} also tabulate the MAP, and \citet{asgari/etal:2021} also tabulate the maximum marginal and associated credible intervals\footnote{We note that the maximum marginal statistic is adopted by HSC for their headline result \citep[see the discussion in][]{li/zhang/etal:2023}.}.  Unfortunately, there are issues related to all three approaches.   The MAP is notoriously challenging to determine requiring significant computational resources to sufficiently decrease the noise on the estimate \citep{muir/etal:2020,joachimi/etal:2021}.   The accurate determination of the corresponding PJ-HPD errors also requires densely sampled chains in order to determine the 68\% credible region around the MAP.  The marginal constraints for multi-dimensional posteriors are subject to projection effects which are known to offset the recovered parameters from the input truth \citep[see for example][and Appendix~\ref{app:projection}]{krause/etal:2021,joachimi/etal:2021,chintalapati/etal:2022}. Whilst the mean marginal estimate is well defined, for skewed posterior distributions the mean can be far from the MAP.  Whilst the maximum marginal is typically closer to the MAP, its estimate depends on the choice of smoothing methodology and scale.  For example, using the {\sc chainconsumer}\footnote{{\sc chainconsumer}:\,\url{https://samreay.github.io/ChainConsumer}} Gaussian kernel density estimation \citep{hinton:2016}, leads to $\sim 10\%$ larger errors for the maximum-marginal constraint, compared to the mean-marginal constraint (see Table~\ref{tab:pipelinemods}).

In our Hybrid pipeline we choose to sample over the KiDS choice of cosmological parameters: $\omega_{\rm c}$, $\omega_{\rm b}$, $h$, $n_{\rm s}$, $S_8$, with the addition of a free neutrino mass parameter, $\sum_\nu m_\nu$. This is a subjective choice that we found helped to minimise projection effects when adopting the NLA model (see Appendix~\ref{app:projection}).  We use the KiDS-like priors with the addition of the DES-like prior on the neutrino mass\footnote{In order to use the {\sc CAMB} python module, we convert the DES prior on $1000 \Omega_\nu h^2 = \bb{0.6,6.44}$ to $\Sigma m_\nu =  \bb{0.055,0.6} {\rm eV}$ using the approximation $\Omega_\nu h^2 = \Sigma m_\nu / 93.4\, {\rm eV}$.}. For our primary parameter $S_8$ we report the maximum and mean marginals along with the MAP and PJ-HPD as determined from oversampled {\sc Polychord} chains\footnote{The oversampled {\sc Polychord} chains contain ten times the number of original sampling points.}.  For all other parameters we report constraints using the standard mean marginal statistic.

\subsection{Goodness of fit}
\label{sec:GoF}
In this analysis we follow \citet{joachimi/etal:2021} to estimate a goodness of fit statistic, $p(\chi^2 > \chi^2_{\rm min} | \nu)$, the probability that a weighted least-square, $\chi^2$, will exceed the measured minimum $\chi^2_{\rm min}$, assuming a $\chi^2$-distribution with $\nu$ degrees of freedom, where $\nu = N_{\rm data} - N_{\Theta}$.  Here, $N_{\rm data}$ is the number of data points\footnote{For the DES, KiDS and joint-survey data vectors, $N_{\rm data}=273, 75, 348$ respectively.}, and $N_{\Theta}$ is the effective number of parameters, which for a prior-informed cosmic shear analysis with correlated sampling parameters, is smaller than the total number of free parameters.  $N_{\Theta}$ is estimated from the average of a likelihood-based and posterior-based estimate \citep[see equations 45 and 46 of][]{joachimi/etal:2021}.  This approach was found to reproduce an unbiased estimate of the true $N_{\Theta}$, on average, with a standard deviation $\sigma \simeq 0.2 N_{\Theta}$, as determined through the analysis of 100 noisy mock KiDS-1000 cosmic shear data vectors (see appendix B.2 of \citealt{asgari/etal:2021}, and section 6.3 of \citealt{joachimi/etal:2021}).  

KiDS define an acceptable fit as $p \ge 0.001$, a $3.1\sigma$ event \citep*{heymans/etal:2021}.  DES define an acceptable fit as $p \ge 0.01$, a $2.3\sigma$ event \citep{3x2ptDES/etal:2021}.  In Section~\ref{sec:results} we show that all data sets and analysis setups recover a fit that meets these criteria.  

We note that the \citet{joachimi/etal:2021} method to determine the goodness of fit is slightly more conservative than an alternative approach adopted, for illustrative purposes, in \citet{amon/etal:2022}; \citet*{secco/etal:2022}.  In that analysis $N_{\Theta}$ was estimated using a `Gaussian linear model' \citep[see equation 29 of][]{raveri/hu:2019}, finding $N_{\Theta} = 5$ for the DES-like analysis.  With the \citet{joachimi/etal:2021} approach we find $N_{\Theta} = 6.7$ for the same setup (for more details see Appendix~\ref{app:extras}).  Neither approach is, however, as accurate as adopting a posterior predictive distribution (PPD) goodness of fit estimate \citep{kohlinger/etal:2019, doux/etal:2021}, which removes the assumption that the distribution of weighted least-squares, between noisy data realisations and the model, follows a $\chi^2$ distribution.  \citet{3x2ptDES/etal:2021} implement this preferred, but more computationally expensive PPD goodness of fit estimate, finding $p=0.21$ for the fiducial DES-Y3 cosmic shear data vector.  Comparing this to the \citet*{secco/etal:2022} Gaussian linear model estimate of $p=0.22$, for the same data vector, gives us confidence in utilising these faster goodness of fit estimates for this analysis.

\subsection{Consistency and tension metrics}
\label{sec:tension}

When comparing cosmological parameter constraints from different surveys or probes, there are a range of different statistical tools to define the degree of consistency, or inconsistency. These can be grouped into methods that focus on differences in a single parameter or in multiple model parameters (parameter-space methods), methods that quantify differences in the data vector space, and methods that summarise the full likelihood or posterior into a single metric, such as the Bayes factor (see appendix G of \citealt*{heymans/etal:2021}, \citealt{lemos/etal:2021}, and references therein).   In the KiDS-1000 and DES Y3 analyses, a wide range of consistency results are presented, both for internal consistency tests and external comparisons with the CMB constraints from \citet{planck/etal:2020}.  Each method tells us about a different aspect of how well the model and the different sections of the data match or are in tension with each other. It is therefore beneficial and prudent to consider more than one method.   

Beyond the methodological choices, there is also a decision to be made on each metric's threshold where data sets are considered to be consistent or inconsistent with each other.  In this analysis we follow \citet{3x2ptDES/etal:2021} who consider there to be evidence of inconsistency between probes when a tension metric results in a probability-to-exceed $p < 0.01$, corresponding to a $> 2.3\sigma$ event.

We limit this analysis to three complementary tension estimators motivated by the consistency methodologies previously tested by each survey.  For the single-parameter tension metric we adopt the Hellinger distance, $d_{\rm H}$, \citep[see for example][]{beran:1977}, to compare the overlap between two 1D probability density distributions, $q(x)$ and $p(x)$;
\be
\label{eqn:hellinger}
d_{\rm H}^2 = \frac{1}{2} \int \d x \left[ \sqrt{p(x)} - \sqrt{q(x)} \right]^2  = 1 - \int \d x  \sqrt{p(x)q(x)} \, .
\ee
When the posteriors are perfectly matched, the Hellinger distance $d_{\rm H} = 0$.  For non-overlapping posteriors $d_{\rm H} = 1$.  We follow \citet*{heymans/etal:2021}, measuring $d_{\rm H}$ from discrete marginal posteriors by taking the average result from two different approaches. For the first measurement, $p(x)$ and $q(x)$ are defined using a binned histogram. The second measurement uses a smoothed kernel density estimate (see appendix G.1 of \citealt*{heymans/etal:2021} for further details).  

The calculation of the Hellinger distance between two marginal posterior distributions makes no assumption about their Gaussianity.  In order to present $d_{\rm H}$ in terms of a familiar $H\sigma$ `tension metric', however, we choose to recast the measured value in terms of two Gaussian posteriors that exhibit the same Hellinger distance, $d_{\rm H}$, when their variance is fixed to match the measured variance of the non-Gaussian posteriors $p$, $\sigma^2_p$, and $q$, $\sigma^2_q$ (see equation G.2 of \citealt{heymans/etal:2021}).  We report the `tension' offset, $H\sigma$, with $H =\delta \mu / \sqrt{\sigma_p^2+\sigma_q^2}$, where $\delta \mu$ is the mean-offset between the two $d_{\rm H}$-separated Gaussian posteriors.

For a `multi-dimensional parameter' tension metric we adopt the Monte Carlo exact parameter shift method\footnote{We note that the \citet{raveri/etal:2020} parameter shift method is mathematically equivalent to the `tier 2' method of \citet{kohlinger/etal:2019}, but differs in the implementation strategy.} from \citet{raveri/etal:2020}.  We define the parameter difference probability density, 
$\mathcal{P}(\Delta \varphi)$, between two uncorrelated data sets, A and B, 
\be
\mathcal{P}(\Delta \varphi) = \int_{V_p}\d\varphi\, \mathcal{P}_{\rm A}(\varphi) \mathcal{P}_{\rm B}(\varphi - \Delta \varphi) \, ,
\ee
where $\mathcal{P}_{\rm X}(\varphi)$ is the parameter posterior distribution from experiment X, over a multi-dimensional parameter space $\varphi$, evaluated within the whole parameter space volume $V_p$. The statistical significance that a shift exists between the underlying parameters of experiment A and B, is then given by
\be
\Delta_{\rm tension} = \int_{\mathcal{P}(\Delta \varphi)>\mathcal{P}(0)}\d\Delta\varphi \, \mathcal{P}(\Delta \varphi) \, ,
\label{eqn:delta_tension}
\ee
which we cast into a `tension' offset, $N_\sigma \sigma$, with an error function, ${\rm erf}$, as,
\be
N_\sigma = \sqrt{2} \,{\rm erf}^{-1} (\Delta_{\rm tension}) \, .
\ee
The calculation of this metric from discrete marginal posteriors is non-trivial.  We refer the reader to section VII.B of \citet{raveri/etal:2020} for details on the methodology adopted and the caveats with this approach.  In practice, we calculate $\Delta_{\rm tension}$ using the {\sc tensiometer}\footnote{{\sc tensiometer}: \url{https://github.com/mraveri/tensiometer}} software package \citep{raveri/doux:2021}. We find that the $\Delta_{\rm tension}$ estimate is sensitive to the choice of {\sc tensiometer} settings when including a large number of unconstrained parameter dimensions.  Furthermore we note that the metric is sensitive to the choice of priors.  For example, the informative prior for $h=\bb{0.64,0.82}$, is set with a width of $\pm 5\sigma$ around the \citet{riess/etal:2016} results, 
encompassing constraints from both \citet{planck/etal:2020} and \citet{riess/etal:2022}.  As cosmic shear is currently unable to constrain $H_0$, the $H_0$ marginal posterior from our Hybrid analysis is given by the prior which is significant for $H_0$ values between $-6\sigma$ and $+27\sigma$ from the {\it Planck} best fit.  With significantly more posterior volume above the {\it Planck} $H_0$ best fit than below it, the inclusion of $H_0$ in the tension metric's parameter space $\varphi$, artificially enhances the multi-parameter tension between the cosmic shear and {\it Planck} constraints. 
We therefore choose to report $\Delta_{\rm tension}$ measured from only the cosmic shear constrained $S_8-\Omega_{\rm m}$ parameter space.

For an `evidence-based' tension metric, we adopt the `suspiciousness' metric, $\ln S = \ln R - \ln I$, which quantifies the difference between the Bayes factor, $R$, and the information ratio, $I$ \citep{handley/lemos:2019}.  \citet*{heymans/etal:2021} show that this metric\footnote{We also refer the interested reader to section IV.E of \citet{joudaki/etal:2022}, showing the relationship between a range of concordance statistics.} can be written in terms of the difference between the expectation value of the log-likelihoods, $\mathcal{L}$, comparing a joint analysis of data sets A and B with independent analyses,
\be
\ln S = \langle\ln\mathcal{L}_{\rm A+B}\rangle_{\mathcal{P}_{\rm A+B}} -\langle\ln\mathcal{L}_{\rm A}\rangle_{\mathcal{P}_{\rm A}} -\langle\ln\mathcal{L}_{\rm B}\rangle_{\mathcal{P}_{\rm B}}  \,.
\label{eqn:suspiciousness}
\ee
Under the assumption that the two data sets, A and B, are concordant, and that the posteriors are Gaussian, a suspiciousness probability can be determined as the quantity $d-2\ln S$, which has a $\chi^2$ distribution with $d$ degrees of freedom.  Here  $d=N_{\Theta}^{\rm A} + N_{\Theta}^{\rm B} - N_{\Theta}^{\rm A+B}$, is the difference in the Bayesian model dimensionality determined from the effective number of free parameters for each data set.   As discussed in Section~\ref{sec:GoF}, $N_{\Theta}$ is non-trivial to accurately determine given a degenerate parameter space with informative priors and we adopt the \citet{joachimi/etal:2021} strategy to estimate this quantity.  In contrast to the Hellinger and parameter shift metrics, the suspiciousness metric requires an additional joint probe chain analysis.  We therefore limit the use of this metric to our fiducial analysis. 

\section{Cosmological Results}
\label{sec:results}
We present the cosmological parameter constraints from a joint-survey analysis of the DES Y3 and KiDS-1000 cosmic shear measurements using the Hybrid pipeline summarised in Table~\ref{tab:hybrid}. In Section~\ref{sec:hybrid}, we show the headline results for DES, KiDS, and the combination of the two surveys.  In Sections~\ref{sec:nuresults}-\ref{sec:baryresults}, we assess the sensitivity of our results to changes in the neutrino mass prior, and the baryon feedback and IA models.  Our constraints are compared to those from the CMB \citep{planck/etal:2020} in Section~\ref{sec:tmetrics}. We compare the constraints from our fiducial Hybrid analysis to a DES-like and KiDS-like re-analysis of the two surveys in Section~\ref{sec:pipes}, and to a Hybrid-like re-analysis of HSC Year 3 in Section~\ref{sec:HSCresults}.  We review the constraints from alternative probes of $S_8$ in Section~\ref{sec:otherprobes}.

In Appendices~\ref{app:mocks},~\ref{app:samplers} and~\ref{app:hybrid} we carry out a detailed investigation into the impact of different modelling choices, under the controlled conditions of `noise-free' mock DES and KiDS data.  This mock data analysis was completed and the Hybrid pipeline defined, see Table~\ref{tab:hybrid}, before embarking on the joint-survey data analysis.  The goodness of fit of the model to the data and the consistency between the DES and KiDS surveys were verified before viewing the cosmological constraints.

\subsection{Fiducial analysis}
\label{sec:hybrid}
\begin{figure*}
\centering
\begin{minipage}{.49\textwidth}
  \centering
  \includegraphics[width=\textwidth]{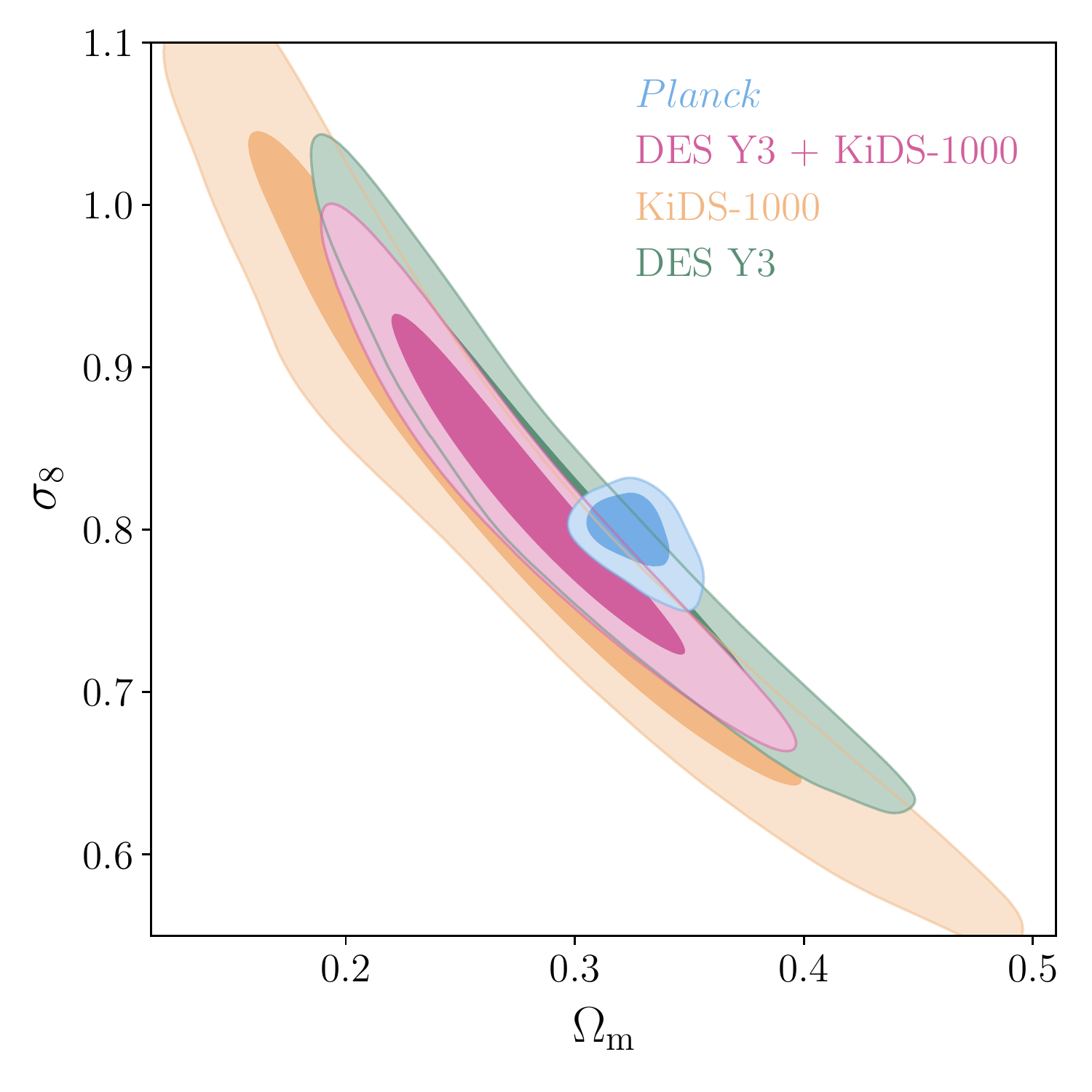}
\end{minipage}
\begin{minipage}{.49\textwidth}
  \centering
  \includegraphics[width=\textwidth]{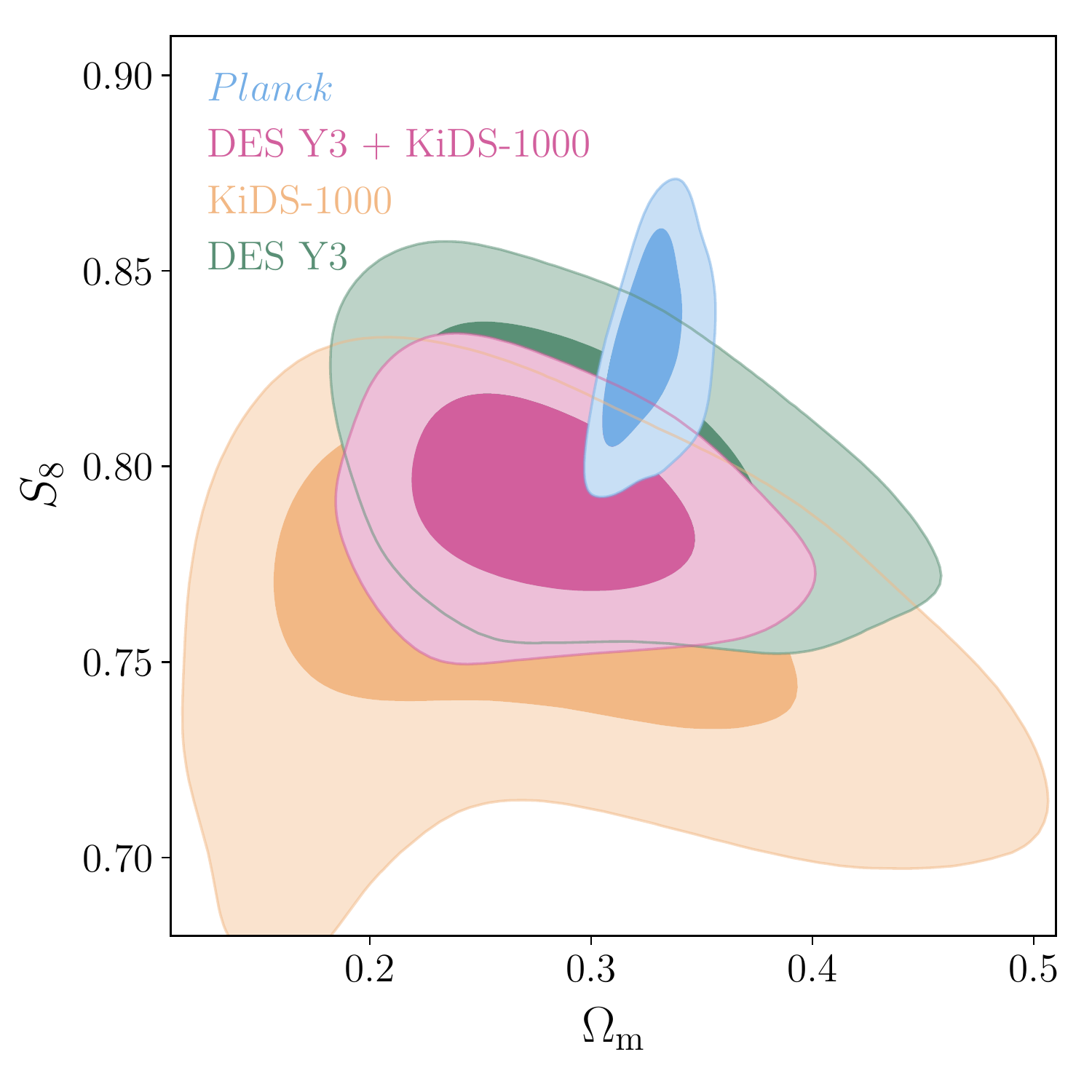}
\end{minipage}
\caption{Cosmological constraints on the cosmological parameters $\sigma_8$ (left) and $S_8$ (right) with the matter density $\Omega_{\rm m}$ in flat-$\Lambda$CDM. We adopt a Hybrid pipeline to re-analyse cosmic shear observations from DES Y3 (green) and KiDS-1000 (yellow) and conduct a joint-survey analysis of DES Y3 + KiDS-1000 (pink).  The cosmic shear constraints can be compared to a re-analysis of the \citet{planck/etal:2020} CMB observations (blue) using a common set of cosmological parameters and priors.  The marginalised posterior contours show the 68\% (inner) and 95\% (outer) credible intervals.
\label{fig:result}}
\end{figure*}

Our fiducial cosmic shear constraints are obtained using the Hybrid pipeline, marginalising over 6 cosmological parameters in the flat-\lcdm model and 18 systematic and astrophysical parameters, as summarised in Tables~\ref{tab:params} and~\ref{tab:hybrid}. In Figure~\ref{fig:result} we present the DES Y3 (green), KiDS-1000 (yellow) and DES Y3+KiDS-1000 (pink) cosmic shear posteriors, projected into a 2D parameter space for $\Omega_{\rm m}$, $\sigma_8$, and $S_8$.  For the individual surveys, the mean marginal values of $S_8$, $\Omega_{\rm m}$ and $\sigma_8$ are found with 68\% credible intervals to be 
\begin{align}
\label{eqn:DKS8results}
\begin{aligned}
{\rm DES:}\,\,\, &S_8 &=&\,\,\,  0.802^{+0.023}_{-0.019} \\  
 &\Omega_{\rm m} &=&\,\,\,  0.297^{+ 0.040}_{-0.060} \\
 &\sigma_8 &=&\,\,\, 0.816^{+ 0.076}_{-0.085} \\
{\rm KiDS:}\,\,\, &S_8 &=&\,\,\,  0.763_{-0.023}^{+0.031}  \\ 
&\Omega_{\rm m} &=&\,\,\,  0.270^{+ 0.056}_{-0.102}  \\
&\sigma_8 &=&\,\,\,  0.833^{+ 0.133}_{-0.146} \,,
\end{aligned}
\end{align}
\noindent 
constituting 2.6\% (DES Y3) and 3.5\% (KiDS-1000) precision measurements of $S_8$.  In Table~\ref{tab:results} we report the maximum marginal and MAP+PJ-HPD statistics for $S_8$. This table also includes constraints from an analysis of the 8\% area-cut DES Y3 data vector that is used in the joint-survey analysis.  Here the overlapping region of the KiDS footprint has been excised to mitigate cross-survey covariance (see Appendix~\ref{app:DESminusKiDS}), which we find has little impact on the DES Y3 results.  

For both surveys, the model provides a good fit to the data, with a goodness of fit probability $p^{\rm DES}(\chi^2 > \chi^2_{\rm min} | \nu) = 0.231$ and $p^{\rm KiDS}(\chi^2 > \chi^2_{\rm min} | \nu) = 0.048$  (see Table~\ref{tab:GoFHybrid}), calculated assuming our data vector is drawn from a multivariate Gaussian likelihood and that our assumed covariance matrix is precisely and fully characterised\footnote{We note that the KiDS goodness of fit probability increases to $p^{\rm KiDS}(\chi^2 > \chi^2_{\rm min} | \nu) = 0.66$ in the \citet{li/etal:2023b} Hybrid analysis of an improved KiDS-1000 shear catalogue that also adopts enhanced shear and redshift calibration techniques.  We note that the \citet{li/etal:2023b} $S_8$ constraints are unchanged from this analysis, with the MAP+PJHPD $S_8=0.776_{-0.027-0.003}^{+0.029+0.002}$.  The second set of errors here account for systematic uncertainties within the shear calibration.}.  Before combining the two surveys we assessed their consistency.  We find a DES-KiDS Hellinger distance offset in $S_8$ of $1.0\sigma$ (Equation~\ref{eqn:hellinger}), and a $\Delta_{\rm tension}$ parameter shift in $S_8-\Omega_{\rm m}$ of $0.8\sigma$ (Equation~\ref{eqn:delta_tension}), thus meeting the $<2.3\sigma$ threshold for consistent data sets. 

\begin{figure*}
\centering 
\includegraphics[width=\textwidth]{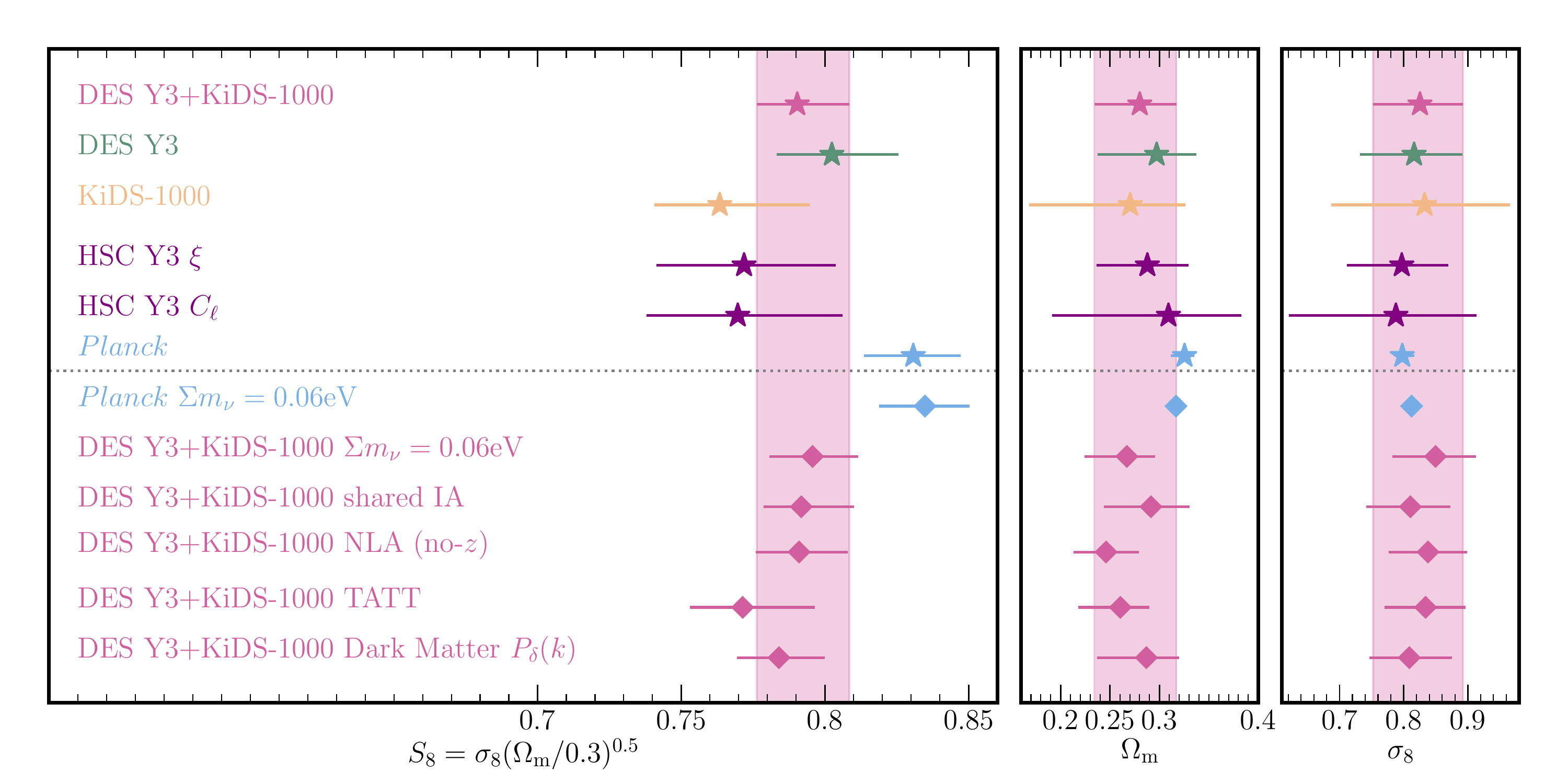}
\caption{Summary of mean marginalised 1D constraints on $S_8$,  $\Omega_{\rm m}$ and $\sigma_8$. The mean of the posterior is indicated by the filled symbol and 68\% credible intervals are shown as horizontal bars. The DES and KiDS Hybrid analyses are represented as the yellow and green stars, respectively, and the joint Hybrid analysis as the pink star and vertical shaded region. Variants of the joint Hybrid analyses are shown as pink diamonds. In descending order, we show variants adopting a fixed neutrino mass, a shared set of intrinsic alignment parameters for the two surveys, an NLA analysis with no redshift evolution fixing $\eta_{\rm IA}=0$, a TATT intrinsic alignment model analysis and a dark matter only analysis with no marginalisation over the effects of baryonic feedback.  The DES and KiDS constraints can be compared to a Hybrid pipeline re-analysis of the HSC Year 3 cosmic shear observations (purple stars) using the shear angular power spectrum, $C_\ell$, \citep{dalal/etal:2023} and the two-point shear correlation function, $\xi_\pm(\theta)$ \citep[][see Section~\ref{sec:HSCresults} for details]{li/zhang/etal:2023}. We also compare to a reanalysis of the \citet{planck/etal:2020} CMB observations (see Section~\ref{sec:tmetrics} for details) using the Hybrid pipeline's set of cosmological parameter priors with a free neutrino mass density (blue star), or a fixed neutrino mass prior with $\Sigma m_\nu = 0.06$eV (blue diamond).  The numerical parameter values for $S_8$ are listed in Table~\ref{tab:results}.}
\label{fig:dataS8}
\end{figure*}

For the DES Y3+KiDS-1000 joint-survey analysis, the mean marginal values of $S_8$, $\Omega_{\rm m}$ and $\sigma_8$ and are found with 68\% credible intervals to be
\begin{align}
\begin{aligned}
S_8 &{} = 0.790_{-0.014}^{+0.018} \\ 
\Omega_{\rm m} &{} =  0.280^{+ 0.037}_{-0.046} \\
\sigma_8 &{} =  0.825^{+ 0.067}_{-0.073} \, ,
\end{aligned}
\end{align}
constituting a 2.0\% precision measurement of $S_8$\footnote{It is interesting to note that the joint-survey constraints on $S_8$ are the same as those estimated through a naive approach of taking the weighted average of the individual survey constraints in Equation~\ref{eqn:DKS8results}.   We do not recommend this naive approach for future survey combinations, especially in cases where the analysis choices differ.  A weighted average of the published constraints from \citet{amon/etal:2022,asgari/etal:2021}; \citet*{secco/etal:2022} is offset from our joint-survey constraints at the level of $1.6\sigma$.  We discuss how the different analysis choices for each survey team impacts the $S_8$ constraints in Section~\ref{sec:pipes}, as quantified through mock survey studies in Appendices~\ref{app:HMCodeTATTtest} and~\ref{app:euclidem}.}.  These constraints are summarised in Figure~\ref{fig:dataS8} and tabulated in Table~\ref{tab:results} including the maximum marginal and MAP+PJ-HPD values for $S_8$.  In all cases the model is found to provide a good fit to the data (see Table~\ref{tab:GoFHybrid}).  For our fiducial joint-survey analysis we find a goodness of fit probability $p(\chi^2 > \chi^2_{\rm min} | \nu) = 0.068$. We also measure the goodness of fit of the DES and KiDS data vector for the best-fit set of parameters from the joint analysis.  The DES goodness of fit is essentially unchanged by the joint analysis. The KiDS goodness of fit degrades slightly, but nevertheless passes the goodness of fit requirement with $p(\chi^2 > \chi^2_{\rm min} | \nu)=0.035$.

\begin{table}  
\centering         
\input{tabletexfiles/Goodness_of_fit_Hybrid_only.tex} 
\caption{Goodness of fit statistics for the Hybrid pipeline: the best-fit $\chi^2_{\rm min}$, the estimated effective number of free parameters, $N_{\Theta}$, the reduced $\chi^2_{\rm red} =  \chi^2_{\rm min}/\nu$, where $\nu$ is the number of degrees of freedom, and the goodness of fit probability $p(\chi^2 > \chi^2_{\rm min} | \nu)$ (see Section~\ref{sec:GoF}).   The number of data points for the DES, KiDS and joint-survey data vectors, are $N_{\rm data}=273, 75, 348$ respectively.  The upper section reports results for the fiducial analysis of the individual and joint surveys.  The lower section varies one aspect of the Hybrid joint-survey analysis: fixing the neutrino mass to $\Sigma m_{\nu}=0.06{\rm eV}$,  sharing the IA parameters between the two surveys, assuming an NLA IA model without redshift evolution (no z), adopting the TATT IA model, and using a dark matter-only correction for the non-linear model of the matter power spectrum, $P_\delta(k)$.}
\label{tab:GoFHybrid}
\end{table}

Reviewing the different mean marginal, maximum marginal and MAP $S_8$ values in Table~\ref{tab:results}, it is worth noting that the $0.6\sigma$ offset between the MAP and mean is expected from our analysis of {\sc EuclidEmulatorv2} mocks in Appendix~\ref{app:euclidem}. This offset reflects the significant skew in the marginal $S_8$ posterior, in addition to a potential projection effect which would arise when marginalising over a neutrino mass prior that is asymmetrical about the truth (see Appendix~\ref{app:projection}). In the discussion that follows we quote the mean marginal values for $S_8$, referring the reader to Table~\ref{tab:results} for the alternative MAP+PJ-HPD or maximum marginal metrics of the posterior.

\begin{table*}   
\centering 
\input{tabletexfiles/MAP_PJHPD_table_all_Hybrid_results.tex} 
\caption{$S_8$ constraints with 68\% credible intervals using the mean 1D marginal posterior, the MAP and PJ-HPD, and the maximum 1D marginal.  Constraints are provided for our fiducial analysis of DES Y3, KiDS-1000 and the joint-survey analysis.  We also report constraints from a series of analysis variants which, in descending order, adopt a fixed neutrino mass, a shared set of intrinsic alignment parameters for the two surveys, an NLA analysis with no redshift evolution fixing $\eta_{\rm IA}=0$, a TATT intrinsic alignment model analysis and a dark matter only analysis with no marginalisation over the effects of baryonic feedback. $\Delta S_8$ quantifies the offset of each statistic's value for $S_8$ relative to the Hybrid joint analysis, mean marginal value $S_8^{\rm Fid} = 0.790$, as a fraction of the $1\sigma$ error for each statistic.  The error is also tabulated as a ratio to the fiducial joint analysis error $\sigma_{\rm Fid} = 0.016$.  The cosmic shear constraints can be compared to a reanalysis of the \citet{planck/etal:2020} CMB observations (see Section~\ref{sec:tmetrics} for details) using the Hybrid pipeline's set of cosmological parameter priors with a free neutrino mass density, or a fixed neutrino mass prior with $\Sigma m_\nu = 0.06$eV.  We do not list $\Delta S_8$ for the {\it Planck} constraints, referring the reader to Section~\ref{sec:tmetrics} for a more appropriate set of metrics to compare independent cosmological probes.
\label{tab:results}}
\end{table*}

\subsection{Fixing the neutrino mass density}\label{sec:nuresults}
In our fiducial analysis we allow the neutrino mass density to vary.  Following \citet{planck/etal:2020} we investigate adopting a fixed neutrino mass with $\Sigma m_{\nu}=0.06$eV, based on the minimum mass allowed by oscillation experiments when assuming a normal mass hierarchy \citep{capozzi/etal:2016}.  We find our constraints to be fairly insensitive to the choice of prior for $\Omega_\nu h^2$, similar to previous studies. Comparing the `DES Y3+KiDS-1000 $\Sigma m_{\nu}=0.06$eV' analysis with the fiducial result, in Figure \ref{fig:dataS8} and Table~\ref{tab:results}, we find the mean value of $S_8$ increases by $0.39\sigma$ and the marginal uncertainty decreases by $2\%$ with:
\begin{equation}
    S^{\Sigma m_{\nu}=0.06{\rm eV}}_8 = 0.797^{+ 0.017}_{-0.014}\, .
\end{equation}
In Table~\ref{tab:results} we find that adopting a fixed neutrino mass brings the mean and maximum marginal estimates in line with the MAP.  This behaviour is also seen in our mock analysis in Appendix~\ref{app:euclidem}.  For the fiducial Hybrid mock analysis, the offset in the marginalised $S_8$ constraints relative to the input truth arises from the projection of a wide and positive $\Sigma m_\nu$ prior, which is heavily skewed about the mock input value of $\Sigma m_\nu = 0.06$eV.

\subsection{Varying the intrinsic alignment model}
\label{sec:iaresults}
\begin{figure}
\centering 
\includegraphics[width=\columnwidth]{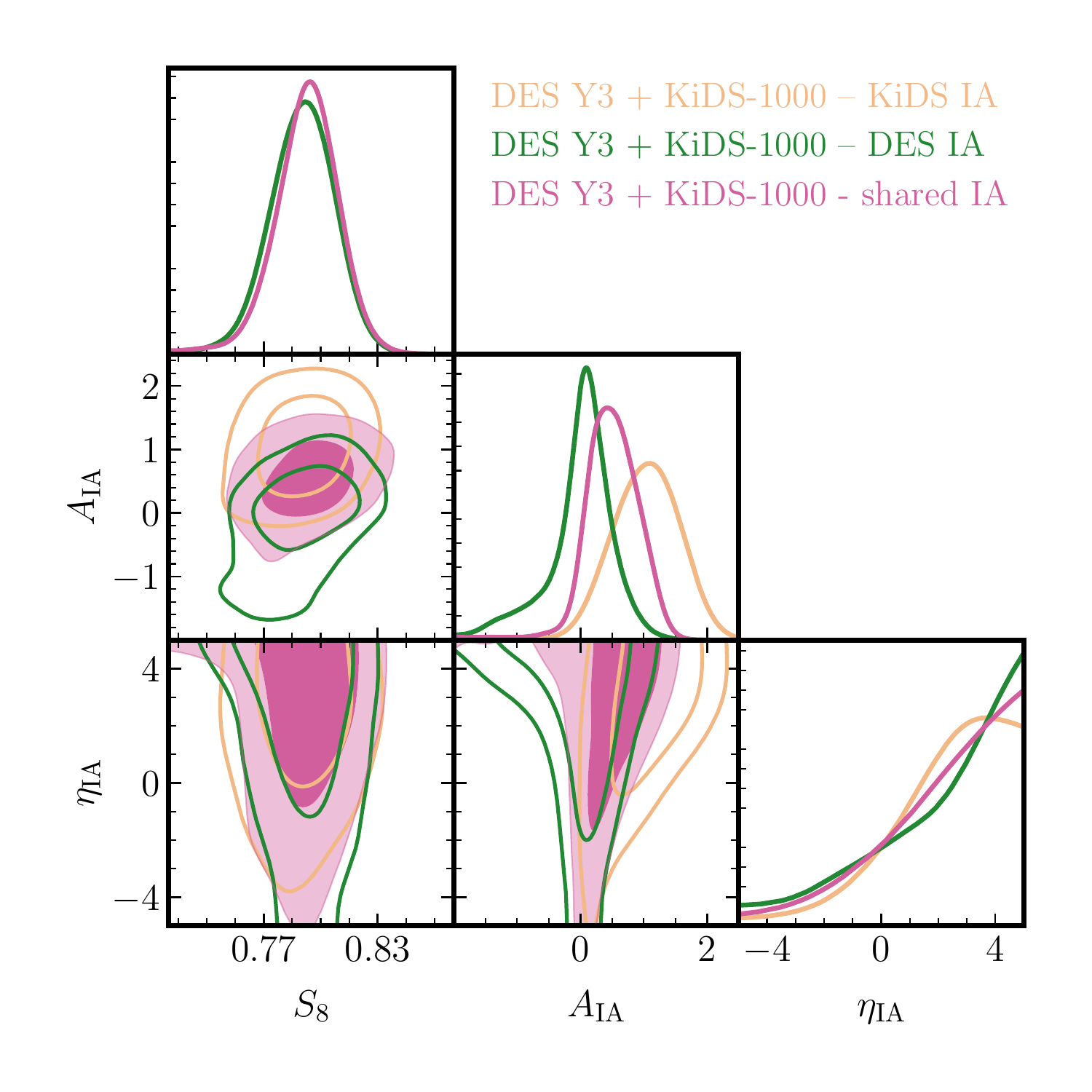}
\caption{Constraints on the NLA-z intrinsic alignment parameters $A_{\rm IA}$ and $\eta_{\rm IA}$ with $S_8$. The marginalised posterior contours (inner 68\% and outer 95\% credible intervals) are shown for the fiducial analysis where the IA parameters are independent for the two surveys.  The DES IA parameters are shown in green with the KiDS IA parameters in yellow.  The fiducial result can be compared to an alternative analysis where the IA parameters are shared (pink).}
\label{fig:2IAvs1IA}
\end{figure}

In our fiducial analysis we adopt the NLA-z IA model, with two independent sets of survey-specific IA parameters: $(A^{\rm DES}_{\rm IA}, \eta^{\rm DES}_{\rm IA})$ and $(A^{\rm KiDS}_{\rm IA},  \eta^{\rm KiDS}_{\rm IA})$.  The mean marginal constraints for these parameters are shown in Figure~\ref{fig:2IAvs1IA} with the DES parameters in green and the KiDS parameters in yellow.  We can compare the constraints from the individual and joint-survey analyses. Analysing the surveys independently we find: 
\begin{equation}
A^{\rm DES}_{\rm IA}=0.32^{+0.43}_{-0.37} \;\;\;\;\; A^{\rm KiDS}_{\rm IA} = 0.65_{-0.54}^{+0.88}\, .  
\end{equation}
\noindent 
In a joint-survey analysis, we find $A^{\rm DES}_{\rm IA}$ reduces to accommodate a reduction in $S_8$ relative to the DES-only preferred value, and the same effect, but in reverse\footnote{In broad terms this can be understood through equation~\ref{eqn:c_ee}.  The amplitude of the shear power spectrum $C_{\rm GG}(\ell)$ is roughly proportional to $S_8^2$ \citep{jain/seljak:1997}.  Given that the total observed signal is unchanged by the analysis, lowering (raising) the value of $S_8$ can be offset by raising (lowering) the amplitude of the IA terms.  As the GI term dominates the IA signal, with $C_{\rm GI}(\ell) \propto -A_{\rm IA}$, a reduction in $S_8$ combined with a reduction in $A_{\rm IA}$ can deliver a good fit to the data.  In reality the situation is more complex with multiple parameters in play, but this effect nevertheless drives the mild degeneracy seen between $S_8$ and $A_{\rm IA}$ in Figure~\ref{fig:2IAvs1IA}.}, for $A^{\rm KiDS}_{\rm IA}$:
\begin{equation}
A^{\rm DES, \; Joint}_{\rm IA} = -0.02_{-0.29}^{+0.58}
\;\;\;\;\;  
A^{\rm KiDS, \; Joint}_{\rm IA} = 1.04^{+0.54}_{-0.52}\, .
\end{equation}
This increases the offset between the survey-preferred $A_{\rm IA}$ amplitude, but the constraints remain formally consistent with a Hellinger offset of $1.47\sigma$. As the intrinsic alignment signal is known to depend on many factors (see the discussion in Appendix~\ref{sec:J2IA}), we do not expect identical intrinsic alignment signals in the two surveys, hence our use of independent IA parameters. That said, as DES and KiDS have broadly similar redshift distributions and depths, with the tomographic samples dominated by fainter bluer galaxies, we do not expect considerable differences between the effective intrinsic alignment contamination of each survey.  The changes to the $A_{\rm IA}$ constraints for each survey in the joint survey analysis may therefore indicate that this flexible nuisance parameter is absorbing more than just the contribution to the tomographic cosmic shear signal of intrinsically aligned galaxies.  When testing a single shared set of IA parameters for the joint-analysis, we find, unsurprisingly, that the IA constraints lie between the best-fits for the two individual surveys, with $A^{\rm Shared-IA}_{\rm IA}=0.43^{+0.37}_{-0.45}$. As shown in Figure~\ref{fig:2IAvs1IA} with the shared-IA constraints in pink, the cosmological parameter constraints are not impacted by the choice of shared or independent IA modelling.  We find negligible differences: $S_8$ increases by $0.1\sigma$, the marginal uncertainty decreases by $2\%$ and the goodness of fit probability decreases by $1\%$.  

\begin{figure}
\centering 
\includegraphics[width=\columnwidth]{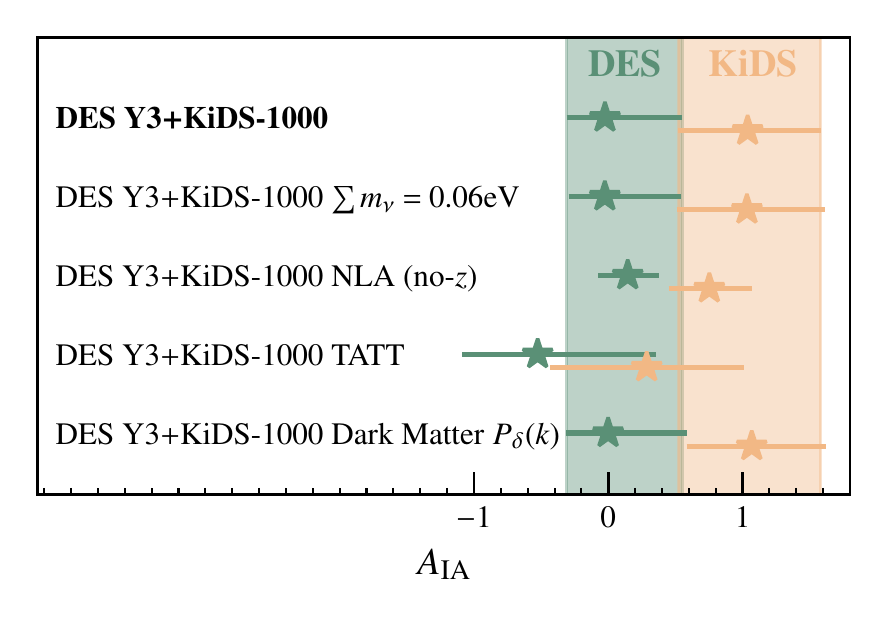}
\caption{Summary of mean marginalised 1D constraints for the NLA-z $A_{\rm IA}$ parameter. The top row and shaded bands show the posterior mean and $1\sigma$ error bars from our fiducial Hybrid pipeline analysis of the joint DES Y3 + KiDS-1000 data vector. In descending order, we show analysis variants with fixed neutrino mass,  an NLA analysis with no redshift evolution fixing $\eta_{\rm IA}=0$, a TATT intrinsic alignment model analysis and an analysis using a dark matter non-linear matter power spectrum with no marginalisation over the effects of baryonic feedback. In the case of the TATT analysis we plot constraints for the $a_1$ parameter, which is equal to $A_{\rm IA}$ in the limit where the TATT parameters $a_2$, $\eta_2$, $b_{\rm TA} \rightarrow 0$. See Appendix~\ref{app:extras} for marginal constraints on each TATT parameter.}
\label{fig:dataAIA}
\end{figure} 

We are unable to constrain the redshift-dependent parameters $\eta_{\rm IA}$, but it is nevertheless interesting to note the high MAP-values for $\eta^{\rm DES, MAP}_{\rm IA} = 4.5$ and $\eta^{\rm KiDS, MAP}_{\rm IA}= 4.1$ which are unexpected\footnote{Direct observations of galaxy position-shape correlations for a sample of LRGs constrain $\eta_{\rm IA} = -0.3 \pm 0.8$ over the redshift range $0<z<0.7$ \citep{joachimi/etal:2011}, and $\eta_{\rm IA}=-0.05 \pm 0.73$ over the redshift range $0.2<z<1.1$ \citep{samuroff/etal:2022}. Null detections of intrinsic alignments for late-type galaxies at $z \sim 0.6$ \citep{mandelbaum/etal:2011}, $z \sim 0.8$ \citep{samuroff/etal:2022} and $z\sim1.4$ \citep{tonegawa/etal:2018} suggest there is also no strong evolution for spiral galaxy alignment, albeit with a fairly large degree of uncertainty on these null results.  Using a halo model, \citet{fortuna/etal:2021} demonstrate that the effective intrinsic alignment signal for a magnitude limited galaxy survey is nevertheless expected to evolve as a result of the changing average luminosity and central/satellite fraction in each redshift bin.  Using observation-informed halo model parameters, they find the halo model prediction for the intrinsic alignment signal from a KiDS-like survey can be represented by an NLA-z model with $A_{\rm IA}=0.42\pm0.02$ and $\eta_{\rm IA}=2.21\pm 0.22$.}.  We find the single-survey $\eta_{\rm IA}$ posteriors also skew towards the upper edge of the $\eta_{\rm IA}$ prior, matching a similar finding in the \citet{amon/etal:2022}; \citet*{secco/etal:2022} DES Y3 analysis, where posteriors for the corresponding $\eta_1$ and $\eta_2$ TATT parameters also skew towards the high values allowed by the prior.   This result could be indicative of the flexibility of the IA model adapting to absorb some tension in the photometric redshift evolution of the tomographic cosmic shear signals \citep[see the discussion in Appendix~\ref{sec:J2IA} and][]{fischbacher/etal:2022}.  We note, however, that as we find $S_8$ to be insensitive to changes in $\eta_{\rm IA}$ (see Figure~\ref{fig:2IAvs1IA}), this result should not impact confidence in the cosmological constraints.

In our fiducial analysis we adopt the NLA-z intrinsic alignment model.  Changing to the one-parameter NLA model, which removes the redshift dependence in the NLA-z framework we find negligible differences with the $S_8$ constraint increasing by $0.1\sigma$ and the marginal uncertainty decreasing by $6\%$ with:
\begin{equation}
    S^{\rm NLA}_8 = 0.792^{+0.016}_{-0.013}  \, .
\end{equation}
As shown in Figure~\ref{fig:dataAIA}, removing freedom in the redshift evolution reduces the uncertainty on the $A_{\rm IA}$ constraints for DES and KiDS, but they remain consistent with a Hellinger offset of $1.53\sigma$.

Changing to the five-parameter TATT model (see Section~\ref{sec:IA}), with DES-like IA priors from Table~\ref{tab:params}, we find the joint-survey cosmic shear $S_8$ constraint lowers by $0.9\sigma$ and the marginal uncertainty widens by $35\%$ with:
\begin{equation}
    S^{\rm TATT}_8 = 0.771^{+0.018}_{-0.025}  \,.
\end{equation}

The choice for our Hybrid pipeline of NLA over TATT therefore introduces the largest impact on our results.  We note that the increased uncertainty in $S_8$ when adopting the TATT model is in contrast to the $\sim 10\%$ reduction in uncertainty on $\Omega_{\rm m}$ and $\sigma_8$ (see Figure~\ref{fig:dataS8}).  Appendix~\ref{app:extras} compares the multi-dimensional posteriors from the TATT and NLA-z analyses finding a similar degeneracy in the $\sigma_8-\Omega_{\rm m}$ plane for the two distributions, but with a significantly broader width in the case of the TATT analysis.  This demonstrates that it is non-trivial to predict estimates of $S_8$ from the marginal distributions of $\Omega_{\rm m}$ and $\sigma_8$.

The trends seen here in the uncertainty and value of $S_8$, are similar to the trends reported by previous DES and KiDS analyses\footnote{We note that the impact of using different IA models in the cosmic shear analysis of HSC is less pronounced than in our joint-survey analysis.  \citet{li/zhang/etal:2023,dalal/etal:2023} report a $\sim0.2-0.4\sigma$ $S_8$ offset when changing from NLA-z to TATT, and a $\sim 15\%$ increase in the uncertainty.  Given the interplay between photometric redshift nuisance parameters, $\Delta z$, and IA parameters \citep{fischbacher/etal:2022}, different IA behaviour is, however, expected for HSC who use wide uninformative priors, $\Delta z = \bb{-1,1}$, for bins with $z_{\rm phot} > 0.9$.}.  \citet{asgari/etal:2021} find switching from an NLA-z to a single-parameter NLA model impacts the $S_8$ uncertainty at the level of $\sim 10-30\%$ dependent on the two-point statistic and has a negligible impact on the value of $S_8$. \citet{samuroff/etal:2019,amon/etal:2022}; \citet*{secco/etal:2022} find switching from NLA-z to TATT decreases $S_8$ by $\sim0.6-1.0\sigma$ depending on the scale cuts, and the cosmic shear-only analysis (without shear ratio data\footnote{Whilst not directly comparable to the cosmic shear only analysis in this study, we note that the DES Y3 $S_8$ uncertainty with TATT is only $6\%$ larger than in the NLA-z analysis when the cosmic shear data is analysed in combination with the shear ratio data.  For the DES Y3 $3\times2$pt analysis, the difference is $\sim13\%$.}) in \citet{amon/etal:2022} find a $\sim 30\%$ increase in the $S_8$ uncertainty when adopting TATT compared to NLA-z. 

To understand the differences between the NLA and TATT analyses, exploring the multi-dimensional posterior shows that the TATT IA model allows freedom for the cosmological model to explore low-$S_8$ values at large-$a_2$ values (see Appendix~\ref{app:extras}).  This introduces a significant skew in the marginal posterior lowering the mean $S_8$ value relative to the maximum marginal value. This is not only a skewness effect, however, as we find the MAP $S_8$ estimate is also low with $S_8=0.761^{+0.024}_{-0.036}$.  The difference between the MAP estimates for the NLA-z and TATT analysis is $1.3\sigma$, demonstrating that the $0.9\sigma$ offset found between the mean $S_8$ marginals is not solely a result of prior volume or projection effects.   

In our mock survey study in Appendices~\ref{app:euclidem} and~\ref{app:HMCodeTATTtest} we quantify the impact of adopting a TATT or NLA cosmic shear analysis for different input IA models.   When the input truth model is NLA (no-z), we find a $0.45\sigma$ reduction in $S_8$ and a $10\%$ widening of the marginal uncertainty when changing between the Hybrid NLA-z and TATT analysis of the same simulated data vector.  When the input truth model is a strong TATT model\footnote{Our `strong' TATT model is given by the best-fit parameters from the \citet{amon/etal:2022}; \citet*{secco/etal:2022} DES Y3 cosmic shear with shear-ratio analysis.  See Table \ref{tab:mock_input} and Appendix \ref{app:HMCodeTATTtest} for details.}, we find a $1.3\sigma$ increase in the mean marginal $S_8$ value, relative to the input cosmology in an NLA-analysis.  The impact of choosing TATT or NLA-z for the data lies between these two cases.  At this point, with the information available to us, it is not possible to say whether the differences we find between the TATT and NLA analyses arises from a significant bias of the NLA-z model relative to the true underlying IA mechanism, or a numerical effect that will disappear as future data becomes more constraining. New observational constraints are required to help distinguish between these IA models, to set tighter priors on the free parameters, and to better inform the analysis choices for future weak lensing surveys.

\subsection{Varying the baryon feedback model}\label{sec:baryresults}
In our fiducial analysis we employ two schemes to mitigate our uncertainty on the impact of baryon feedback on the non-linear matter power spectrum: scale cuts (see Appendix~\ref{app:scalecuts}) and marginalisation over the {\sc HMCode2020} $T_{\rm AGN}$ parameter (see Appendix~\ref{app:euclidem}).   Changing to use the {\sc HMCode2020} dark matter-only correction for the non-linear $P_\delta(k)$, but retaining the scale cuts, we find the joint-survey cosmic shear $S_8$ constraint lowers by $0.4\sigma$ and the marginal uncertainty decreases by $5\%$ with:
\begin{equation}
    S^{\rm{Dark\,Matter}\,P_\delta(k)}_8 = 0.784^{+0.016}_{-0.015}  \, .
\end{equation}
With the inclusion of scale cuts, our constraints are therefore robust at the $0.4\sigma$-level to the use of baryon feedback marginalisation as an analysis choice.  This finding is consistent with the baryon feedback sensitivity analysis in \citet{asgari/etal:2021} and \citet*{secco/etal:2022} using $A_{\rm bary}$ and {\sc HMCode2016}.  It also matches the $0.5\sigma$ reduction in $S_8$ found when switching off the $T_{\rm AGN}$ marginalisation in our mock survey analysis, where the input baryon feedback was modelled using the {\sc OWLS-AGN} simulation (see Appendix~\ref{app:euclidem}).  In Appendix~\ref{app:extras} we show that we are unable to set constraints on the amplitude of the baryon feedback model.

\subsection{Quantifying consistency/tension with {\it Planck}}
\label{sec:tmetrics}
Figure~\ref{fig:result} compares our cosmic shear constraints to the \textit{Planck} satellite CMB temperature and polarisation measurements \citep{planck/etal:2020}.  Specifically we use the \textit{Planck} measurements of the auto power spectra of temperature $C_{\ell}^{\rm TT}$, of $E$-modes $C_{\ell}^{\rm EE}$, and their cross-power spectra $C_{\ell}^{\rm TE}$, using the `Plik' version for $\ell$ >30.   In the range 2< $\ell$ <29, we only analyse the measurements of $C_{\ell}^{\rm TT}$ and $C_{\ell}^{\rm EE}$.  We choose to not include the CMB lensing data that is sensitive to a wide range of redshifts, extracting cosmological information solely from the high redshift primary CMB anisotropies\footnote{The inclusion of CMB lensing in the {\it Planck} data vector decreases the uncertainty on $S_8$ by 23-33\% without influencing the mean value \citep{planck/etal:2020,efstathiou/gratton:2021}.}.  In order to assess consistency, we reanalyse {\it Planck} using the Hybrid set of cosmological priors (Table~\ref{tab:hybrid}), primarily to allow for variations in the sum of neutrino masses, a quantity which is fixed to $\Sigma m_{\nu}=0.06$eV in the fiducial {\it Planck} analysis.

The sensitivity of the CMB constraints to the choice of neutrino mass prior is shown by comparing the `{\it Planck} $\Sigma m_{\nu}=0.06$eV' analysis with our Hybrid-prior re-analysis\footnote{For the fixed neutrino mass re-analysis of {\it Planck}, adopting Hybrid priors for all other parameters, we choose to use the {\sc Multinest} sampler for speed.  In this specific case, the posteriors are more Gaussian and constrained and therefore less sensitive to the {\sc Multinest} issues that affect non-Gaussian cosmic shear posteriors (see Appendix~\ref{app:samplers}).  From this analysis we recover the same $S_8$ constraints as \citet{planck/etal:2020}, within the expected $0.1\sigma$ chain-to-chain variance.  These constraints are $0.38\sigma$ higher than the $S_8$ constraint from \citet{efstathiou/gratton:2021} in their {\it Planck} re-analysis.  The measurement of an $S_8$ tension metric between our joint survey analysis and the \citet{efstathiou/gratton:2021} CMB constraints would therefore be reduced relative to the \citet{planck/etal:2020} CMB tension metrics in Table~\ref{tab:Hellinger}.  We note that had we chosen to use the \citet{efstathiou/gratton:2021} {\it Planck} re-analysis and also include  CMB lensing observations, enhancing the overall constraining power, the resulting tension metrics would be fairly similar to our quoted values.} of {\it Planck} in Figure \ref{fig:dataS8} and Table~\ref{tab:results}.  Fixing the neutrino masses increases the {\it Planck} mean marginal $S_8$ value by $0.27\sigma$ with the marginal uncertainty decreasing by $10\%$.

\begin{table}   
\centering  
\input{tabletexfiles/DKP_tension_table.tex}
\caption{Tension metrics comparing the cosmological constraints from cosmic shear with those from \citet{planck/etal:2020} CMB observations.  We tabulate the Hellinger distance offset $d_{\rm H}(S_8)$, and the multi-dimensional parameter shift offset $\Delta_{\rm tension}$, for the cosmic-shear constrained parameter set $(S_8,\Omega_{\rm m})$. Metrics are reported for the fiducial Hybrid pipeline analysis of DES Y3 (both the full area and the KiDS-excised area), KiDS-1000 and the joint-survey analysis.  We also report metrics for the Hybrid analysis variants where we adopt a fixed neutrino mass, a shared set of intrinsic alignment parameters for the two surveys, an NLA analysis with no redshift evolution fixing $\eta_{\rm IA}=0$, a TATT intrinsic alignment model analysis and a dark matter-only correction for the non-linear modelling of the matter power spectrum $P_\delta(k)$. In the case of the fixed neutrino mass analysis, we compare the cosmic shear constraint to a {\it Planck} analysis which also fixes the sum of neutrino masses to $\Sigma m_{\nu}=0.06$eV. }
\label{tab:Hellinger}
\end{table}

In Table~\ref{tab:Hellinger} we report the Hellinger distance offset $d_{\rm H}(S_8)$ (Equation~\ref{eqn:hellinger}) between the Hybrid pipeline cosmic shear $S_8$ constraints and {\it Planck} $S_8$ constraints.  We also report the multi-dimensional parameter shift offset $\Delta_{\rm tension}$ (Equation~\ref{eqn:delta_tension}) for the cosmic-shear constrained parameter set $(S_8,\Omega_{\rm m})$, finding similar results for the two tension metrics.  In all cases we find consistency between DES, KiDS and {\it Planck}.  We find that the joint Hybrid constraint differs from the \textit{Planck} CMB result by $1.7\sigma$, for both our fiducial setup and an analysis where the neutrino mass is fixed.  The tension between the observations is driven by the KiDS survey with an $S_8$ Hellinger distance offset of $2.1\sigma$, compared to the DES offset of $1.0\sigma$.    For our fiducial analysis we also quantify the Suspiciousness metric, Equation~\ref{eqn:suspiciousness}, finding a probability of $p=0.28$ to observe the measured offset between two concordant data sets.  This corresponds to consistency between DES, KiDS and {\it Planck} at the level of $0.6 \sigma$. 

The adoption of the Hybrid pipeline reduces the previously reported tension between the DES Y3 and KiDS-1000 cosmic shear observations and {\it Planck}.  In the case of KiDS, this reduction is primarily driven by an increase in the uncertainty on $S_8$ arising from the use of {\sc Polychord} over {\sc Multinest} and the inclusion of additional flexibility with the NLA-z model.  In the case of DES, the reduction is primarily driven by an upward shift in $S_8$ which we find in our mock studies is to be expected when changing both the IA and non-linear matter power spectrum models (see Appendices~\ref{app:HMCodeTATTtest} and ~\ref{app:euclidem}).  These differences are also highlighted by the range of constraints from our variants of the Hybrid analysis.  Using a TATT model increases the offset with {\it Planck}, to the level of $2.2\sigma$ (Hellinger) and $2.3\sigma$ ($\Delta_{\rm tension}$), bringing us to the limit where we would consider there to be evidence of inconsistency.  Analysing the data vector with a dark matter only model for the matter power spectrum also increases the offset relative to the fiducial case, with a $2.0\sigma$ Hellinger distance offset between the joint-survey constraints and {\it Planck}.

\subsection{A DES-like and KiDS-like re-analysis}
\label{sec:pipes}
\begin{figure*}
\centering
\begin{minipage}{.49\textwidth}
  \centering
  \includegraphics[width=\textwidth]{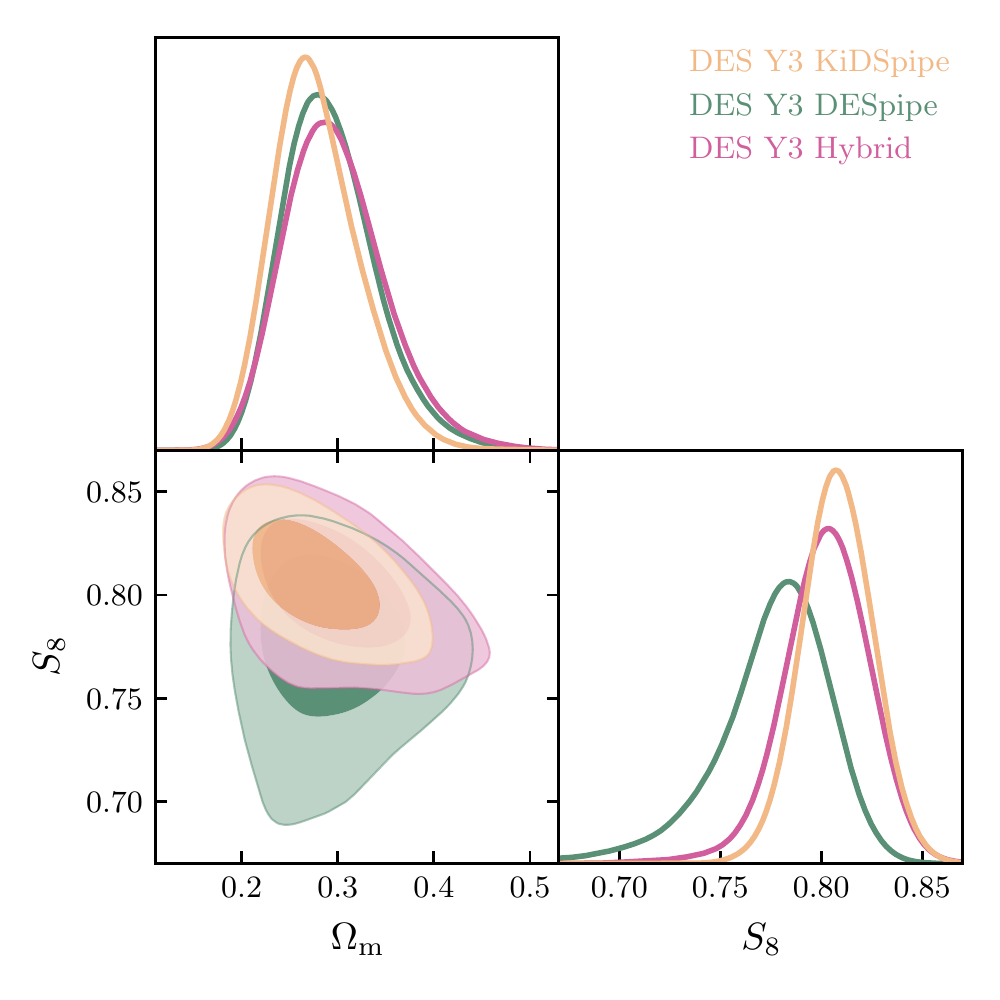}
\end{minipage}
\begin{minipage}{.49\textwidth}
  \centering
  \includegraphics[width=\textwidth]{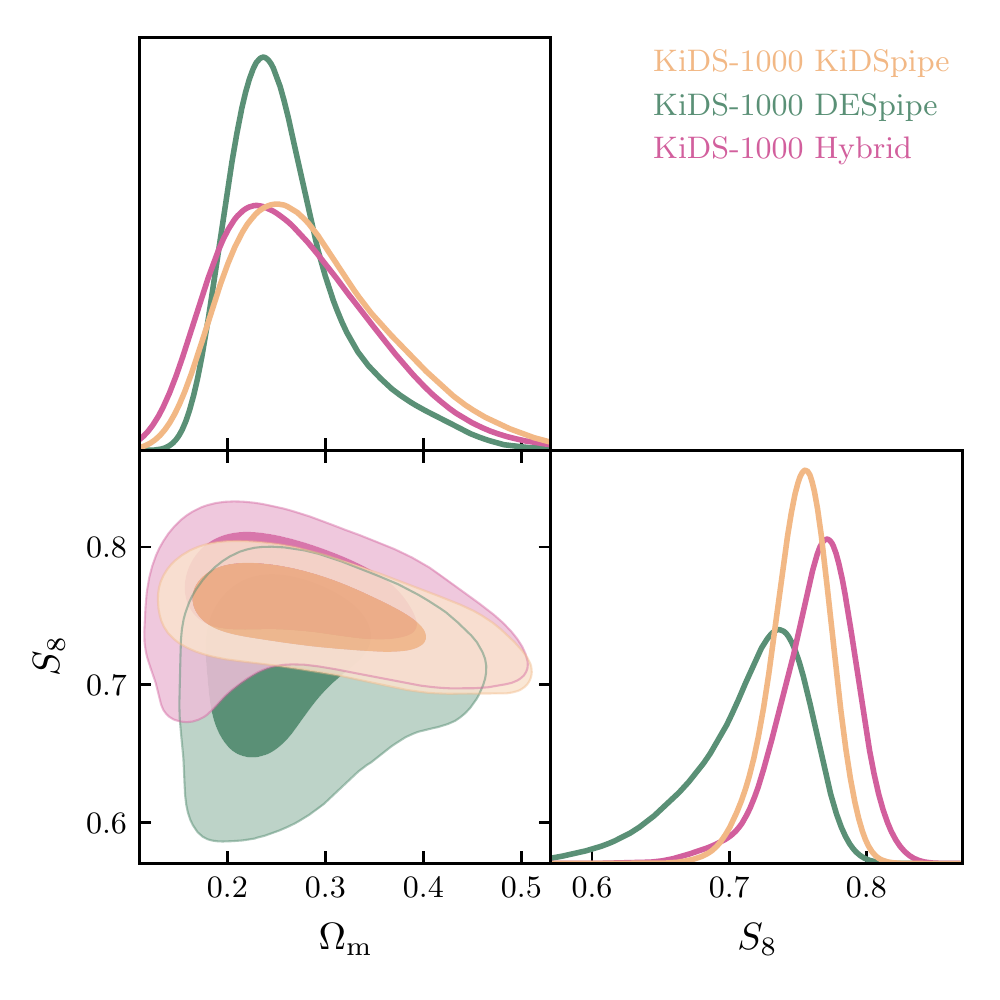}
\end{minipage}
\caption{Cosmological constraints on $S_8$ and the matter density $\Omega_{\rm m}$ from DES Y3 (left) and KiDS-1000 (right): comparing our fiducial Hybrid re-analysis of the two cosmic shear surveys (pink) to analyses that adopt the original DES-like pipeline (green) and KiDS-like pipeline (yellow).  The DES-like analysis of DES Y3 (green left) replicates the $\Lambda$CDM-optimised cosmic shear only constraints from \citet{amon/etal:2022}; \citet*{secco/etal:2022}.   The KiDS-like analysis of KiDS-1000 (yellow right) replicates the COSEBIs cosmic shear constraints from \citet{asgari/etal:2021}. The marginalised posterior contours show the 68\% (inner) and 95\% (outer) credible intervals.}
\label{fig:pipesresult}
\end{figure*}

In this section, we return to the original DES-like and KiDS-like pipelines, summarised in Table~\ref{tab:hybrid}, comparing constraints with our fiducial Hybrid pipeline in the re-analysis of the DES Y3 and KiDS-1000 cosmic shear observations.  Figure~\ref{fig:pipesresult} compares constraints in the $S_8-\Omega_{\rm m}$ plane for DES Y3 (left) and KiDS-1000 (right), using a DES-like analysis (green), a KiDS-like analysis (yellow) and the Hybrid analysis (pink).

For both surveys, the $S_8$ uncertainty relative to the Hybrid analysis increases by $\sim 20\%$ for the DES-like analysis and decreases by $\sim 20\%$ for the KiDS-like analysis, in line with expectations from our mock analysis.  We also see offsets between the DES-like and KiDS-like constraints for the same data set. For our re-analysis of DES Y3 we find an offset between the results from the two analysis pipelines with $\Delta S_8^{\rm DES} = 0.033$.  We can write this offset as a factor of the DES-like analysis error on $S_8$, with $\Delta S_8^{\rm DES} = 1.40\sigma^{\rm DES}_{\rm DES-like}$, or as a factor of the more constraining KiDS-like analysis error on $S_8$, with $\Delta S_8^{\rm DES} = 1.99\sigma^{\rm DES}_{\rm KiDS-like}$. Similar differences are seen when using the two pipelines to analyse KiDS-1000 with $\Delta S_8^{\rm KiDS} = 0.044$.  Casting this offset again as a factor of the DES-like or KiDS-like pipeline's reported error we find  $\Delta S_8^{\rm KiDS} =1.19\sigma^{\rm KiDS}_{\rm DES-like} = 1.97\sigma^{\rm KiDS}_{\rm KiDS-like}$.  In Appendices~\ref{app:HMCodeTATTtest} and~\ref{app:euclidem}, we find that this level of offset is to be expected when analysing the same data set with the modelling combination of TATT and {\sc Halofit} (DES-like) versus NLA (no-z) and {\sc HMCode} (KiDS-like).  

Comparing the Hybrid and KiDS-like analyses, we find similar $S_8$ constraints for the re-analysis of the DES survey in the left panel of Figure~\ref{fig:pipesresult}.  This is expected given that the main difference for the Hybrid pipeline is the use of the {\sc Polychord} sampler, in place of {\sc Multinest}, with additional parameter freedoms of a varied neutrino mass and redshift-dependence in the IA modelling.  These modifications are only expected to significantly impact the constraining power.   In the re-analysis of the KiDS survey, however (see the right panel of Figure~\ref{fig:pipesresult}), we find a larger offset between the Hybrid and KiDS-like constraints, arising from the addition of scale cuts to the COSEBIs data vector in the Hybrid case. Appendix~\ref{app:KiDSCOSEBIs} examines this offset in more detail, finding a $0.7\sigma$ offset in $S_8$ when adopting scale cuts with an {\sc HMCode2016} KiDS-like analysis.  Using a series of mocks, we show that there is a $23$\% chance of such an offset arising from random shape noise.  We also show that an offset of this size is unlikely to be fully attributable to small-scale baryon feedback effects using the \citet{lebrun/etal:2014} {\sc Cosmo-OWLS}:8.7 hydrodynamical simulation as an example of extreme feedback.  For the KiDS-like analyses we find a DES-KiDS Hellinger distance offset in $S_8$ of $2.0\sigma$, and a $\Delta_{\rm tension}$ parameter shift in $S_8-\Omega_{\rm m}$ of $1.8\sigma$, within the $<2.3\sigma$ threshold for consistent data sets.   Given that we have shown in Appendix~\ref{app:samplers} that the KiDS-like pipeline's use of {\sc Multinest} leads to a systematic underestimate of the constraining power, we choose to not present joint-survey constraints using the KiDS-like pipeline. 

\begin{figure}
\centering 
\includegraphics[width=0.5\textwidth]{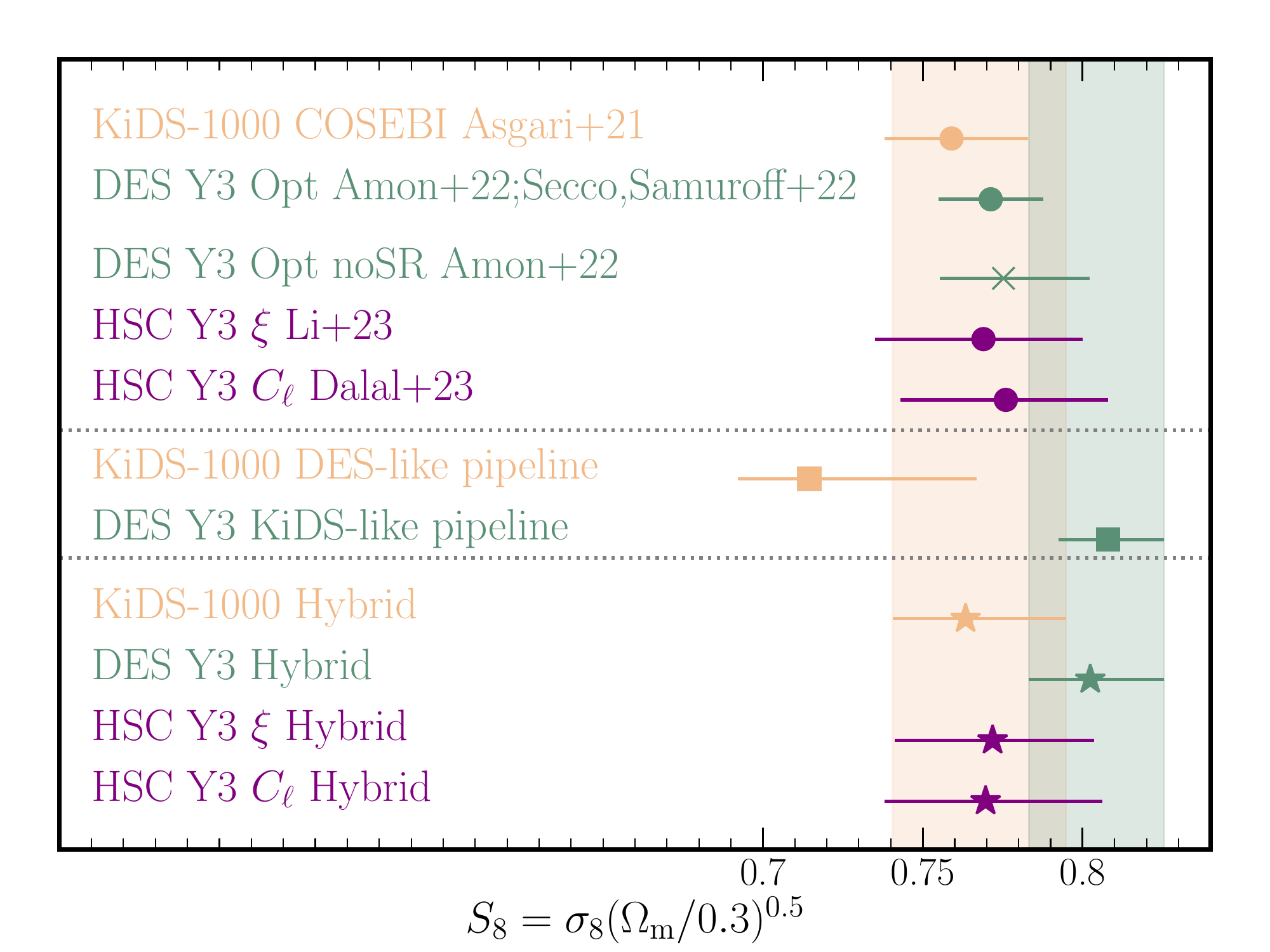}
\caption{Constraints on $S_8$ from the KiDS-1000 (yellow), DES Y3 (green) and HSC Year 3 (purple) surveys, with 68\% credible intervals shown as horizontal bars. In the upper section we compare the published headline results from each survey (circles).  In the case of HSC we show constraints from two cosmic shear statistics, labelled $\xi_\pm(\theta)$ \citep{li/zhang/etal:2023} and $C_\ell$ \citep{dalal/etal:2023}. In the case of DES we show both the primary `\lcdm optimised' cosmic shear result from \citet{amon/etal:2022}; \citet*{secco/etal:2022}, which includes additional shear ratio data (green circle), and the cosmic shear-only constraint (`no SR', green cross).  In the middle section we present constraints from a DES-like re-analysis of KiDS-1000 (yellow square) and a KiDS-like re-analysis of DES Y3 (green square), demonstrating the differences between a TATT with {\sc Halofit} analysis (DES-like) and an NLA (no-z) with {\sc HMCode2016} analysis (KiDS-like).  The lower section presents the results from a unified Hybrid pipeline re-analysis of each survey (stars), along with a shaded vertical bar for reference.}
\label{fig:pipessummary}
\end{figure}   

Comparing the Hybrid and DES-like analyses, we find higher $S_8$ values with the Hybrid setup for both surveys.  This is predicted by our {\sc EuclidEmulatorv2} mock survey analysis in Appendix~\ref{app:hybrid} where changing the non-linear power spectrum model from {\sc Halofit} to {\sc HMCode2020} combined with a change in the IA model from TATT to NLA-z increases $S_8$ by $\sim 1\sigma$. Including a free baryon feedback parameter in the analysis raises $S_8$ by an additional $\sim 0.5\sigma$ when the underlying baryon feedback model is given by OWLS-AGN  \citep[see Table~\ref{tab:EE2mocks} and also][where the impact of these analysis choices are documented for the fiducial DES Y3 analysis]{secco/etal:2022}.  In Figure~\ref{fig:pipessummary} we show that the 34\% increase in $S_8$ uncertainty in the DES-like re-analysis of DES Y3 compared to the headline results from \citet{amon/etal:2022}; \citet*{secco/etal:2022} arises from our decision to focus on a cosmic shear-only analysis, excluding the additional shear ratio data used in the original studies (denoted no SR).  For the DES-like analyses we find a DES-KiDS Hellinger distance offset in $S_8$ of $1.2\sigma$, and a $\Delta_{\rm tension}$ parameter shift in $S_8-\Omega_{\rm m}$ of $0.7\sigma$, within the $<2.3\sigma$ threshold for consistent data sets.   Given the offset between the recovered and input $S_8$ parameters in our DES-like analysis of joint-survey {\sc EuclidEmulatorv2}-based mocks, however, we choose to not present joint-survey constraints using the DES-like pipeline.    In our conclusions, we discuss the implications of these differences for future survey analyses.

\subsection{A comparison of constraints with HSC Year 3}
\label{sec:HSCresults}

In Figure~\ref{fig:pipessummary} we compare the DES Y3 and KiDS-1000 $S_8$ constraints to the cosmic shear analysis of HSC Year 3 data \citep{li/etal:2022}.  The upper section presents the headline HSC results from the two-point shear correlation function analysis, $\xi_\pm(\theta)$ \citep{li/zhang/etal:2023}, and the shear angular power spectrum analysis, $C^{\rm EE}_{\epsilon \epsilon}(\ell)$ \citep{dalal/etal:2023}.  These results are in good agreement with the headline results\footnote{A reminder that the published headline results presented in Figure~\ref{fig:pipessummary} use different statistics to define an $S_8$ value from each analysis. The HSC team chooses to quote the maximum marginal $S_8$ value. The KiDS team chooses to quote the MAP+PJ-HPD.  The DES team chooses to quote the mean marginal $S_8$ value.   Table~\ref{tab:results} demonstrates how these three value estimates differ for the joint DES+KiDS analysis and we refer the reader to the discussion in Section~\ref{sec:inference} on the merits and challenges associated with estimating each value statistic.} from DES Y3 and KiDS-1000.  As demonstrated throughout this paper, however, caution is required when comparing headline results as the offsets induced by different analysis choices can be significant.
   
We compare the HSC Year 3 fiducial analysis choices to the Hybrid approach adopted in this study.
Both analyses use the {\sc Polychord} sampler and to mitigate the impact of baryon feedback, both analyses incorporate scale cuts and model-marginalisation with a free nuisance parameter.  The HSC team uses {\sc HMCode2016} for their headline result, however, in contrast to the Hybrid choice of {\sc HMCode2020}.  The HSC team chooses TATT for their IA model, in contrast to NLA-z.  The $A_{\rm s}$ parameter is chosen to sample over, along with the inclusion of a correction weight \citep{sugiyama/etal:2020}, in contrast to the $S_8$ parameter sampling chosen in the Hybrid setup.  The HSC team also choose to fix the sum of the neutrino masses to $\Sigma m_{\nu}=0.06$eV.  The projection effects and impact of each of these analysis choices is quantified in detail in \citet{dalal/etal:2023,li/zhang/etal:2023}, finding any offsets induced in $S_8$ are less than $\sim 0.5 \sigma$.

In the lower section of Figure~\ref{fig:pipessummary}, we compare DES Y3, HSC Year 3 and KiDS-1000 $S_8$ constraints analysed with the same Hybrid pipeline and reported using mean marginal values. As in the cases of DES and KiDS, in the Hybrid re-analysis of HSC we preserve the observational calibration parameters determined by the survey. We discuss the differences between the published and Hybrid DES and KiDS results in Section~\ref{sec:pipes}.  For HSC Year 3 we find little impact\footnote{When using the Hybrid setup for the $\xi_\pm(\theta)$ HSC analysis, $S_8$ increases by $0.1\sigma$ and the error on $S_8$ decreases by 4\%. For the $C^{\rm EE}_{\epsilon \epsilon}(\ell)$ analysis, $S_8$ decreases by $0.2\sigma$ with the error on $S_8$ increasing by 5\%.  We note that these results are subject to chain-to-chain variance noise which we estimate to be at the level of $\sim 0.1\sigma$ \citep{joachimi/etal:2021}.} on the reported constraints when adopting the Hybrid pipeline in place of the HSC fiducial analysis pipeline.  In contrast to the DES+KiDS joint-survey analysis in Section~\ref{sec:iaresults}, a significant offset is not seen in the HSC analysis when changing IA models from TATT to NLA-z \citep{dalal/etal:2023,li/zhang/etal:2023}.  The differing response to the intrinsic alignment model might be understood given the HSC team's adoption of wide uninformative priors on the $\Delta z$ redshift calibration parameter for bins with $z_{\rm phot}>0.9$.  The HSC data self-calibrates these parameters with a resulting uncertainty that is a factor of 3-9 times larger than the corresponding informative priors adopted by the DES and KiDS teams.

With a Hybrid analysis evaluated for each survey we can directly compare the results within a unified framework.  The constraints are consistent between the three surveys, with all three recovering a lower value for $S_8$ compared to the CMB result from \citet{planck/etal:2020}.  DES Y3 yields the tightest 2.6\% precision measurement on $S_8$ and the highest $S_8$ value out of the three surveys.  KiDS-1000 reports the lowest $S_8$ value from the set with 3.5\% precision.  The HSC Year 3 result lies between DES Y3 and KiDS-1000 with a 4\% precision measurement.  We remind the reader that as almost half of the HSC footprint overlaps with KiDS, with most of the other half overlapping with DES, these surveys are not independent.

We note that the 2\% precision constraints on $S_8$ from our joint Hybrid analysis of DES Y3 and KiDS-1000 are less precise than the 1.6-1.9\% precision constraints reported from the unified analysis of DES Y1, HSC Year 1 and KiDS-1000 by \citet{longley/etal:2022}.   
Excluding HSC Year 1 from their analysis, \citet{longley/etal:2022} report a DES Y1 and KiDS-1000 $S_8$ precision of 2.4\%.  Compared to this analysis, the area of DES Y3 triples relative to DES Y1, but the resulting constraints are systematics limited. In a DES Y3-like analysis of DES Y1, \citet{amon/etal:2022} finds the increased area leads to a $\sim 50\%$ improvement in the DES constraining power.    The main differences between the \citet{longley/etal:2022} fiducial analysis framework and the Hybrid setup include the use of the {\sc Multinest} sampler and the adoption of scale cuts without marginalisation over baryon feedback.  From Appendices~\ref{app:samplers} and~\ref{app:hybrid} we would expect this combination of choices to decrease the \citet{longley/etal:2022} reported uncertainty on $S_8$ by $\sim 20\%$ relative to a Hybrid analysis.  Factoring these known differences together, we conclude that our findings are in line with those from \citet{longley/etal:2022}.

\subsection{Alternative large scale structure observations}
\label{sec:otherprobes}
To date, the most precise constraints on $S_8$ come from the analysis of cosmic shear or the CMB.  Other large scale structure observations also constrain this parameter including galaxy-galaxy lensing; galaxy clustering; redshift space distortions; peculiar velocities; X-ray, optical and thermal SZ cluster counts; CMB lensing; with data sometimes combined in cross-correlation or in a multi-probe analysis \citep[see the discussion in][and references therein]{abdalla/etal:2022,madhavacheril/etal:2023,amon/efstathiou:2022}.  In common with our constraints in Section~\ref{sec:hybrid}, many results in the large scale structure literature favour $S_8$ values that are formally consistent, but low in comparison to {\it Planck}.  

We leave a detailed comparison to other cosmological probes for future work as these results adopt a range of cosmological parameter priors which must be homogenised for a robust assessment.  It is worth noting, however, the exceptions to the low-$S_8$ trend: the CMB lensing results from the Atacama Cosmology Telescope \citep[ACT,][]{madhavacheril/etal:2023} and the South Pole Telescope \citep[SPT,][]{bianchini/etal:2020}.  A joint analysis of ACT CMB lensing with baryon acoustic oscillation observations,  finds a mean marginal value for $S_8 = 0.840 \pm 0.028$, with similarly high but less constraining results from SPT. These results challenge the hypothesis that an $S_8$-tension exists between the {\it Planck} and large scale structure analyses as a result of a non-\lcdm redshift evolution between the CMB early Universe prediction and direct observations of the late-time Universe.  CMB lensing probes linear scales and is sensitive to structure at $0.5<z<5$ with sensitivity peaking at $z\sim 2$. This can be contrasted to cosmic shear, which probes non-linear scales and is currently sensitive to structure at $z<1$.

\section{Conclusions}
\label{sec:conclusions}
In a joint cosmic shear analysis of DES Y3 (\citealt{amon/etal:2022}; \citealt*{secco/etal:2022}) and KiDS-1000 \citep{asgari/etal:2021} we find consistent cosmological constraints with CMB observations from \citet{planck/etal:2020}, at the $1.7\sigma$ level. We constrain cosmological parameters in flat-$\Lambda$CDM, while also varying the neutrino mass, to find a 2\% fractional uncertainty on $S_8$, with a mean-marginal value of $S_8 = 0.790^{+0.018}_{-0.014}$ and a MAP+PJ-HPD value of $S_8=0.801^{+0.011}_{-0.023}$.  Our results are also consistent with cosmic shear constraints from the Hyper Suprime Camera Survey \citep[HSC,][]{dalal/etal:2023,li/zhang/etal:2023}.  This analysis adopts a hybrid of the DES Y3 and KiDS-1000 pipelines.  Through a detailed mock analysis, comparing the impact of each modelling choice, we determine that the Hybrid pipeline is robust for the statistical power of the joint survey cosmic shear analysis. Further work will explore whether the Hybrid setup is also suitable for a $3\times2$pt analysis.

Reflecting on two decades of cosmic shear constraints, it is clear that the significant advances made in terms of survey area have not yet been matched by dramatic improvements in constraining power.  This is because with each enhancement in statistical power, as in this joint-survey analysis, it is necessary for modelling choices, nuisance parameters and priors to be re-assessed.  Where there is systematic uncertainty on par with the newly reduced statistical noise, a more conservative approach must be adopted.  As a result, the full statistical power of modern cosmic shear surveys has yet to be realised.  In this analysis we have quantified the impact of making different analysis choices to mitigate systematics arising from astrophysical processes: intrinsic galaxy alignments, non-linear modelling of the dark matter power spectrum and the impact of baryon feedback on the total matter distribution.  We have also investigated different choices for the cosmological parameter priors and samplers for the inference.  

In both our mock and data studies, the most significant changes arise from the choice of IA model: NLA or TATT.  Replacing our fiducial NLA-z analysis with a TATT analysis reduces the cosmological consistency with {\it Planck} to $2.2\sigma$ with the Hellinger distance metric, and $2.3\sigma$ with the parameter-shift metric $\Delta_{\rm tension}$.  As such, the TATT $S_8$ constraints are on the borderline $2.3\sigma$ limit above which we consider there to be evidence of inconsistency.  This result therefore highlights the critical importance of future research to distinguish between these IA models and others \citep[see for example][]{vlah/etal:2020,fortuna/etal:2021}, better inform the parameter priors and develop alternative strategies such as a self-calibrated halo modelling approach \citep{asgari/etal:2023}.  Direct measurements of intrinsic alignments, comparing high resolution large volume hydrodynamical simulations \citep[see for example][]{delgado/etal:2023} with observations from overlapping spectroscopic and imaging surveys (see for example \citealt{mandelbaum/etal:2006}; \citealt*{johnston/etal:2019}; \citealt{samuroff/etal:2022}) will be central to this development.  Significant advances are anticipated with direct-IA analyses of the Physics of the Accelerating Universe Survey \citep[PAUS,][]{johnston/etal:2021}, the Dark Energy Spectroscopic Instrument Bright Galaxy Survey \citep[DESI-BGS,][]{hahn/etal:2022}, the Javalambre-Physics of the Accelerated Universe Astrophysical Survey \citep[J-PAS,][]{benitez/etal:2014} and the 4-metre Multi-Object Spectrograph Telescope Wide Area Vista Extragalactic Survey \citep[4-MOST WAVES,][]{driver/etal:2019}.  It may also become necessary for more complex cosmic shear studies that separate the red and blue galaxy populations to become the future standard \citep{heymans/etal:2013, samuroff/etal:2019, li/etal:2021}, given that these galaxy populations exhibit clear differences in their alignment properties. 

Our fiducial mitigation strategy for baryon feedback combines two complementary approaches.  First, led by the findings of the {\sc BAHAMAS} hydrodynamical simulations \citep{mccarthy/etal:2017,vandaalen/etal:2020}, we set a prior on the likely magnitude range for baryon feedback. Using the {\sc BAHAMAS}-calibrated {\sc HMCode2020} model we then marginalise over our uncertainty on the non-linear suppression of power. Secondly, we adopt a set of scale cuts to reduce our sensitivity to baryon feedback effects, especially those that are not represented by the {\sc BAHAMAS}-calibrated model and prior range. Comparing our fiducial results with an analysis that adopts a non-linear matter power spectrum with no baryon feedback, we find the $S_8$ constraint changes by only $0.4\sigma$.   We note that these findings would likely differ, however, if we had explored more extreme feedback models which tend to raise the mean marginal value found for $S_8$ and decrease the overall constraining power of the analysis \citep{amon/efstathiou:2022,preston/etal:2023}.  Future research is therefore essential to determine robust simulation-data comparisons and thus feedback limitations from a range of hydrodynamical simulations.  The direct calibration of baryonic feedback models through mass-gas cross-correlation analyses \citep[see for example the shear-Sunyaez Zel'dovich cross-correlation constraints from][]{troester/etal:2022,gatti/etal:2022b, pandey/etal:2023} and galaxy-gas cross-correlation analyses \citep{vikram/etal:2017,pandey/etal:2019,koukoufilippas/etal:2019,chiang/etal:2020,amodeo/etal:2021,yan/etal:2021,schneider/etal:2022,sanchez/etal:2023} will also be critical if we are to robustly access cosmic shear information on scales $k>0.1 h {\rm Mpc}^{-1}$ in the {\it Euclid-Roman-Rubin} era.

The choice of the non-linear matter power spectrum model and sampler are found to make less of an impact on our conclusions.  In these two cases the best analysis choice for the Hybrid pipeline was, however, unambiguous.  The {\sc Polychord} sampler was found to accurately recover the wings of the non-Gaussian posterior.  In terms of speed it is, however, significantly slower than the less accurate {\sc Multinest} sampler, for our cosmic shear constraints.  Future speed enhancements are likely attainable, however, with the inclusion of accurate likelihood emulators \citep{spurio-mancini/etal:2022}.
Using the large-scale high resolution suite of dark matter simulations from \citet{EuclidEmv2/etal:2021} we found 1\% level accuracy in predictions of the DES and KiDS cosmic shear observables when adopting {\sc HMCode2020}.  Accurate recovery was found across the full emulator parameter space.  Similar tests are now warranted for $3\times2$pt studies, which are sensitive to different $k$-scales and may therefore reach different conclusions.  This study should also be extended to include the recently released {\sc Aemulus} $\nu$ numerical simulations \citep{derose/etal:2023} which span a broad range of $\sigma_8$ values. 

When reporting the marginalised constraints for constrained parameters, informative priors on unconstrained parameters can introduce an offset relative to the MAP, given the degeneracies between cosmological and nuisance parameters \citep{joachimi/etal:2021,krause/etal:2021,chintalapati/etal:2022}.
Using mock analyses, we have quantified these offsets for different sets of priors, concluding that they are significant enough to caution against the direct comparison of marginal statistics from surveys that adopt different primary parameter sets and priors \citep[see similar conclusions from][]{chang/etal:2019,joudaki/etal:2020,longley/etal:2022}.  With the nuisance parameter space likely to further grow to allow for new flexibility in the modelling of astrophysical systematics, this issue is unlikely to naturally resolve.  The MAP statistic is insensitive to projection effects, but with the MAP being challenging to accurately determine and a standardised method to report credible intervals around the MAP yet to be widely adopted, the MAP is traditionally underutilised.  On the question of consistency or tension, we therefore encourage future studies to continue publishing chains, data vectors, covariances and software to allow for the homogenisation of priors in comparison analyses and the subsequent direct comparison of posteriors, rather than marginal statistics.

This is the first joint collaboration cosmic shear analysis.  We have found it to be a highly productive learning experience. Previous independent cosmic shear analyses of public survey data certainly served to drive the field forward.  By enabling a direct and sustained interaction between the two teams in this joint study, we found our collective understanding has been accelerated in a way that would not have been feasible from literature reviews alone.  We recommend this approach to future cosmology experiments with independent but complementary teams to promote innovation and allow for the accurate comparison of results.

\section*{Acknowledgements}
{\bf Software}: The figures in this work were created with {\sc matplotlib} \citep{hunter:2007} and {\sc getdist} \citep{lewis:2019}, making use of the {\sc CosmoSIS} \citep{zuntz/etal:2015}, {\sc ChainConsumer} \citep{hinton:2016}, {\sc numpy} \citep{harris/etal:2020}, {\sc scipy} \citep{jones:2001} and {\sc tensiometer} \citep{raveri/doux:2021} software packages. 

{\bf Authors}: 
We thank the anonymous referee for their thorough review, positive comments and constructive remarks.  We acknowledge support from: the European Research Council under grant numbers 647112 (MA, BG, CH, CL, TT) and 770935 (AD, HHi, CM, RR, JLvdB, AHW); the UK Science and Technology Facilities Council (STFC) under grant numbers ST/V000594/1 (CH, JP, AP, NR, JZ) and ST/V000780/1 (BJ); the Max Planck Society and the Alexander von Humboldt Foundation in the framework of the Max Planck-Humboldt Research Award endowed by the Federal Ministry of Education and Research (CH, BS, MY); the Netherlands Organisation for Scientific Research (NWO) under grant numbers 621.016.402 (JTAdJ) and 639.043.512 (HHo). AA acknowledges the support of a Kavli Fellowship. AC acknowledges support from the SPHEREx project under a contract from NASA/Goddard Space Flight Center to the California Institute of Technology. MB is supported by the Polish National Science Center through grants 2020/38/E/ST9/00395, 2018/30/E/ST9/00698, 2018/31/G/ST9/03388 and 2020/39/B/ST9/03494, and by the Polish Ministry of Science and Higher Education through grant DIR/WK/2018/12. RD acknowledges support from the NSF Graduate Research Fellowship Program under grant number DGE-2039656. BG acknowledges the support of the Royal Society through an Enhancement Award (RGF/EA/181006) and the Royal Society of Edinburgh through the Saltire Early Career Fellowship (ref. number 1914). HHi is supported by a Heisenberg grant of the Deutsche Forschungsgemeinschaft (Hi 1495/5-1). HHi and AHW acknowledge funding from the German Science Foundation DFG, via the Collaborative Research Center SFB1491 ``Cosmic Interacting Matters - From Source to Signal''. KK acknowledges support from the Royal Society and Imperial College. SL is supported by NOVA, the Netherlands Research School for Astronomy. XL is supported in part by the Department of Energy grant DE-SC0010118 and in part by a grant from the Simons Foundation (Simons Investigator in Astrophysics, Award ID 620789). NRN acknowledges financial support from the National Science Foundation of China, Research Fund for Excellent International Scholars (12150710511), and from the China Manned Space Project research grant CMS-CSST-2021-A01. SS and JB are supported in part by NSF grant AST-2206563. HYS acknowledges the support from CMS-CSST-2021-A01 and CMS-CSST-2021-A04, NSFC of China under grant 11973070, and Key Research Program of Frontier Sciences, CAS, Grant No. ZDBS-LY-7013 and Program of Shanghai Academic/Technology Research Leader. CS acknowledges support from the Agencia Nacional de Investigaci\'on y Desarrollo (ANID) through FONDECYT grant no. 11191125 and BASAL project FB210003. TT acknowledges funding from the Swiss National Science Foundation under the Ambizione project PZ00P2\_193352. AHW is supported by the Deutsches Zentrum für Luft- und Raumfahrt (DLR), made possible by the Bundesministerium für Wirtschaft und Klimaschutz. For the purpose of open access, the authors have applied a Creative Commons Attribution (CC BY) licence to any Author Accepted Manuscript version arising from this submission.

{\bf Dark Energy Survey}: This project used public archival data from the Dark Energy Survey (DES). Funding for the DES Projects has been provided by the U.S. Department of Energy, the U.S. National Science Foundation, the Ministry of Science and Education of Spain, the Science and Technology Facilities Council of the United Kingdom, the Higher Education Funding Council for England, the National Center for Supercomputing Applications at the University of Illinois at Urbana-Champaign, the Kavli Institute of Cosmological Physics at the University of Chicago, the Center for Cosmology and Astro-Particle Physics at the Ohio State University, the Mitchell Institute for Fundamental Physics and Astronomy at Texas A\&M University, Financiadora de Estudos e Projetos, Funda{\c c}{\~a}o Carlos Chagas Filho de Amparo {\`a} Pesquisa do Estado do Rio de Janeiro, Conselho Nacional de Desenvolvimento Cient{\'i}fico e Tecnol{\'o}gico and the Minist{\'e}rio da Ci{\^e}ncia, Tecnologia e Inova{\c c}{\~a}o, the Deutsche Forschungsgemeinschaft, and the Collaborating Institutions in the Dark Energy Survey. The Collaborating Institutions are Argonne National Laboratory, the University of California at Santa Cruz, the University of Cambridge, Centro de Investigaciones Energ{\'e}ticas, Medioambientales y Tecnol{\'o}gicas-Madrid, the University of Chicago, University College London, the DES-Brazil Consortium, the University of Edinburgh, the Eidgen{\"o}ssische Technische Hochschule (ETH) Z{\"u}rich,  Fermi National Accelerator Laboratory, the University of Illinois at Urbana-Champaign, the Institut de Ci{\`e}ncies de l'Espai (IEEC/CSIC), the Institut de F{\'i}sica d'Altes Energies, Lawrence Berkeley National Laboratory, the Ludwig-Maximilians Universit{\"a}t M{\"u}nchen and the associated Excellence Cluster Universe, the University of Michigan, the National Optical Astronomy Observatory, the University of Nottingham, The Ohio State University, the OzDES Membership Consortium, the University of Pennsylvania, the University of Portsmouth, SLAC National Accelerator Laboratory, Stanford University, the University of Sussex, and Texas A\&M University. The DES results in this paper are based on observations at Cerro Tololo Inter-American Observatory, National Optical Astronomy Observatory, which is operated by the Association of Universities for Research in Astronomy (AURA) under a cooperative agreement with the National Science Foundation. 

{\bf Hyper Suprime-Cam Survey}:  The HSC collaboration includes the astronomical communities of Japan and Taiwan, and Princeton University. The HSC instrumentation and software were developed by the National Astronomical Observatory of Japan (NAOJ), the Kavli Institute for the Physics and Mathematics of the Universe (Kavli IPMU), the University of Tokyo, the High Energy Accelerator Research Organization (KEK), the Academia Sinica Institute for Astronomy and Astrophysics in Taiwan (ASIAA), and Princeton University. Funding was contributed by the FIRST program from the Japanese Cabinet Office, the Ministry of Education, Culture, Sports, Science and Technology (MEXT), the Japan Society for the Promotion of Science (JSPS), Japan Science and Technology Agency (JST), the Toray Science Foundation, NAOJ, Kavli IPMU, KEK, ASIAA, and Princeton University.  The HSC results in this paper make use of software developed for Vera C. Rubin Observatory. We thank the Rubin Obsertevatory for making their code available as free software at \url{http://pipelines.lsst.io/}. The HSC results in this paper are based on data collected at the Subaru Telescope and retrieved from the HSC data archive system, which is operated by the Subaru Telescope and Astronomy Data Center (ADC) at NAOJ. Data analysis was in part carried out with the cooperation of Center for Computational Astrophysics (CfCA), NAOJ. We are honoured and grateful for the opportunity of observing the Universe from Maunakea, which has the cultural, historical and natural significance in Hawaii. 

{\bf Kilo-Degree Survey}: The KiDS results in this paper are based on observations made with ESO Telescopes at the La Silla Paranal Observatory under programme IDs 177.A-3016, 177.A-3017, 177.A-3018 and 179.A-2004, and on data products produced by the KiDS consortium. The KiDS production team acknowledges support from: Deutsche Forschungsgemeinschaft, ERC, NOVA and NWO-M grants; Target; the University of Padova, and the University Federico II (Naples).  Data processing for VIKING has been contributed by the VISTA Data Flow System at CASU, Cambridge and WFAU, Edinburgh.

\section*{Author contributions}  
All authors contributed to this paper, to the data products, the data calibration and/or the scientific analysis and interpretation.  A core team of DES and KiDS members brought together the wide body of expertise from the two survey teams along with two extensive banks of software to create the Hybrid pipeline and conduct the cosmological analysis: Alexandra Amon, Marika Asgari, Ami Choi, Catherine Heymans, Anna Porredon, Simon Samuroff and Joe Zuntz.  Additional support was provided by Felipe Andrade-Oliveira, Gary Bernstein, Jonathan Blazek, Hugo Camacho, Roohi Dalal, Konrad Kuijken, Xiangchong Li, Chieh-An Lin and Jessie Muir and from the internal reviewers Chihway Chang, Scott Dodelson, Hendrik Hildebrandt and Michael Troxel. 

\section*{Data and Software Availability}
The \href{https://des.ncsa.illinois.edu/releases/y3a2}{DES Y3} and \href{https://kids.strw.leidenuniv.nl/DR4/lensing.php}{KiDS-1000} catalogues and data products are freely available for download. The fiducial chains from the DES Y3, KiDS-1000 and Joint analysis can be accessed at \url{https://des.ncsa.illinois.edu/releases/y3a2/Y3key-joint-des-kids}.  In {\sc CosmoSIS} \href{https://github.com/joezuntz/cosmosis-standard-library/releases/tag/v3.3}{v3.3} we include our joint survey likelihood with the example file des-y3\_and\_kids-1000.ini to allow for the easy reproduction of this Hybrid analysis.  This version of {\sc CosmoSIS} allows for $S_8$ sampling, the use of correlated priors for nuisance parameters, and includes a subset of the software from the \href{https://github.com/maricool/2pt_stats}{Asgari COSEBIs library}.

\bibliographystyle{mnras}
\bibliography{DESKiDS_cosmic_shear} 
\normalsize
\section*{Affiliations}
\noindent
\input{affiliations}

\appendix
\section{Survey overlap: excising the KiDS area from the DES footprint}
\label{app:DESminusKiDS}

\begin{figure*}
\centering 
\includegraphics[width=0.8\textwidth]{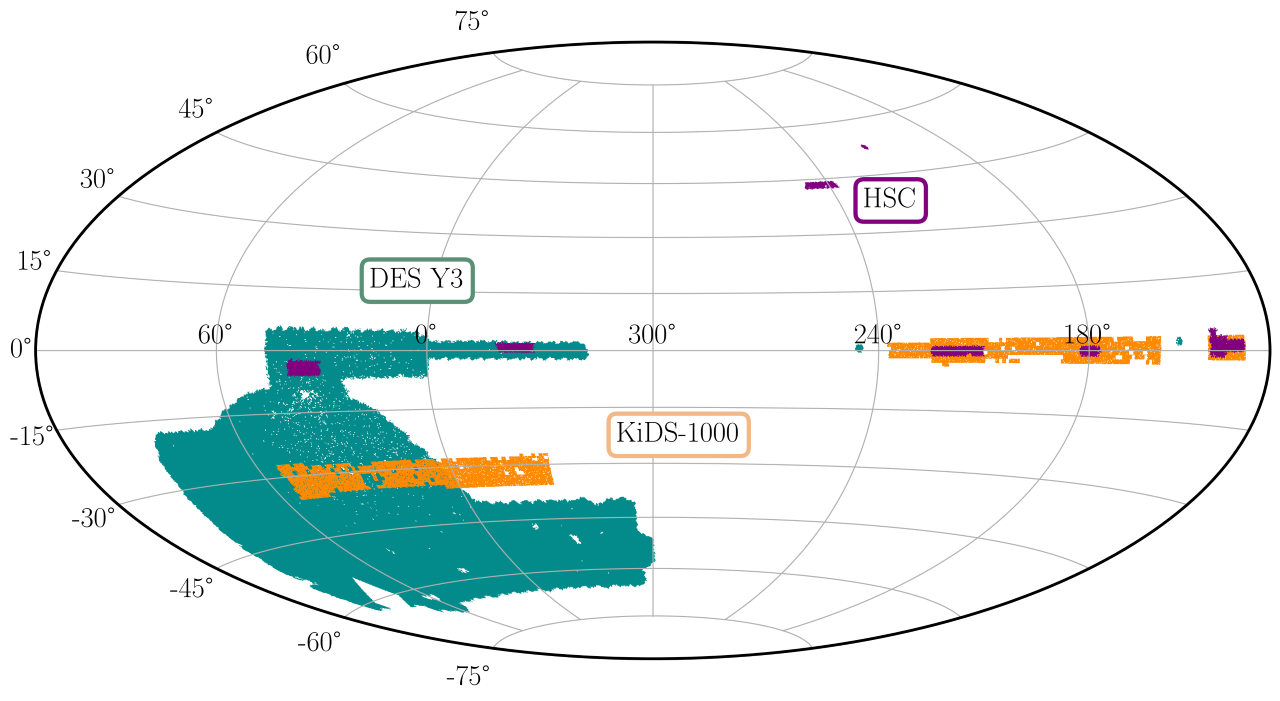}
\caption{Survey footprints from DES Y3 (green) and KiDS-1000 (orange).  The HSC-Y1 footprint (purple) overlaps KiDS in the North and DES in the South complicating the modelling of cross-survey covariance.  For this reason, we limit our joint-survey analysis to DES and KiDS.}
\label{fig:footprint}
\end{figure*}   

{\bf Summary}:  In this appendix we use simulations to estimate the cross-covariance between the DES and KiDS surveys when removing the KiDS-overlap region from the DES footprint.  We conclude that the cross-covariance is sufficiently small and can therefore be neglected in the joint survey cosmic shear analysis.  We confirm that the DES footprint modification, which reduces the total DES area by 8\%, does not significantly impact the DES Y3 cosmological parameter constraints. \\

In Figure~\ref{fig:footprint} we compare the DES and KiDS on-sky footprints, highlighting the Southern overlap region covering 471 square degrees, 337/246 of which are unmasked in the DES/KiDS area respectively.   Data in the overlap region represents $8\%$ of DES Y3, and $32\%$ of KiDS-1000, introducing a cross-covariance between the two surveys.  Given the different redshift distributions of the two surveys, it is non-trivial to analytically calculate the cross-covariance.  We therefore choose to reduce the cross-covariance by excising the KiDS-overlap region from the DES survey\footnote{To excise the KiDS-overlap, we remove DES galaxies with $-0.63^{\circ} <{\rm RA} < 53.94^{\circ} $ and  $ -35.62^{\circ} <{\rm Dec} < -26.98^{\circ}$.}, given the relatively smaller impact on the total DES area.   We assume that DES is sufficiently homogenous that the redshift distributions are unchanged by the reduction in area. We re-measure the effective number density, the ellipticity shape noise and the mean shear and selection response, for every tomographic source bin\footnote{The effective number density, response and shape noise are weighted quantities.  The DES Y3 per-galaxy weights are defined from empirical measurements of the observed ellipticity variance across the survey, binned by galaxy-to-PSF size ratio and signal-to-noise \citep[see section 4.3 of][]{gatti/etal:2021}.  It is not necessary to recalculate these galaxy weights for the reduced survey area as the binned measurements automatically allow us to account for the effects of varying PSF size and survey depth in our re-defined footprint.}.  We find $<0.1$\% changes for $\sigma_e$ in all bins, $<1$\% changes for the mean shear response and  $<1$\%  changes for the mean selection response in all but the second bin where the change is 2.6\%.  There are
small changes in $n_{\rm eff}$, which is systematically higher in all bins ranging from 2\% in the highest redshift source bin to 4\% in the lowest redshift source bin. We then re-calculate the analytical covariance for the revised DES footprint following \citet{friedrich/etal:2021}.

\begin{figure*}
\centering 
\includegraphics[width=\textwidth]{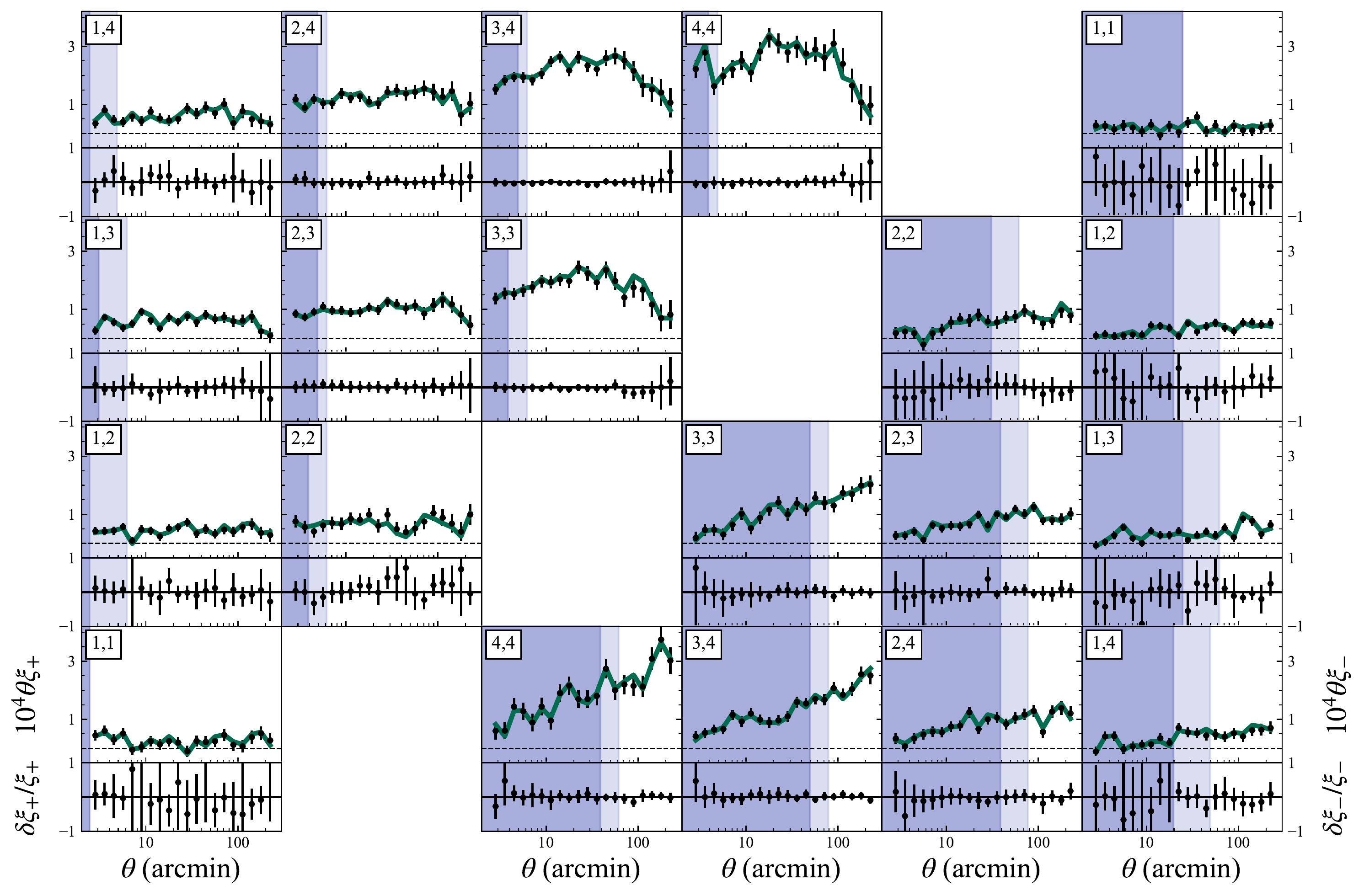}
\caption{The DES Y3 cosmic shear two-point correlation function, $\xi_+(\theta)$ (left) and $\xi_-(\theta)$ (right), for 10 tomographic bin combinations (see label).  The fiducial DES Y3 scale cuts are indicated in pale blue, with the $\Lambda$CDM-optimised scale cuts that are adopted in this cosmic shear analysis indicated in dark blue.  In the upper section of each panel we compare measurements from the full DES Y3 footprint (green) and the footprint with the KiDS-overlap region excised (black) which covers 8\% less area.  The lower section of each panel shows the signal difference as a fraction of the best-fit theory.  The analytical covariance is calculated following \citet{friedrich/etal:2021}, accounting for the reduction in area.}
\label{fig:DESDV_comp}
\end{figure*}

In Figure~\ref{fig:DESDV_comp} we compare the DES data vector from \citet{amon/etal:2022}; \citet*{secco/etal:2022} to the excised DES data vector and covariance that we analyse in Section~\ref{sec:results}.  We find that excising the KiDS area has little impact on the DES cosmological parameter constraints with $S_8^{\rm original} - S_8^{\rm excised} = -0.0004$, using a DES-like analysis and $S_8^{\rm original} - S_8^{\rm excised} = -0.004$, using the Hybrid pipeline.  The constraining power of the reduced survey area decreases when using a DES-like analysis with the 68\% credible interval $\sigma_{S_8}^{\rm original}/\sigma_{S_8}^{\rm excised} = 0.91$. For the Hybrid pipeline analysis, however, we find little impact on the constraining power with $\sigma_{S_8}^{\rm original}/\sigma_{S_8}^{\rm excised} = 1.02$.  We note these numbers will be subject to chain-to-chain variance which we expect to be at the level of $\sim 0.1\sigma$ \citep{joachimi/etal:2021}.

To assess the residual cross-survey covariance that remains after excising the overlap region, we estimate the cross-covariance from 1250 lognormal simulations created with {\sc FLASK} \citep{xavier/etal:2016,friedrich/etal:2021}.   For simplicity, we only simulate the properties of DES Y3, measuring the $\xi_{\pm}(\theta)$ statistic\footnote{For this analysis we measure $\xi_{\pm}(\theta)$ using {\sc TreeCorr} with the bin slop parameter $b=0.1$  \citep{Jarvis/etal:2004} and adopt the fiducial DES Y3 scale cuts. These cuts are more conservative than the $\Lambda$CDM-optimised scale cuts adopted in our cosmic shear analysis (see Figure~\ref{fig:DESDV_comp}).  This difference is unlikely to impact our conclusions, however, as any cross-survey covariance will predominantly impact the larger angular scales that feature in both analyses.} in 4 tomographic bins for the primary DES area where the KiDS-overlap region is excised, and the remaining DES data within the KiDS-overlap region.  The additional KiDS depth and different sampling of $k$-space with the COSEBIs statistic should serve to reduce the cross-survey covariance estimated here. This analysis therefore provides an upper limit of the cross-covariance between the DES and KiDS data vectors analysed in Section~\ref{sec:results}. 

\begin{figure}
\centering 
\includegraphics[width=0.48\textwidth]{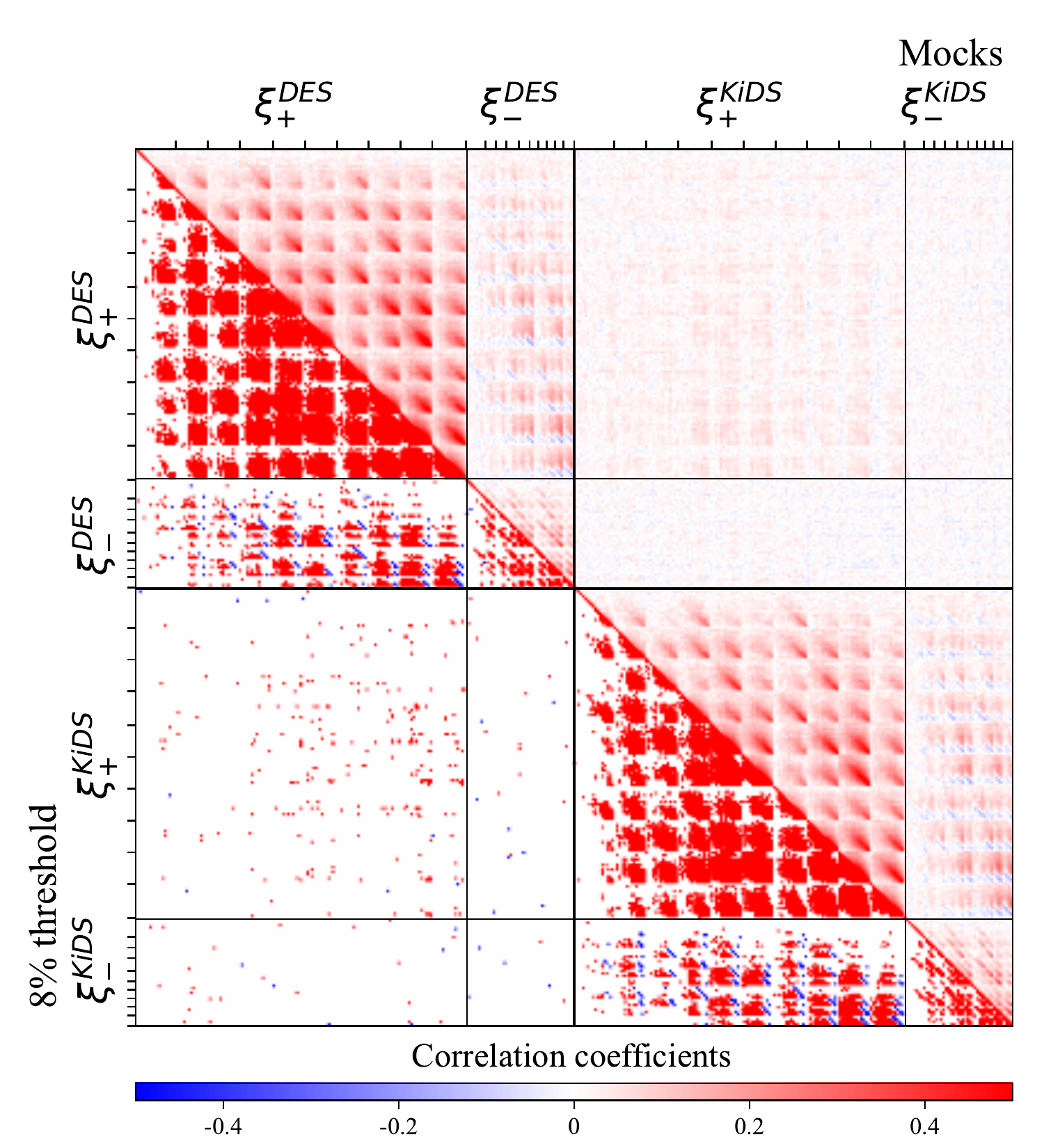}
\caption{Correlation coefficient matrix for the DES Y3 tomographic cosmic shear data vector measured in two areas: the DES Y3 footprint with the KiDS overlap region excised, denoted $\xi_\pm(\theta)^{\rm DES}$, and the KiDS-excised region, denoted $\xi_\pm(\theta)^{\rm KiDS}$.  The upper triangle shows the correlation matrix as calculated from {\sc FLASK} mocks; the lower triangle shows the correlation coefficients above $\pm 8\%$.  The low-levels of correlation between the main DES area and the KiDS-excised area, in the lower-left quadrants, supports our decision to neglect the cross-survey covariance between DES and KiDS in our joint-survey analysis.} 
\label{fig:corrmatrix}
\end{figure}

Figure~\ref{fig:corrmatrix} presents the {\sc FLASK}-simulation estimated joint-survey correlation matrix.   We find the expected level of intra-survey covariance between the different $\xi_+(\theta)$ and $\xi_-(\theta)$ tomographic bins.  For the cross-survey covariance, the majority of correlation coefficients are within a threshold of $\pm 8\%$.  These levels are similar to the study in \citet{joachimi/etal:2021} which concludes the cross-survey covariance can be neglected in a $3\times2$pt analysis when the majority of correlation coefficients are within a threshold of $\pm 5\%$.

We follow \cite{fang/etal:2020,andrade-oliveira/etal:2021} to assess the impact of neglecting cross-survey covariance through a cosmic shear $\chi^2$ analysis of $i=[1..500,000]$ mock joint-survey data vectors with correlated noise randomly sampled from the {\sc FLASK}-estimated covariance $\bf{C}_{\rm full}$.  We measure
\be
\chi^2_{\rm A,i}= [\bf{d}_{i} - \bf{d}_{\rm th}]^{\rm T} \bf{C}_{\rm A}^{-1} [\bf{d}_{i} - \bf{d}_{\rm th}] \,,
\ee
where $\bf{d}_i$ is the data vector for realisation $i$ and $\bf{d}_{\rm th}$ is the theoretical data vector.  The label A indicates the covariance matrix with A=`full' or A=`no-cross' where we zero the cross-survey blocks shown in Figure~\ref{fig:corrmatrix}.  We calculate $\Delta\chi^2_i= \chi^2_{{\rm no-cross},i} - \chi^2_{{\rm full},i}$ finding $\langle \Delta\chi^2 \rangle= 0 \pm 11$ for $N_{\rm d}=454$ data points\footnote{There are $N_{\rm d}=227$ data points in the fiducial DES Y3 cosmic shear analysis, which doubles to $N_{\rm d}=454$ with the inclusion of the KiDS-overlap region as an additional data vector, analysed in the same way as the primary DES-patch.}.  These values are consistent with analytical estimates of $\Delta \chi^2$ from the different covariance matrices using equations 20 and 21 from \citet{andrade-oliveira/etal:2021}.  We therefore conclude that the use of an inaccurate covariance matrix that neglects cross-survey covariance will not introduce any significant effects on the inferred goodness of fit or the best-fit parameter values.   Our joint-survey analysis therefore adopts a covariance matrix for the KiDS and excised-DES data vectors with zero cross-survey correlation.

Our approach differs from that of \citet{longley/etal:2022} who account for the cross-survey covariance between KiDS and HSC in their joint cosmic shear analysis by conservatively assuming 100\% correlation between the Northern stripe of KiDS and half of HSC. They then enlarge either survey's covariance by the ratio of the full survey footprint area and the Northern footprint area.  In lieu of an accurate analytical model, we find our data excision approach to be better suited for KiDS and DES, given the relatively small fractional area of DES which overlaps with KiDS.  In principle, we could have adopted the same approach to also include the public HSC-Y1 catalogues into this joint-survey analysis, excising the smaller HSC-Y1 footprint from the DES equatorial stripe and KiDS-North (see Figure~\ref{fig:footprint}). Given that this would not increase our overall joint-survey area, however, and that the uncertainty in the $S_8$ constraints from this first HSC data release \citep{hikage/etal:2019} are $\sim 40\%$ larger than the KiDS-1000 or DES Y3 results, we choose not to extend our analysis with the inclusion of HSC-Y1.  As half of the complete HSC footprint has full overlap with half of the complete KiDS footprint, any future combination of the final data releases from KiDS and HSC will require an analytical solution to account for cross-survey covariance.

\renewcommand{\thesection}{\Alph{section}}
\section{Mitigating the uncertain impact of baryon feedback with scale cuts}
\label{app:scalecuts}

{\bf Summary}: In this appendix we determine scale cuts for the KiDS COSEBIs data vector, following the baryon feedback mitigation strategy of \citet{krause/etal:2021}. When used in a joint-survey analysis alongside the $\Lambda$CDM-optimised scale cuts for the DES $\xi_\pm(\theta)$ data vector, the Hybrid pipeline is shown to be robust to the presence of baryon feedback at the level of $<0.22 \sigma_{\rm 2D}$.  Here $\sigma_{\rm 2D}$ is the 68\% credible interval in the 2D $(S_8, \Omega_{\rm m})$ marginalised posterior for a joint survey cosmic shear analysis.  \\

We compare constraints in the $(S_8, \Omega_{\rm m})$ plane between a joint-survey analysis of a mock dark-matter only and a mock \citet{vandaalen/etal:2011} {\sc OWLS}-AGN baryon-feedback contaminated data vector\footnote{We note that the {\sc OWLS} AGN model introduces less baryon feedback  suppression to the matter power spectrum than the \citet{lebrun/etal:2014} {\sc Cosmo-OWLS} AGN simulations that were included in the \citet{huang/etal:2021} baryon feedback analysis of DES Y1.  We choose to exclude the {\sc Cosmo-OWLS} models in this analysis as, in contrast to {\sc BAHAMAS} which guides our chosen $T_{\rm AGN}$ prior range in the Hybrid pipeline analysis, they are unable to recover both the observed global stellar mass function \citep{mccarthy/etal:2017} and the observed baryon fraction on group scales \citep{vandaalen/etal:2020}.}.  We then progressively remove small-scale information until the two analyses agree to within $0.3 \sigma_{\rm 2D}$. We conduct this study following \citet{krause/etal:2021} by using the DES methodology, summarised in Table~\ref{tab:hybrid}, and adopting the best fit flat-$\Lambda$CDM cosmological parameters from \citet{planck/etal:2020}.

In contrast to the DES-adopted shear correlation function, $\xi_{\pm}(\theta)$, the KiDS-adopted COSEBIs statistic does not straightforwardly allow for redshift-dependent scale cuts.  The theoretical prediction for each of the $E_{\rm n}$ and $B_{\rm n}$ tomographic modes is dependent on a fixed angular range $\theta_{\rm min} < \theta < \theta_{\rm max}$ (see Equation~\ref{eqn:cosebis}).  Whilst predictions for varied angular ranges for each tomographic bin combination, and the corresponding analytical covariance calculation, can in principle be calculated, this approach would require an extensive update to the existing COSEBIs software. We therefore take a pragmatic approach in this study, fixing the redshift-dependent DES scale cuts to the \citet{amon/etal:2022}; \citet*{secco/etal:2022} defined $\Lambda$CDM-optimised scales, which predict a baryon feedback bias in a DES Y3 $\Lambda$CDM cosmic shear analysis of $<0.14 \sigma_{\rm 2D}$.  We then progressively increase $\theta_{\rm min}$ for the KiDS-COSEBIs analysis, in all tomographic bins, until the joint-survey analysis meets the \citet{krause/etal:2021} robustness criteria.

\begin{figure}
\centering 
\includegraphics[width=0.48\textwidth]{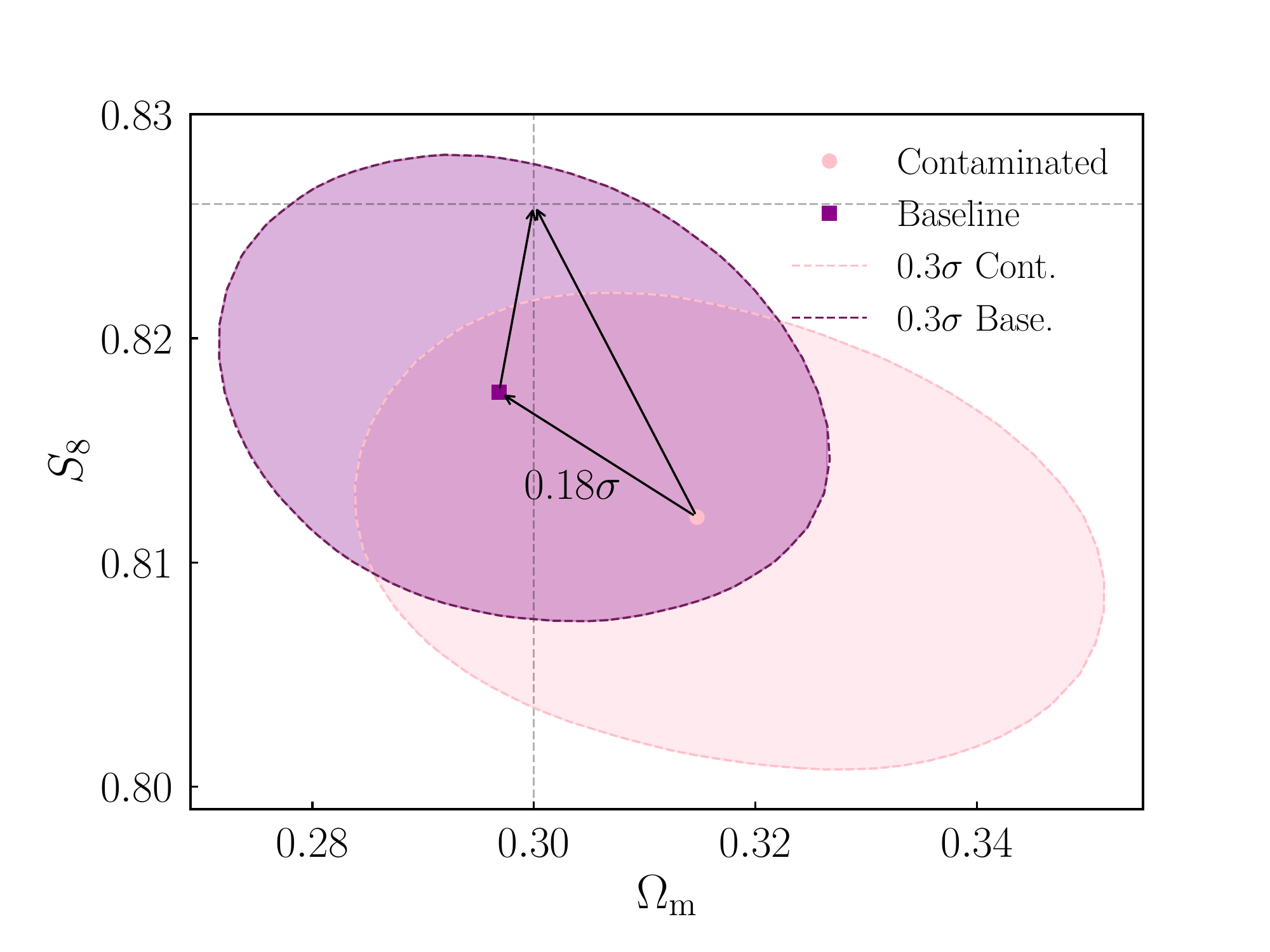}
\caption{Cosmological constraints in the $\Omega_{\rm m} - S_8$ plane from a mock DES-like joint-survey analysis of data vectors from a dark-matter only Universe (`Baseline', purple), and an {\sc OWLS}-AGN baryon-feedback contaminated Universe (`Contaminated', pink), for an input cosmology $\Omega_{\rm m} = 0.3$ and $S_8 = 0.826$ (grey-dashed).  The contours are set at $0.3 \sigma_{\rm 2D}$, where $\sigma_{\rm 2D}$ is the 68\% credible interval in the 2D marginalised posterior.  In this analysis the KiDS COSEBIs data vector has been created with an increased minimum angular scale of $\theta_{\rm min}=$ 2\decimalarcmin0, leading to an offset from the Baseline dark-matter only case of $0.18 \sigma_{\rm 2D}$.  We discuss the projection effects that impact both the Baseline and Contaminated analysis in Appendix~\ref{app:projection}.}
\label{fig:scale_cuts}
\end{figure}

For the KiDS-adopted $\theta_{\rm min}=$ 0\decimalarcmin5, we do not meet the $<0.3 \sigma_{\rm 2D}$ offset criteria for a KiDS-only analysis using the DES framework.  We find that raising $\theta_{\rm min}=$ 2\decimalarcmin0 is sufficient to mitigate the bias from {\sc OWLS}-AGN baryon feedback, increasing the error on $S_8$ by only $\sim 10\%$.  In Figure~\ref{fig:scale_cuts} we show that with this limit for KiDS, in addition to the redshift-dependent DES scale cuts from \citet{amon/etal:2022}; \citet*{secco/etal:2022}, the \citet{krause/etal:2021} methodology predicts a baryon feedback bias of $<0.18 \sigma_{\rm 2D}$ in a DES-like joint-survey cosmic shear analysis.   We note that raising $\theta_{\rm min}>$ 2\decimalarcmin0 was found to significantly reduce the joint survey constraining power, without any significant reduction in the bias. For example with $\theta_{\rm min}=$5\decimalarcmin0, the fractional error on $S_8$ increased, relative to the $\theta_{\rm min}=$ 0\decimalarcmin5 case, by $\sim 30\%$, with the bias remaining at $\sim 0.2 \sigma_{\rm 2D}$.

We find that this scale cut combination is also sufficient for our Hybrid set-up, using the mock analysis in Appendix~\ref{app:hybrid}.  Here we measure a baryon feedback bias for the joint cosmic shear analysis of $<0.14 \sigma_{\rm 2D}$ with scale cuts alone, and $<0.22 \sigma_{\rm 2D}$ for the combination of scale cuts and marginalisation over the $T_{\rm AGN}$ parameter in the {\sc HMCode2020} non-linear model.  We remind the reader that these results will be subject to chain-to-chain variance.  For {\sc Multinest}, marginal constraints are found to differ between successive runs at the level of $\sim 0.1\sigma$ \citep{joachimi/etal:2021}.  We have not carried out a systematic analysis to quantify the level of chain-to-chain variance for {\sc Polychord}, but find $\sim 0.1\sigma$ variations in the handful of {\sc Polychord} chains that have been repeated in this analysis.

\section{Mock data analysis}
\label{app:mocks}

{\bf Summary}: In this appendix we quantify the impact of the different analysis choices in the DES and KiDS fiducial pipelines using a series of mock data vectors. More specifically:
\begin{list}{\labelitemi}{\leftmargin=1em}
\item In Appendix~\ref{app:constrainingpower} we find that a DES-like analysis increases the $1\sigma$ uncertainty on $S_8$ by up to $\sim 60\%$ compared to the corresponding KiDS-like analysis of the same data set.  This distinction primarily arises from the TATT or NLA choice of IA model, with little difference seen in the $S_8$ constraints when switching between each team's baryon feedback mitigation strategy.
\item In Appendix~\ref{sec:J2IA} we review the intrinsic alignment literature, selecting survey-independent IA parameters for our fiducial analysis.  We show this does not degrade the expected constraining power of the analysis.
\item In Appendix~\ref{app:projection} we quantify projection effects in the multi-dimensional parameter space.  We measure the offset of the mean and maximum value of the marginal $S_8$ posterior from the true input $S_8$ value.  When using KiDS priors for a KiDS-like analysis, and DES priors for a DES-like analysis the offset introduced by projection effects on the $S_8$ marginal is negligible. Adopting the KiDS cosmological parameter priors for a DES-like analysis or the DES cosmological parameter priors for KiDS-like analysis, however, results in a $ \sim \pm0.5\sigma$ offset. 
\item In Appendix~\ref{app:HMCodeTATTtest} we analyse a mock data vector created with a TATT IA model and an {\sc HMCode2020} non-linear matter power spectrum with baryon feedback.  In this scenario the different astrophysical modelling choices of the two teams bias the $S_8$ constraints in opposite directions leading to a $\sim 2\sigma$ level tension between a DES-like and a KiDS-like and analysis of the same joint-survey data vector.
\end{list}

\begin{table}
\centering                                      
\begin{tabular}{llll}          
\toprule
Parameter & Mock Input & DES Prior &  KiDS Prior \\    
\midrule
\multicolumn{3}{l}{\bf Cosmological parameters:}\\                             
\midrule
$10^{-9}\As$ & 2.41 & $\bb{0.5,\,5.0}$ &  - \\
$S_8$ & $0.759$ & - & $\bb{0.1,1.3}$ \\
$h$ & 0.767 & $\bb{0.55,0.91}$ & $\bb{0.64,0.82}$ \\
$\Om$ & 0.246 & $\bb{0.1,0.9}$ & - \\
$\omega_{\rm c}$& 0.118 & - & $\bb{0.051,0.255}$\\
$\Ob$ & 0.044 & $\bb{0.03,0.07}$ & - \\
$\ob$ & 0.026 & - & $\bb{0.019,0.026}$\\
$\ns$ & 0.90 & $\bb{0.87,\,1.07}$ & $\bb{0.84,\,1.1}$\\
$1000\,\Omega_\nu h^2$ & 0.64 & $\bb{0.6,6.44}$ & - \\
\midrule   
\multicolumn{3}{l}{\bf Astrophysical systematic model parameters:}\\      
\midrule
\multicolumn{3}{l}{\bf TATT Intrinsic Alignments: `Strong'}\\                             
$a_1$ & -0.44 & $\bb{-5,5}$ & - \\
$\eta_1$ & 4.33 & $\bb{-5,5}$ & - \\
$a_2$ & 1.25 & $\bb{-5,5}$ & - \\
$\eta_2$ & 2.12 & $\bb{-5,5}$ & - \\
$b_{\rm TA}$ & 1.81 & $\bb{0,2}$ &  - \\ 
\multicolumn{3}{l}{\bf TATT Intrinsic Alignments: `Weak'}\\                             
$a_1$ & 0.15 & $\bb{-5,5}$ & - \\
$\eta_1$ & -4.09 & $\bb{-5,5}$ & - \\
$a_2$ & -0.02 & $\bb{-5,5}$ & - \\
$\eta_2$ & 2.31 & $\bb{-5,5}$ & - \\
$b_{\rm TA}$ & 0.04 & $\bb{0,2}$ &  - \\
\multicolumn{3}{l}{\bf NLA Intrinsic Alignments:}\\                             
$A_{\rm IA}$ & 0.26 & $\bb{-6,6}$ & $\bb{-6,6}$ \\
\multicolumn{3}{l}{\bf Baryon Feedback:}\\                             
$\log_{10}(T_{\rm AGN}/{\rm K})$ & 7.8 & - & $\bb{7.6,8.0}$ \\\bottomrule
\end{tabular}
\caption{Cosmological and astrophysical systematic parameters adopted for the mock survey analysis.  The input parameters can be compared to the top-hat prior ranges of the DES-like and KiDS-like  pipelines. The TATT `Strong' parameters are the mean marginal values from the DES Y3 cosmic shear with shear ratio analysis (\citealt{amon/etal:2022}; \citealt*{secco/etal:2022}).  The TATT `Weak' parameters are the mean marginal values from the DES Y3 $3\times2$pt without shear ratio analysis \citep{3x2ptDES/etal:2021}.  The TATT `Strong' model is used to create the mocks analysed with the DES-like pipeline in Appendices~\ref{app:constrainingpower},~\ref{sec:J2IA} and~\ref{app:projection}.  The impact of using the TATT `Weak' or `Strong' parameters as an input is quantified for both a DES-like and KiDS-like mock survey analysis in Appendix~\ref{app:HMCodeTATTtest}.   
\label{tab:mock_input}}          
\end{table}

This study was motivated as an alternative safety measure to a blinded data analysis, as the standard DES and KiDS blinding procedures became redundant when each survey's result entered the public domain.  Given the wide range of conclusions drawn from this mock survey analysis, we recommend that in addition to existing blinding tools, future studies use mocks to validate pipelines in both the fiducial setup \citep[as in][]{joachimi/etal:2021,krause/etal:2021}, and any proposed alternative combinations of models and priors.  

The suite of mock data was created using the cosmological parameters listed in Table~\ref{tab:mock_input} which are in agreement with both the DES Y3 and KiDS-1000 cosmic shear constraints.    Specifically, we choose $S_8=0.759$ to match the identical reported best-fit $S_8$ values from the fiducial cosmic shear analysis from  \citet{amon/etal:2022}; \citet*{secco/etal:2022}, and the fiducial COSEBIs cosmic shear analysis from \citet{asgari/etal:2021}.  The choice is well within $1\sigma$ of the best-fit $S_8$ value from the DES Y3 optimised $\Lambda$CDM cosmic shear analysis where $S_8 = 0.772^{+0.018}_{-0.017}$, and the best-fit $S_8$ from the KiDS-1000 band power spectra cosmic shear analysis where $S_8 = 0.760^{+0.016}_{-0.038}$.

Throughout Appendix~\ref{app:mocks}, when we refer to the `KiDS-like' pipeline, we use the KiDS-1000 \citet{asgari/etal:2021} pipeline with one update: the {\sc HMCode2016} non-linear model has been updated to the improved {\sc HMCode2020} model adopted in the \citet{troester/etal:2021} KiDS-1000 cosmic shear analysis.  At the outset of this project it was our intention to only report constraints from a DES-like and KiDS-like analysis and the KiDS team wished to use the version of their pipeline that they considered to be the most accurate.  They chose a $\log_{10}(T_{\rm AGN}/{\rm K})$ prior of $\bb{7.6,8.0}$, to match the allowed baryon feedback range where {\sc BAHAMAS} reproduces the observed baryon fraction at group scale.   In the process of using this updated pipeline to conduct the mock survey study documented in this appendix, we appreciated the necessity to extend the scope of project and define a new Hybrid setup for our headline joint-survey results.  For consistency with the most well known KiDS-1000 cosmic shear results, we chose to revert to the original \citet{asgari/etal:2021} pipeline for the KiDS-like data analysis reported in Section~\ref{sec:pipes}.  In Appendix~\ref{app:KiDSCOSEBIs} we find the KiDS-1000 constraints using {\sc HMCode2016} and {\sc HMCode2020} differ by $0.2\sigma$.  The conclusions that we draw in this appendix are therefore also applicable to the original \citet{asgari/etal:2021} pipeline.

In Appendices~\ref{app:constrainingpower},~\ref{sec:J2IA} and~\ref{app:projection}, when reporting DES-like constraints, the mock surveys have been created and analysed using {\sc Halofit}, with no baryon feedback and a TATT IA model with 5 free parameters.  For the same appendices, when reporting KiDS-like constraints, the mock surveys have been created and analysed using {\sc HMCode2020}, including baryon feedback and an NLA IA model without redshift dependence.   We choose to use a `noise-free' mock data vector, such that for these mocks which match the expectation of each pipeline, the maximum a posteriori (MAP) is found at the set of input parameters.  This data vector is then analysed, adopting the DES Y3, KiDS-1000 and DES Y3+KiDS-1000 cosmic shear covariances, assuming zero correlation between KiDS-1000 and DES Y3 in the joint-survey analysis. 

The parameters encoded for each astrophysical systematic are listed in Table~\ref{tab:mock_input}, as informed by the best-fit values for these parameters in \citet{amon/etal:2022}; \citet*{secco/etal:2022}; \citet{3x2ptDES/etal:2021}; \citet{asgari/etal:2021,troester/etal:2021}. 

\begin{figure}
\centering 
\includegraphics[width=0.48\textwidth]{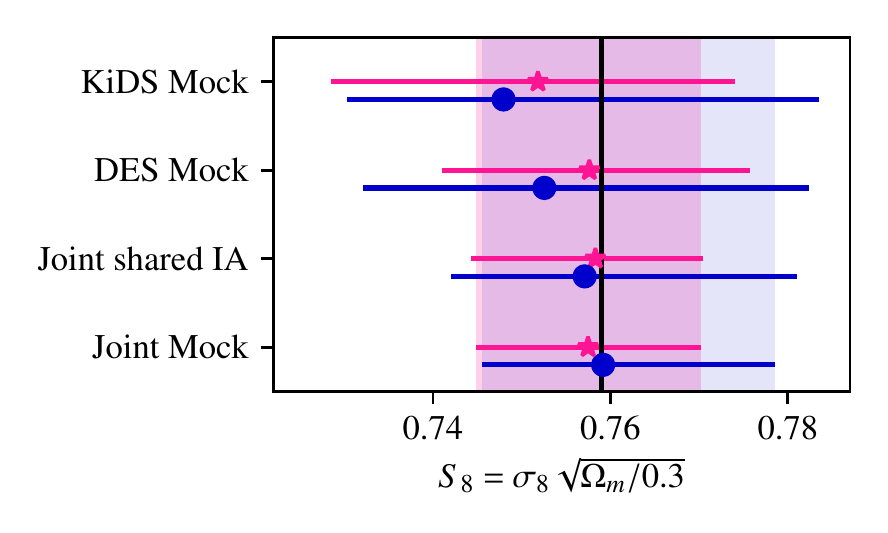}
\caption{$S_8$ constraints reported from a DES-like (blue circle) and KiDS-like (pink star) analysis of mock survey data representing DES, KiDS and a joint survey.   In this figure the noise-free mock survey data {\it differs} for each analysis, as they are created to match the astrophysical systematic framework within which they are analysed. For the DES-like analysis (blue), the mock surveys are created using {\sc Halofit} and a TATT `Strong' IA model (see Table~\ref{tab:mock_input}).  For the KiDS-like analysis (pink), the mock surveys are created using {\sc HMCode} and an NLA IA model.  The input cosmological model and MAP at $S_8=0.759$ is marked in black.   Any offset in $S_8$ is caused by projection effects (see Appendix~\ref{app:projection}).  For the joint-survey analysis we compare an analysis where the IA nuisance parameters are shared between the surveys (Joint shared IA), to an analysis where the IA signal for each survey is considered to be independent, doubling the number of IA parameters (Joint Mock and colour bars).  For the KiDS-like analysis we display the survey-preferred maximum-marginal $S_8$ and 68\% credible intervals.  For the DES-like analysis we display the survey-preferred mean-marginal $S_8$ and 68\% credible intervals.    These data are tabulated in Table~\ref{tab:singlemocks}.}
\label{fig:ineqout}
\end{figure}   

\begin{table}   
\centering         
\input{tabletexfiles/single_survey_table_mock_analysis.tex} 
\caption{$S_8$ constraints for a DES-like and KiDS-like analysis of noise-free mock survey DES Y3 and KiDS-1000 data.  In all cases the mock is created to match the non-linear, IA and baryon feedback astrophysical systematic framework within which they are analysed (see upper panel).  The different rows present the results for the analysis of the KiDS, DES and Joint mocks.  The columns present the results from each pipeline, reporting both the maximum-marginal $S_8$ (Max) and the mean-marginal $S_8$ (Mean) both with 68\% credible intervals.  $\Delta S_8$ quantifies the bias of the recovered $S_8$ relative to the true $S_8 = 0.759$, as a fraction of the $1\sigma$ error.  This offset is induced by a projection effect when reporting one-dimensional marginal constraints from prior-dominated and multi-dimensional degenerate posteriors (see Appendix~\ref{app:projection}).  As the signal-to-noise of each mock data vector is identical across trials, an inspection of the relative $S_8$ constraining power, $\sigma^{\rm survey}/\sigma^{\rm survey}_{\rm min}$, reflects the impact of different analysis choices.  Here $\sigma^{\rm survey}_{\rm min}$ is the minimum reported 68\% credible interval for each mock survey across the four analysis trials (KiDS-like, DES-like, and Max and Mean defined constraints). For the joint-survey analysis we compare an analysis where the IA nuisance parameters are shared between the surveys (J1IA), to an analysis setup where the IA parameters for each survey are independent (Joint).   For reference the lower section compares the constraining power for the KiDS-only, DES-only and joint shared IA analyses (J1IA), to the joint-survey analysis in each scenario. Some of this data is displayed in Figure~\ref{fig:ineqout}.
\label{tab:singlemocks}}
\end{table}

\subsection{The constraining power of a KiDS-like and DES-like cosmic shear analysis}
\label{app:constrainingpower}
\begin{figure}
\centering 
\includegraphics[width=0.48\textwidth]{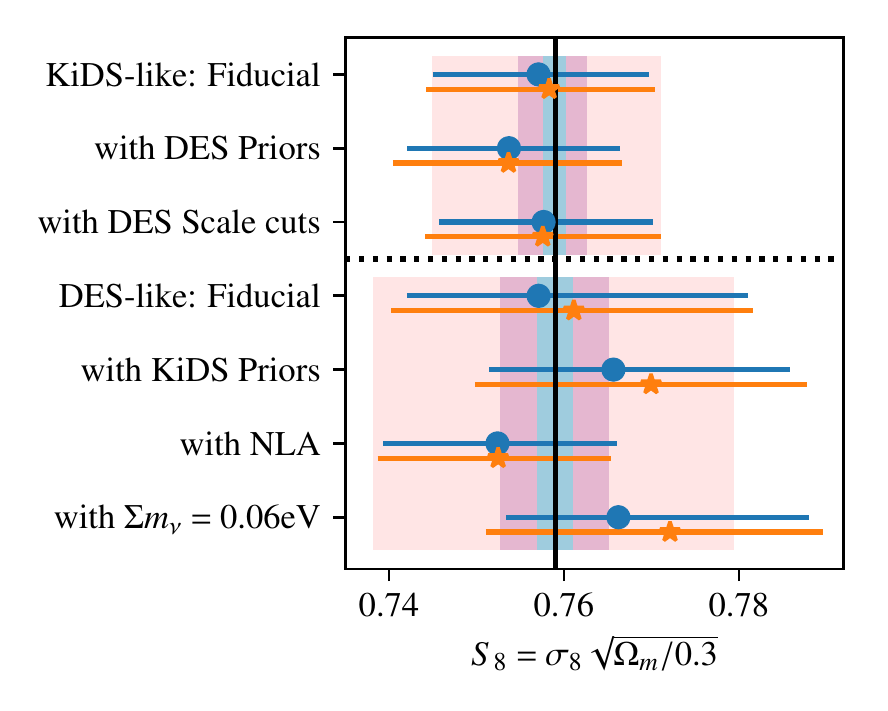}
\caption{Mock joint-survey $S_8$ constraints, comparing the KiDS-like (upper section) and DES-like (lower section) fiducial analyses with modifications: adopting an alternative set of cosmological parameter priors (as listed in the second panel of Table~\ref{tab:params}); adopting alternative astrophysical systematic mitigation strategies with the KiDS-like analysis adopting scale cuts to mitigate baryon feedback (see Appendix~\ref{app:scalecuts}), and the DES-like analysis adopting the NLA model to mitigate IA.   We present both the maximum-marginal $S_8$ and 68\% credible intervals (blue circle), and the mean-marginal $S_8$ and 68\% credible intervals (orange star).  In this figure the mock data matches each survey's chosen astrophysical systematic framework, with an identical input cosmology and signal-to-noise.   For reference the colour bars show $0.1\sigma, 0.3\sigma$ and $1\sigma$ bands from the fiducial analysis, centred on the input $S_8$, marked in black. These data are tabulated in Table~\ref{tab:pipelinemods}.}
\label{fig:pipelinemods}
\end{figure}

In Table~\ref{tab:singlemocks} and Figure~\ref{fig:ineqout} we present $S_8$ constraints from a KiDS-like and DES-like analysis of mock cosmic shear data from DES Y3, KiDS-1000 and a joint-survey analysis.   We find that the DES-like constraints increase the $1\sigma$ uncertainty on $S_8$ by up to $\sim 60\%$ compared to the corresponding KiDS-like analysis.   From the NLA-modified DES-like analysis in Table~\ref{tab:pipelinemods} and Figure~\ref{fig:pipelinemods}, where the difference with the KiDS-like fiducial $S_8$ uncertainties is reduced to $\sim 10\%$, we can conclude that the contrast in constraining power between the KiDS and DES pipelines is mainly driven by the choice of IA model\footnote{We remind the reader that \citet{amon/etal:2022}; \citet*{secco/etal:2022} show that the decrease in constraining power when adopting TATT can be mitigated with the inclusion of additional data from the shear ratio test.}.   The remaining 10\% difference is likely driven by the chosen samplers, where the KiDS-chosen {\sc Multinest} sampler is known to underestimate uncertainty at this level, in comparison to the DES-chosen {\sc Polychord} sampler (see Appendix~\ref{app:samplers} and \citealt*{lemos/etal:2022}). 

The different mitigation strategies for baryon feedback were not found to significantly impact the constraining power in the KiDS-like analysis.  Table~\ref{tab:pipelinemods} and Figure~\ref{fig:pipelinemods} present constraints from a KiDS-like analysis that also adopted DES-like scale cuts to mitigate baryon feedback (see Appendix~\ref{app:scalecuts}).  Given the similar results from these two analyses we can conclude that the small-scale information removed by the DES scale cuts is already being effectively removed by the baryon feedback, $T_{\rm AGN}$, nuisance parameter marginalisation in the KiDS pipeline.  

Comparing the constraints from the individual mock surveys and the joint mock constraints in Table~\ref{tab:singlemocks}, we draw similar conclusions on the expected fractional gain for constraining power, irrespective of the framework used in the analysis.   In the mocks we find the $1\sigma$ uncertainty on $S_8$ from a joint-survey analysis decreases by a factor of $1.4-1.7$ relative to KiDS-only, and a factor of $\sim 1.3$ relative to DES-shear-only.  These results are replicated in our Hybrid pipeline analysis of the DES Y3, KiDS-1000 and joint-survey data in Section~\ref{sec:results}.

\begin{table*}   
\centering                                     

\input{tabletexfiles/table_mock_analysis.tex}
\caption{Quantifying the impact of analysis choices in a joint mock survey analysis where the noise-free mock survey data matches the non-linear, IA and baryon feedback astrophysical systematic framework within which they are analysed.  We report both the maximum-marginal $S_8$ and 68\% credible intervals, and the mean-marginal $S_8$ and 68\% credible intervals.  $\Delta S_8$ quantifies the offset of the recovered $S_8$ from the true $S_8 = 0.759$, as a fraction of the $1\sigma$ error (see appendix~\ref{app:projection}).  The fiducial constraints from the KiDS-like and DES-like joint mock analysis can be compared to a modified analysis where the alternative set of cosmological parameter priors are adopted (as listed in Table~\ref{tab:hybrid}).  We also explore the impact of using an alternative astrophysical systematic mitigation strategy, with the KiDS-like analysis adopting scale cuts to mitigate baryon feedback (see Appendix~\ref{app:scalecuts}), and the DES-like analysis adopting the NLA model to mitigate IA.  As the signal-to-noise of each mock data vector is identical for all tests, an inspection of the $S_8$ constraining power between the trials, $\sigma/\sigma_{\rm min}$, reflects the impact of each survey's analysis choices.  These data are displayed in Figure~\ref{fig:pipelinemods}. \label{tab:pipelinemods} }        
\end{table*}

\subsection{Adopting shared or independent intrinsic alignment parameters}
\label{sec:J2IA}
In a DES-like analysis, five nuisance parameters are used to marginalise over uncertainty in the intrinsic alignment of galaxies.  In a KiDS-like analysis a single parameter is adopted (see Section~\ref{sec:IA} for details).  In direct IA measurements significant variation is seen as a function of galaxy type and luminosity (see for example \citealt{mandelbaum/etal:2006}; \citealt*{johnston/etal:2019}, and references therein).  Furthermore \citet*{singh/mandelbaum:2016,georgiou/etal:2019} measure a dependence of the IA amplitude on the isophote used to define galaxy ellipticity.  This means that the IA nuisance parameters in our cosmic shear analyses constrain a shape-measurement dependent model of the average intrinsic alignment contamination for the source galaxy population of each survey\footnote{Note that some cosmic shear analyses also include a red/blue galaxy split in their intrinsic alignment analysis \citep{heymans/etal:2013, samuroff/etal:2019,li/etal:2021}.}.  

Both \citet{amon/etal:2022}; \citet*{secco/etal:2022} and \citet{asgari/etal:2021} measure an effective intrinsic alignment amplitude that is consistent with zero.  Given the broad similarities between the DES and KiDS survey depths, redshift distributions and shape measurement approaches, we do not expect to detect any statistically significant differences between the effective intrinsic alignment contamination of the two surveys.  As such, it could be preferable to adopt a single set of intrinsic alignment parameters to coherently model IA contamination in a joint-survey analysis, and we denote this approach, `Joint shared IA'.

It has been noted that the IA parameter constraints are somewhat sensitive to systematic errors in the photometric redshift distributions (see for example the $\lesssim 0.5\sigma$ parameter variations in Figure D.1 of \citealt{wright/etal:2020cs}, Figure 18 of \citealt{amon/etal:2022} and Figure 15 of \citealt*{secco/etal:2022}, when different redshift distributions are adopted in the cosmic shear analysis). \citet{fischbacher/etal:2022} quantify the interplay between the free parameters of the IA model and photometric redshift calibration corrections which can lead to biases that are sensitive to prior volume effects. Given the differences between the DES and KiDS approach to redshift calibration, it could therefore be preferable to adopt independent sets of IA parameters, accounting for both the extra freedom that IA-marginalisation affords to absorb redshift errors and the uncertainty over the impact of variations in galaxy population\footnote{See also figure 16 of \citet{amon/etal:2022} which finds significant variation  in the recovered TATT amplitude parameters when different tomographic bins are excluded from the analysis.} and shape measurement methods.  In our fiducial analyses we adopt this `Joint' approach, doubling the number of IA parameters.

In Table~\ref{tab:singlemocks} and Figure~\ref{fig:ineqout} we compare the two scenarios, finding the $1\sigma$ constraints on $S_8$ from the joint analysis to be $\sim 10\%$ smaller than the constraints from the joint shared IA analysis.   This result is counter-intuitive, given that a larger number of free nuisance parameters would typically lead to a less constrained cosmological model.  This behaviour is, however, not replicated in our Hybrid analysis of the data in Section~\ref{sec:iaresults} where the $S_8$ constraints from the shared IA analysis and fiducial joint analysis are indistinguishable. We therefore consider this unexpected result to be an artefact related to the noise-free nature of the mock data vector.

\subsection{Quantifying projection effects and the offset introduced on $S_8$ marginals}
\label{app:projection}  
For the mocks analysed so far in this appendix, the MAP is the input parameter set by definition, as 
we do not apply noise and we use priors that are either flat, or peaked at the input values. We therefore use these mocks to quantify the bias\footnote{Throughout Appendix~\ref{app:mocks} we use the term `bias', defined as the measured offset between the reported $S_8$ value and the input truth.  As this word commonly implies inaccuracy, it is worth highlighting that an offset induced by a projection effect is not a reflection of any error with the measured posterior.  Instead these offsets demonstrate the non-Gaussian nature of the cosmic shear multi-dimensional posterior: the conventional expectation that the mean and maximum marginal distribution will recover the MAP no longer applies.  This analysis is intended to draw attention to the problematic practice of directly comparing 1D marginal values from a range of different surveys and probes where the projection effects will differ.  It also allows for quantification of the marginal offsets inherent to each parameter and prior set, before determining the inaccuracy that arises when the models adopted for the analysis differ from the underlying truth (see Appendices~\ref{app:HMCodeTATTtest} and~\ref{app:euclidem}).} that is introduced, relative to the MAP, when reporting projected 1D marginal constraints \citep[see also][]{chintalapati/etal:2022}.  In Table~\ref{tab:singlemocks} and Figure~\ref{fig:ineqout} we compare $S_8$ values and their 68\% credible intervals defined from the maximum-marginal\footnote{The maximum-marginal posterior is the KiDS-preferred alternative to their fiducial, but computationally expensive, MAP+PJ-HPD estimate.  Here we use {\sc chainconsumer} with the settings statistics=`max' and kde=1.0 \citep{hinton:2016}. The credible intervals are defined using a 1D kernel density estimator (KDE), reporting the asymmetric iso-likelihood levels above and below the KDE maximum.}, and the mean-marginal\footnote{The mean-marginal posterior is the DES-preferred estimator.  Here we use {\sc postprocess} within {\sc CosmoSIS} using the default settings \citep{zuntz/etal:2015}. The credible intervals are defined using a 1D KDE with the equal-posterior asymmetric limits defined to contain the correct number of samples.  This is in contrast to the intervals defining the credible region relative to the smoothed probability volume \citep[see appendix F of][for details]{zuntz/etal:2015}.}.  We find that the reported maximum-marginal credible intervals are $\sim 10\%$ larger than the reported mean-marginal credible intervals for all chains.  This difference is likely a result of the different approaches taken by {\sc chainconsumer} and {\sc CosmoSIS-postprocess} to define credible intervals \citep{zuntz/etal:2015,hinton:2016}.  We find a $\sim 0.2 \sigma$ offset between the mean and maximum marginal for the more skewed {\sc Polychord} chains that include a TATT IA model, which we discuss further in Appendix~\ref{app:samplers}.

In Table~\ref{tab:pipelinemods} and Figure~\ref{fig:pipelinemods} we find that the KiDS-like and DES-like fiducial joint-survey analyses recover the input cosmology within $\sim 0.1 \sigma$, which is the expected run-to-run variance between chains \citep{joachimi/etal:2021}.  Interestingly, however, we find a significant offset when the alternative survey's cosmological priors are adopted (see Table~\ref{tab:params}).  In a modified KiDS-like analysis which adopts the DES cosmological priors, the reported $S_8$ is underestimated by $0.4 \sigma$. In a modified DES-like analysis which adopts the KiDS cosmological priors, the reported $S_8$ is overestimated by $0.4-0.6 \sigma$.  Note in both these modified analyses the astrophysical systematics mitigation strategy for each survey was retained. 

In order to understand the different projection effects with each survey's choice of cosmological priors, we conducted a modified DES-like analysis using the NLA IA model, as both input to the simulated mock and as part of the analysis pipeline, and the DES cosmological priors.  The resulting marginal $S_8$ value is underestimated by $0.5 \sigma$, consistent with the bias found with a KiDS-like NLA analysis adopting DES priors.  We also conducted a modified DES-like analysis adopting a fixed neutrino mass prior in addition to the other DES cosmological priors.  We find the resulting marginal $S_8$ to be overestimated by $0.4-0.7 \sigma$, similar to the bias found for a DES-like analysis which adopts the KiDS cosmological priors with a fixed neutrino mass.  From these modified analyses we conclude that there is a complex interplay between the cosmological parameter priors and the astrophysical systematic prior space, especially the neutrino and IA priors.  As an example of where complexity could arise we note that the non-NLA component of the TATT model, the tidal torque term modulated by the $A_2$ and $\eta_2$ parameters, depends on the linear matter power spectrum.  This means that the calculated contribution to the observed cosmic shear signal from the TATT tidal torque term is insensitive to variations in the neutrino mass or the baryon feedback parameter in an inference analysis.

\citet{chintalapati/etal:2022} show that the picture is further complicated when priors for unconstrained cosmological parameters are asymmetric about the true parameter value, as in the case of the neutrino mass prior in the DES-like analysis of these mocks. Projection effects therefore need to be carefully considered when comparing marginal statistics of surveys that adopt different primary parameter sets and priors \citep[see for example][who find significant shifts in poorly constrained parameters, such as $\Omega_{\rm m}$, with different prior choices]{longley/etal:2022}.

\subsection{Cosmological constraints when the underlying astrophysical model differs from the survey-specific framework}
\label{app:HMCodeTATTtest}

In this section we analyse mock joint-survey data where the underlying astrophysical systematic differs from the theoretical models adopted by each survey, in one way.  In contrast to the mocks previously analysed, the MAP will no longer be found at the set input parameters.  This study therefore quantifies potential systematic biases in the recovered cosmological parameters from a joint-survey analysis which adopts a different astrophysical systematic model from the truth.  This analysis extends and complements existing studies in this area, which we briefly review before presenting the mock analysis. 

In comparing IA models, \citet{blazek/etal:2019} demonstrate that a future {\it Rubin} cosmic shear analysis that adopts the NLA IA model, when the underlying IA truth is TATT, may be subject to a significant bias in the recovered cosmological parameters.  They find an offset $\Delta S_8 \sim 0.08$ for their preferred set of TATT parameters, which would translate into a $\sim 0.7 \sigma$ offset for a joint DES-KiDS analysis.  \citet{fortuna/etal:2021} reach a similar conclusion for a future {\it Euclid}-like survey cosmic shear analysis adopting the NLA IA model, when the underlying IA truth was an aligned analytical halo model. The offset they find is less significant, however, with $\Delta S_8 \sim 0.005$, which would be negligible for this joint-survey study.  \citet*{secco/etal:2022}, analysing mock DES Y3 cosmic shear data created with the DES Y1 best-fit TATT parameters from \citet{samuroff/etal:2019}, find a significant $\sim3\sigma$ offset in $\Omega_{\rm m}$ with a good recovery of the input $S_8$ when adopting an NLA model. \citet{campos/etal:2022} show that the different implications of these studies can be understood by recognising the strong dependence of model tests on the assumed IA parameters.  Defining a range of IA models from 21 samples of the DES Y1 TATT parameter posteriors, they find bias from an NLA analysis of DES Y3-like mocks could range from $0.04 \sigma_{\rm 2D}$ to $5 \sigma_{\rm 2D}$, where $\sigma_{\rm 2D}$ is the 68\% credible interval in the 2D $(S_8, \Omega_{\rm m})$ marginalised posterior. They propose an empirical model selection approach to better inform the choice of IA model in future cosmic shear analyses.

Turning to the non-linear power spectrum model, \citet{krause/etal:2021} use N-body simulations from {\sc Mira Titan}, {\sc Cosmic Emu} \citep{lawrence/etal:2017} and the {\sc EuclidEmulator1} \citep{euclid/knabenhans:2019}, to demonstrate that the {\sc Halofit} non-linear model for the matter power spectrum is sufficiently accurate for the DES Y3 baseline $3\times2$pt analysis with biases at $<0.15 \sigma_{\rm 2D}$. They also conclude that {\sc HMCode2016} fails their requirements for a $2\times2$pt joint galaxy-galaxy lensing and galaxy clustering analysis \citep[see section B1 of][]{krause/etal:2021}.  In contrast, \citet{joachimi/etal:2021} use the {\sc Cosmic Emu} simulations \citep{heitmann/etal:2014} to demonstrate that the {\sc HMCode} non-linear model for the matter power spectrum is sufficiently accurate for the KiDS-1000 baseline analysis with biases at $<0.1 \sigma$ in $S_8$.   They find, however, an unacceptable $0.3 \sigma$ offset in the recovered value of $S_8$ when adopting {\sc Halofit}\footnote{\interlinepenalty10000 The \citet{joachimi/etal:2021} result is consistent with the $0.4 \sigma$ offset in $S_8$ between the {\sc Halofit} and {\sc BaccoEmu} DES Y3 cosmic shear analysis presented in \citet{arico/etal:2023} which they conclude arises from the lower accuracy of {\sc Halofit} over the wide parameter space that is sampled by the posterior.  It is also consistent with the $0.5 \sigma$ offset in $S_8$ between the {\sc Halofit} and {\sc HMCode} DES Y3 cosmic shear analysis presented in \citet*{secco/etal:2022} which they conclude originates from projection effects from the adopted baryon feedback priors which are significantly wider than the {\sc HMCode2020} $T_{\rm AGN}$ priors studied in this analysis.}.  We find that these apparently inconsistent conclusions can be resolved by recognising the differing goals of the two surveys, with DES focussed on optimising the accuracy of the $3\times2$pt analysis and KiDS more focussed on the accuracy of the cosmic shear measurement.  Furthermore, a fixed tolerance level in terms of $\sigma_{\rm 2D}$ in the $\Omega_{\rm m} - S_8$ plane, allows for a less restricted tolerance level in the 1D $S_8$ marginals (see the discussion in Appendix~\ref{app:hybrid}).

\begin{table*}   
\centering                                      

\input{tabletexfiles/mixed_input_and_Euclid_table_mock_analysis.tex}
\caption{Joint survey $S_8$ constraints from mock data where the underlying astrophysical systematic differs from the adopted survey-specific model.   The fiducial analyses can be compared to a scenario where the non-linear matter power spectrum, including baryon feedback, is defined using {\sc HMCode2020}, deviating from the {\sc Halofit} model adopted by DES.  The IA model is then defined using TATT, deviating from the NLA model adopted by KiDS, exploring two strengths for the TATT IA contamination (see Table~\ref{tab:mock_input}).   We report the $S_8$ constraints and 68\% credible interval using both the maximum-marginal and mean-marginal approach, with $\Delta S_8$ quantifying the offset from the true $S_8 = 0.759$, as a fraction of the $1\sigma$ error. These data are displayed in Figure~\ref{fig:Euclidtest}.
\label{tab:Euclidtest}.}        
\end{table*}

\begin{figure}
\centering 
\includegraphics[width=0.48\textwidth]{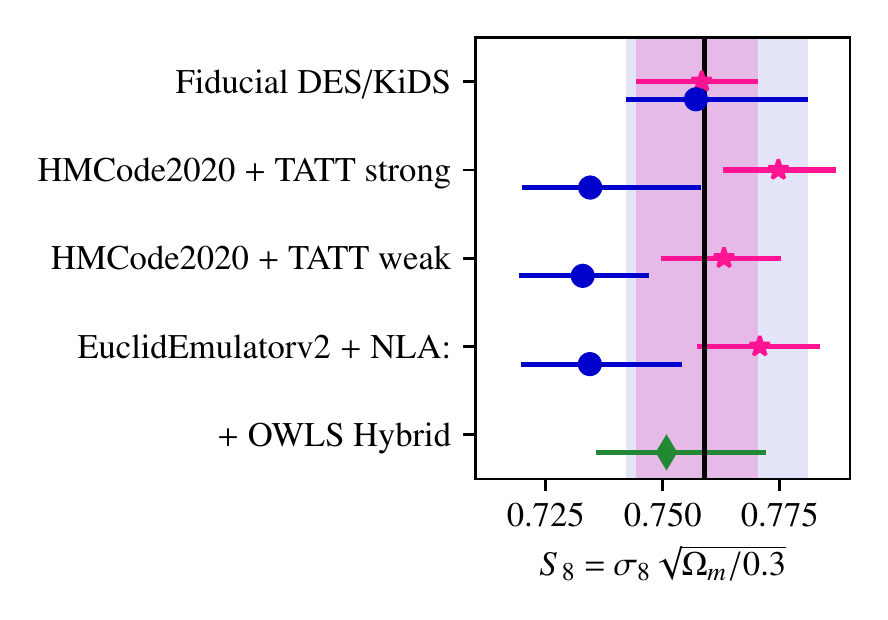}
\caption{Joint survey $S_8$ constraints reported from a DES-like (blue) and KiDS-like (pink) analysis of mock survey data where the underlying astrophysical systematics differ in one way from the theoretical models adopted by each survey: the non-linear matter power spectrum, including baryon feedback, is defined using {\sc HMCode2020} as expected by the KiDS-like pipeline, and the IA contamination is defined using two different strengths of TATT as expected by the DES-like pipeline (see Table~\ref{tab:mock_input}).  For reference the colour bars show the KiDS-like (pink) and DES-like (blue) constraints from the fiducial analyses where the mocks are constructed using the astrophysical systematics expected by each pipeline.  In addition, we present DES-like and KiDS-like constraints for an NLA (no-z) mock where the input power spectrum changes from an {\sc HMCode2020} model including baryon feedback, to a dark matter only {\sc EuclidEmulatorv2} model.  Constraints from our Hybrid analysis are shown in green, analysing the same {\sc EuclidEmulatorv2} mock with the addition of {\sc OWLS}-AGN baryon feedback.  For the KiDS-like analysis we display the survey-preferred maximum-marginal $S_8$ and 68\% credible intervals.  For the DES-like and Hybrid analysis we display the survey-preferred mean-marginal $S_8$ and 68\% credible intervals.    These data are tabulated in Tables~\ref{tab:Euclidtest} and~\ref{tab:EE2mocks}.
\label{fig:Euclidtest}}
\end{figure}

Table~\ref{tab:Euclidtest} and Figure~\ref{fig:Euclidtest} present the $S_8$ constraints from a DES-like and KiDS-like analysis of a joint survey cosmic shear data vector that has been created with one underlying astrophysical systematic that differs from the model adopted in each framework.   The non-linear matter power spectrum is defined using {\sc HMCode2020}.  The baryon feedback is set with an AGN heating temperature, $\log_{10}(T_{\rm AGN}/{\rm K})=7.8$, to match the {\sc BAHAMAS} simulation which best recovers the observed hot gas mass fraction of groups and clusters in addition to the galaxy stellar mass function \citep{mccarthy/etal:2017}. We include intrinsic alignments using the TATT model comparing two sets of TATT parameters.  We take the best-fit TATT parameters\footnote{The $\eta_1$, $\eta_2$, and $b_{\rm TA}$ TATT parameters are formally unconstrained in all DES Y3 analysis combinations.  We denote the DES Y3 cosmic shear with SR parameters TATT `Strong', because the best-fit $\eta_1$ and $b_{\rm TA}$ parameters are close to the allowed upper prior limit for these values. We denote the TATT parameters from the DES Y3 $3\times2$pt without SR MagLim analysis, TATT `Weak' because the $a_2$ and $b_{\rm TA}$ values are close to zero, such that the TATT model is tending towards an NLA model.} from the optimised $\Lambda$CDM cosmic shear analysis of \citet{amon/etal:2022}; \citet*{secco/etal:2022}, denoted  TATT `Strong' in Table~\ref{tab:mock_input}.  We also take the best-fit TATT parameters from the $3\times2$pt analysis in \citet{3x2ptDES/etal:2021}, denoted  TATT `Weak' in Table~\ref{tab:mock_input}. Both of these scenarios are within the range allowed by the most constraining Y3 analysis ($3\times2$pt with shear ratio data), and so are considered plausible given the current data.

Assessing the KiDS-like pipeline, a comparison of the fiducial analysis with the analyses from both {\sc HMCode2020}+TATT mocks reveals the systematic bias in the recovered $S_8$ that is introduced by adopting the NLA model when the underlying truth is a DES Y3-like TATT model.   The mismatch between intrinsic alignment models leads to an overestimate of the true $S_8$ at the level of $1.3\sigma$, in the TATT-strong case, and $0.3\sigma$ in the TATT-weak case.   This result is consistent with the analysis in figure 15 of \citet{amon/etal:2022}, where the $S_8$ NLA constraint is $\sim 0.9\sigma$ higher than the corresponding TATT analysis.   In contrast to the mock analysis of \citet*{secco/etal:2022}, we find that the $\Omega_{\rm m}$ constraints for the {\sc HMCode2020}+TATT mocks are unchanged compared to the fiducial analysis.  This is in agreement with the NLA-TATT analysis comparison in \citet{amon/etal:2022}, demonstrating how sensitive the conclusions of mock studies are to the default TATT parameters adopted for the study.

Assessing the DES-like pipeline, we can compare the fiducial analysis with either of the {\sc HMCode2020}+TATT mock analyses.  This reveals the systematic bias in the recovered $S_8$ that results from adopting a {\sc Halofit} non-linear correction with scale cuts to mitigate baryon feedback, when the underlying truth is a non-linear matter power spectrum from {\sc HMCode2020} including AGN baryon feedback.   The mismatch between non-linear power spectrum models leads to an underestimate of the true $S_8$ at the level of $1.3\sigma$, in the TATT-strong case, and $1.9\sigma$ in the TATT-weak case when reporting mean marginals. As can be seen in Figure~\ref{fig:Euclidtest} the difference between these two results is primarily driven by the reduced $\sigma$ error on $S_8$ in the TATT-weak case.  These offsets reduce to $0.9\sigma$, and $1.6\sigma$, when considering the maximum marginal constraints.  Given the scale-cut analysis in Appendix~\ref{app:scalecuts}, we expect any bias from neglecting baryon feedback to be within $\sim 0.3 \sigma_{\rm 2D}$.  We therefore expect that this offset is primarily caused by the mismatch in the non-linear modelling of the power spectrum.  We explore this further with a dark matter-only mock survey analysis in Appendix~\ref{app:euclidem}.

From this mock joint survey cosmic shear analysis we conclude that neither the KiDS-like nor DES-like analysis choices pass the accuracy requirements set by each survey, when considering an $S_8=0.759$ cosmology with a TATT IA model and an {\sc HMCode2020} non-linear matter power spectrum.   In this specific scenario,  the combination of different systematic modelling choices bias the constraints in opposing directions leading to a $\sim 2\sigma$ level tension between a KiDS-like and DES-like analysis of identical data vectors.  To address this potential issue we developed the concept of the Hybrid pipeline, described in Section~\ref{sec:analysis}, which is validated using the {\sc EuclidEmulatorv2} in Appendix~\ref{app:hybrid}.  The Hybrid setup was finalised before the joint survey data analysis commenced to minimise any confirmation bias in our decision making process.

\section{Sampler comparison}
\label{app:samplers}

{\bf Summary}: In this appendix we find that cosmic shear posteriors sampled with {\sc Multinest} underestimate the width of the 68\%/95\% marginal credible interval for $S_8$ by 12\%/15\%. {\sc Polychord} is shown to provide an accurate recovery, within a few percent, of both the 68\% and 95\% $S_8$ marginal credible interval when compared to a Metropolis-Hastings sampler and {\sc Emcee}.\\

In this appendix we compare the posteriors recovered from a DES Y3 cosmic shear analysis\footnote{We note that this cosmic shear analysis does not include additional information from the shear ratio test.} using two nested sampling algorithms, {\sc Multinest} \citep{feroz/etal:2009} and {\sc Polychord} \citep{handley/etal:2015}, and two Markov Chain Monte Carlo (MCMC) algorithms, Metropolis-Hastings and {\sc emcee} \citep{metropolis/etal:1953, GoodmanWeare:2010, foreman-mackey/etal:2013}.  {\sc Polychord} and {\sc Multinest} vary in their implementation of nested sampling. {\sc Multinest} uses a series of multivariate ellipses to define likelihood thresholds which \citet*{lemos/etal:2022} show can lead to an undersampling of the posterior tails for many-parameter unimodal non-Gaussian posteriors. {\sc Polychord} uses one-dimensional slice sampling, starting at a given set of parameter values to determine the posterior.  Samples are then taken across one parameter only to find a new parameter set with a higher posterior, continuing the iteration one dimension at a time. This approach makes {\sc Polychord} significantly slower\footnote{The {\sc Polychord} DES Y3 cosmic shear analysis adopting a TATT (NLA) IA model, consumed a factor of $\sim 3 (\sim 7)$ times as many CPU hours as the equivalent {\sc Multinest} analysis.} than {\sc Multinest}, but it was found to be more accurate in the DES Y1 lensing and clustering sampling study of \citet*{lemos/etal:2022}.  This appendix extends this study by analysing DES Y3 cosmic shear data from \citet{amon/etal:2022}; \citet*{secco/etal:2022} within the flat $\Lambda$CDM model. Our analysis duplicates and agrees with the recent HSC Y3 cosmic shear sampler comparison \citep{li/zhang/etal:2023}. 

We adopt the DES Y3 sampler settings\footnote{{\sc Polychord} settings:  $n_{\rm live}=500$, $n_{\rm repeats}=60$, tolerance=0.01, fast\_fraction = 0.1, fast\_slow = True.} for {\sc Polychord}, and the KiDS-1000 sampler settings\footnote{{\sc Multinest} settings:  $n_{\rm live}=1000$, efficiency = 0.3, tolerance = 0.01, constant efficiency = False, max\_iterations = 1,000,000.} for {\sc Multinest} (see section 4.2 and 4.3 of \citealt*{lemos/etal:2022} for details of the nested sampling hyper parameters and convergence diagnostics for each software package). To benchmark the results from these two nested samplers we use both a classical Metropolis-Hastings sampler\footnote{We use the {\sc CosmoSIS} Metropolis sampler, running eight independent chains.  We follow \citet*{lemos/etal:2022} in burning the first 20\% of each chain and determine convergence using the Gelman-Rubin test with a tolerance of 0.01.} and the {\sc emcee} affine invariant MCMC sampler\footnote{{\sc emcee} simultaneously evolves an ensemble of walkers, where the position in parameter space of each walker depends on the positions of the remaining walkers. We combine four independent {\sc emcee} chains run with the parameter settings walkers = 160, samples = 12000.  The resulting chains have a total of $\sim 6$M samples with $\sim 2$M burn-in samples removed using {\sc Chainconsumer} diagnostic statistics to indicate convergence according to both the Gelman-Rubin and Geweke metrics \citep{hinton:2016}.  We note that these metrics provide only approximate estimates of convergence for {\sc emcee} given the correlations between the walkers.  A more appropriate metric is the integrated autocorrelation time \citep{GoodmanWeare:2010}.  Unfortunately this statistic was found to be highly sensitive to variations in the methodology chosen to estimate it, suggesting that our chains are relatively short compared to the quantity. Our best estimate suggests that after burn-in removal, there are $\sim 5$ autocorrelation times in each of the four independent chains being combined to produce {\sc emcee} posteriors.}.  Both algorithms start from a proposal distribution which we take from the output of the {\sc Multinest} or {\sc Polychord} analysis.  We do not use these two samplers for our cosmological analysis with the Hybrid pipeline due to the lack of Bayesian evidence estimation.  Furthermore, when analysing large-dimensional parameter spaces and non-Gaussian posterior distributions, the run times to reach convergence can become prohibitive\footnote{Even when started with accurate proposal distributions, the {\sc emcee} (Metropolis-Hastings) DES Y3 cosmic shear analysis consumed a factor of $\sim 10$ ($\sim 1.5$) times as many CPU hours as the equivalent {\sc Polychord} analysis.}. 

\begin{figure}
\centering 
\includegraphics[width=\linewidth]{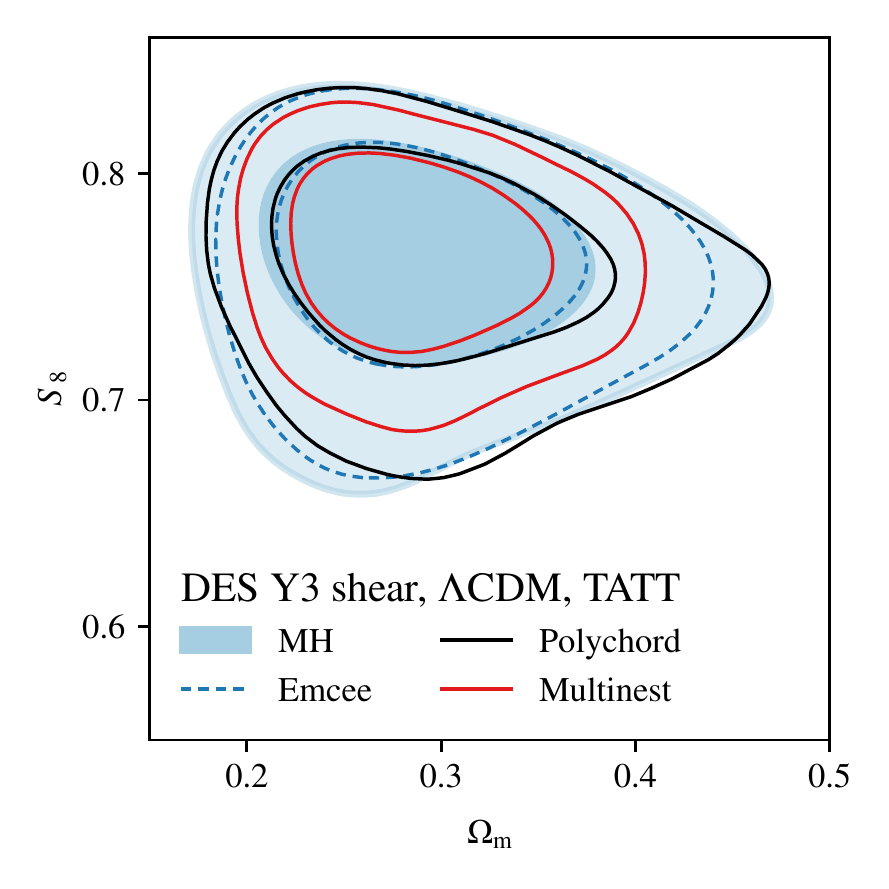}
\caption{Comparing the {\sc Polychord} (black), {\sc Multinest} (red), {\sc emcee} (blue, dashed) and Metropolis-Hastings (MH, light blue, filled) samplers for a DES Y3 cosmic shear analysis: the 68\% and 95\% credible intervals are shown for the two-dimensional posterior distribution in the $\Omega_{\rm m}-S_8$ plane. }
\label{fig:2Dsamp_comp}
\end{figure}

\begin{figure}
\centering 
\includegraphics[width=\linewidth]{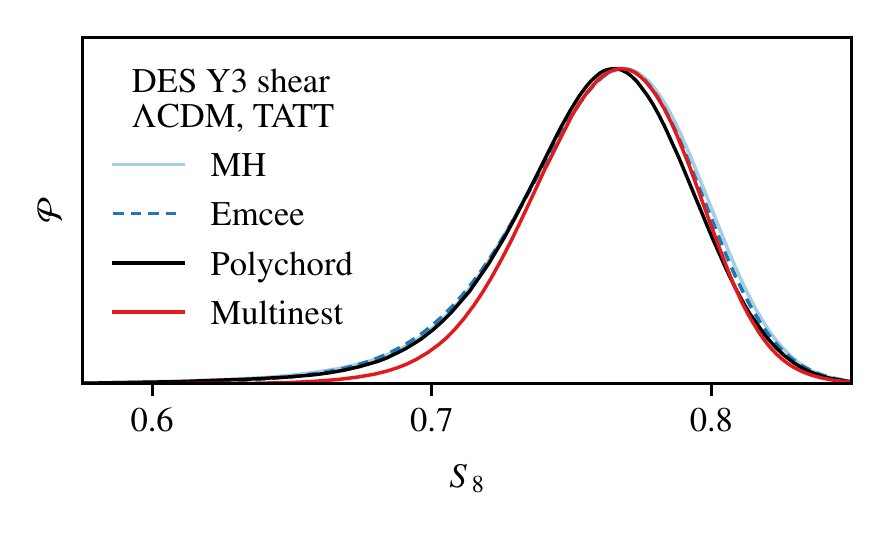}\\
\includegraphics[width=\linewidth]{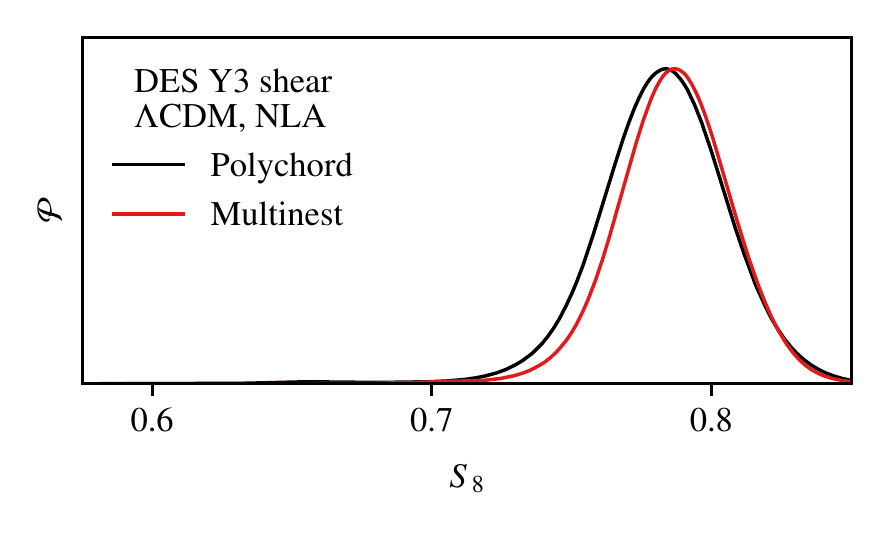}
\caption{Comparing DES Y3 cosmic shear one-dimensional marginalized posteriors for $S_8$ obtained using {\sc Polychord} (black), {\sc Multinest} (red), {\sc emcee} (blue, dashed) and Metropolis-Hastings (MH, light blue) samplers. Results from an analysis using the TATT IA model, matching the posteriors shown in Figure~\ref{fig:2Dsamp_comp}, are shown in the upper row, while the lower row adopts an NLA IA model.} 
\label{fig:1Dsamp_comp}
\end{figure}

Figure~\ref{fig:2Dsamp_comp} shows the 68\% and 95\% credible intervals of the sampled posteriors, marginalised in the $\Omega_{\rm m}-S_8$ plane, where we have adopted the DES methodology framework summarised in Table~\ref{tab:hybrid}.  We find significant differences between the credible intervals, with a narrower {\sc Multinest} posterior compared to the {\sc Polychord}, {\sc emcee} and Metropolis-Hastings posterior.  Specifically we find the width of the 68\% credible interval, $\Delta \sigma_{68} = \sigma_{68}^{\rm upper} - \sigma_{68}^{\rm lower}$, for the $S_8$ marginal, to be $\sim 12\%$ ($\sim 9\%$) narrower compared to the Metropolis-Hastings ({\sc emcee} or {\sc Polychord}) posterior (see Figure~\ref{fig:1Dsamp_comp} and Table~\ref{tab:samplers}).   

Inspecting the width of the 95\% marginal credible interval, the $\sim 15\%$ narrowing of the {\sc Multinest} posterior relative to the `truth' is even more pronounced.  The $S_8$ credible interval from {\sc Polychord} is, however, accurate to a few percent.  We note the {\sc emcee} and Metropolis-Hastings analyses require a proposal distribution of starting guesses, finding a few percent variation between two analyses using initial starting points drawn from the {\sc Polychord} or {\sc Multinest} posteriors (see Table~\ref{fig:1Dsamp_comp} for details, where Figures~\ref{fig:2Dsamp_comp} and~\ref{fig:1Dsamp_comp} display the {\sc Polychord}-starting option).  As such, even though we use the {\sc emcee} and Metropolis-Hastings result as our `truth', there is still some degree of uncertainty in it. 

In Figure~\ref{fig:1Dsamp_comp} we compare the 1D marginalised posteriors for $S_8$ for both the fiducial DES analysis which adopts the TATT IA model (upper panel), and an alternative analysis adopting the NLA (no-z) IA model (lower panel).   We note that the narrowing of the posterior is asymmetrical about the maximum, which is likely a side-effect of the {\sc Multinest} ellipsoidal sampling (see Figure 1 of \citealt*{lemos/etal:2022} for an illustration).  The skew of the {\sc Polychord} $S_8$ posterior, relative to the {\sc Multinest} posterior leads to significant differences between the lower $S_8$ credible interval, and larger offsets between mean $S_8$ marginal estimates, compared to estimates of the maximum of the marginalised $S_8$ posterior.

\begin{table}   
\centering                                     
\begin{tabular}{lccc}         
\toprule
 & $S_8(\rm mean)$ & $\Delta \sigma_{68}$ & $\Delta \sigma_{95}$ \\ \hline
{\bf TATT:} \\ \hline
{\sc Polychord} & 0.759 & 0.063 & 0.134\\ 
{\sc Multinest} & 0.763 & 0.058 & 0.115\\ 
{\sc emcee} (PC) & 0.760 & 0.064 &0.134 \\
{\sc emcee} (MN) & 0.761 & 0.064 & 0.132 \\
Metropolis-Hastings (PC) & 0.760 & 0.065 & 0.135 \\
Metropolis-Hastings (MN) & 0.760 & 0.067 & 0.137 \\\hline
{\bf NLA:} \\ \hline
{\sc Polychord} & 0.782 & 0.043 & 0.091\\ 
{\sc Multinest} & 0.786 & 0.039 & 0.081\\ 
\bottomrule
\end{tabular}
\caption{The impact of sampler choice on one-dimensional marginalized mean $S_8$ constraints from the DES Y3 cosmic shear analyses. For the {\sc emcee} and Metropolis-Hastings analyses, `PC' and `MN' indicate whether the initial starting points were drawn from the {\sc Polychord} or {\sc Multinest} posteriors, respectively. Columns show the {\sc CosmoSIS-postprocess}-defined posterior mean, the full width of the 68\% credible interval ($\Delta \sigma_{68} = \sigma_{68}^{\rm upper} - \sigma_{68}^{\rm lower}$) and the full width of the 95\% credible interval ($\Delta \sigma_{95}= \sigma_{95}^{\rm upper} - \sigma_{95}^{\rm lower}$).}
\label{tab:samplers}
\end{table}

\section{Verifying the Hybrid analysis pipeline}
\label{app:hybrid}
{\bf Summary}: In this appendix we quantify the accuracy of the Hybrid pipeline analysis through a series of mock analyses using the {\sc EuclidEmulatorv2} \citep{EuclidEmv2/etal:2021}.
\begin{list}{\labelitemi}{\leftmargin=1em}
\item In Appendix~\ref{app:NLmodels} we show {\sc HMCode2020} predictions of the DES $\xi_\pm(\theta)$ and KiDS $E_n$ cosmic shear observables agree with the {\sc EuclidEmulatorv2} model at the percent level across a wide range of cosmological parameters.
\item In Appendix~\ref{app:euclidem} we find that using the Appendix~\ref{app:scalecuts}-defined scale cuts alone results in a $0.5\sigma$ bias on $S_8$ in the presence of {\sc OWLS} AGN baryon feedback.  Combining scale cuts with marginalisation over a free AGN feedback parameter $T_{\rm AGN}$, however, reduces the bias to $0.0\sigma$.  We find the inclusion of an free neutrino mass parameter introduces a projection offset in the marginal value of $S_8$ at the level of $0.3 \sigma$.
\end{list}

\subsection{Quantifying the accuracy of non-linear models for DES and KiDS cosmic shear}
\label{app:NLmodels}

\begin{figure*}
\centering 
\includegraphics[width=\textwidth]{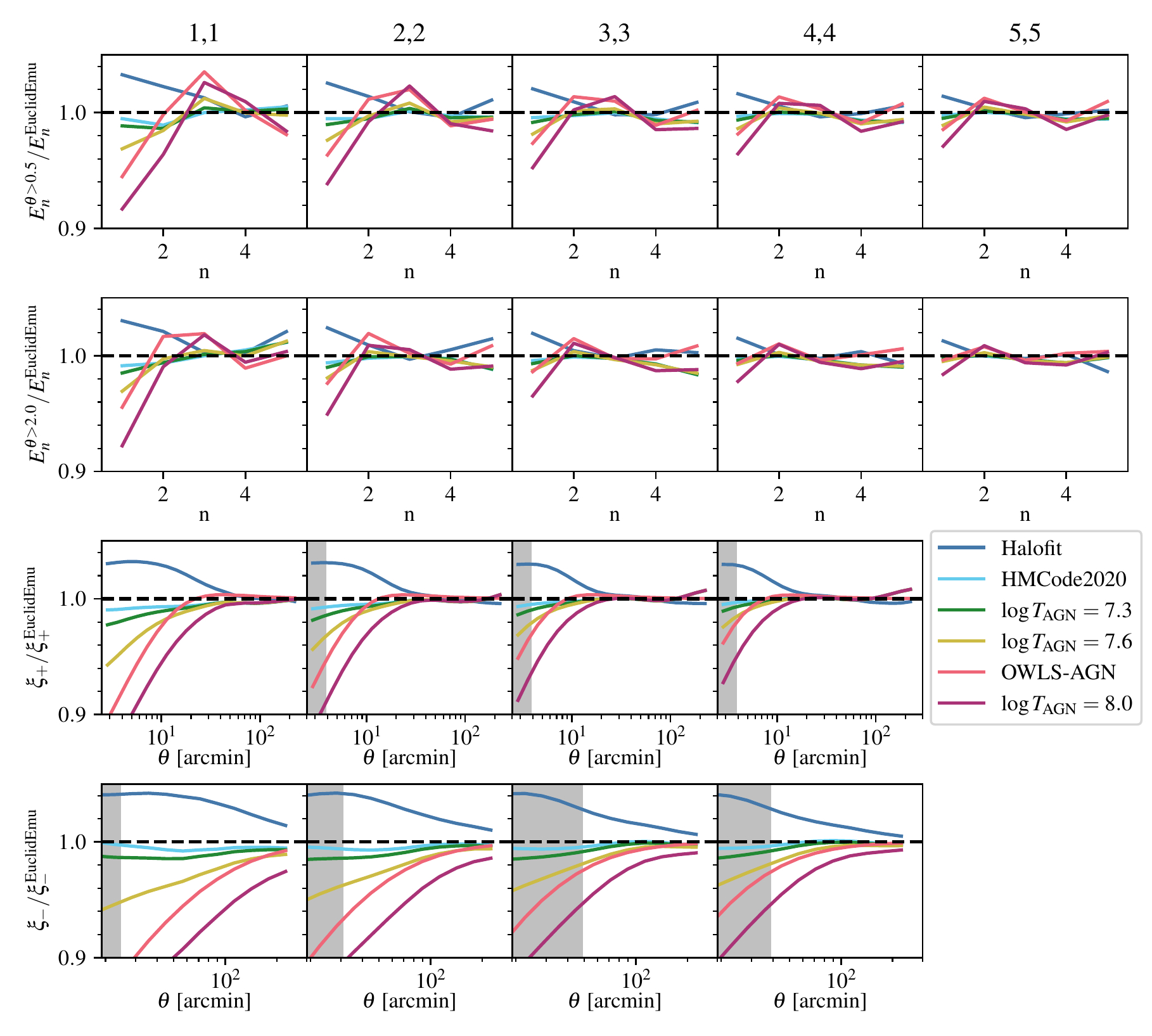}
\caption{The ratio for a range of KiDS $E_n$ (upper) and DES $\xi_\pm(\theta)$ (lower) and cosmic shear models to the prediction from {\sc EuclidEmulatorv2}.  Each auto-correlation tomographic bin is shown, with similar trends found for the cross-correlations (not shown).  The dark matter-only predictions for {\sc Halofit} (blue) and {\sc HMCode2020} (cyan) can be compared to models that include baryon feedback using {\sc OWLS}-AGN (pink) and the {\sc HMCode2020} $T_{\rm AGN}$ parameter with $\log_{10}(T_{\rm AGN}/{\rm K}) = 7.3$ (green), $\log_{10}(T_{\rm AGN}/{\rm K}) = 7.6$ (yellow)  and $\log_{10}(T_{\rm AGN}/{\rm K}) = 8.0$ (purple).  We have verified that cosmic shear signals predicted by the {\sc HMCode2020} $T_{\rm AGN}$ model (green, yellow, purple), agree at the percent level to the signals predicted when using the equivalent {\sc BAHAMAS} power spectrum to directly suppress the {\sc EuclidEmulatorv2} dark matter power spectrum. The grey shaded regions indicate the adopted DES scale cuts.   The upper two rows compare the KiDS fiducial COSEBIs $E_n$ with $\theta_{\rm min}=$ 0\decimalarcmin5, to the scale cut version with $\theta_{\rm min}=$ 2\decimalarcmin0.}
\label{fig:2ptOWLS}
\end{figure*}

In Figure~\ref{fig:2ptOWLS} we compare the DES and KiDS cosmic shear statistics predicted from the {\sc EuclidEmulatorv2} to predictions adopting different non-linear models\footnote{We refer readers interested in a direct comparison between the different $P_\delta(k,z)$ models to figures 2 and D.1 of \citet{mead/etal:2021}, figures 13 and 14 of \citet{EuclidEmv2/etal:2021} and figure 7 of \citet{adamek/etal:2022}.}. The {\sc EuclidEmulatorv2} has been shown to be accurate to $<1$\% for $k \lesssim 10 h {\rm Mpc}^{-1}$ and $z \lesssim 3$ \citep{EuclidEmv2/etal:2021,adamek/etal:2022}.  For the mock survey analysis in this appendix we have modified our input set of cosmological parameters to lie within the emulator's parameter range (see Table~\ref{tab:Euclid_input}), retaining the input $S_8=0.759$, to match the fiducial $S_8$ cosmic shear result from both KiDS-1000 and DES Y3.  With this set of cosmological parameters we find significant offsets in the small-scale cosmic shear signal when adopting {\sc Halofit}.

\begin{figure}
\centering 
\includegraphics[width=0.48\textwidth]{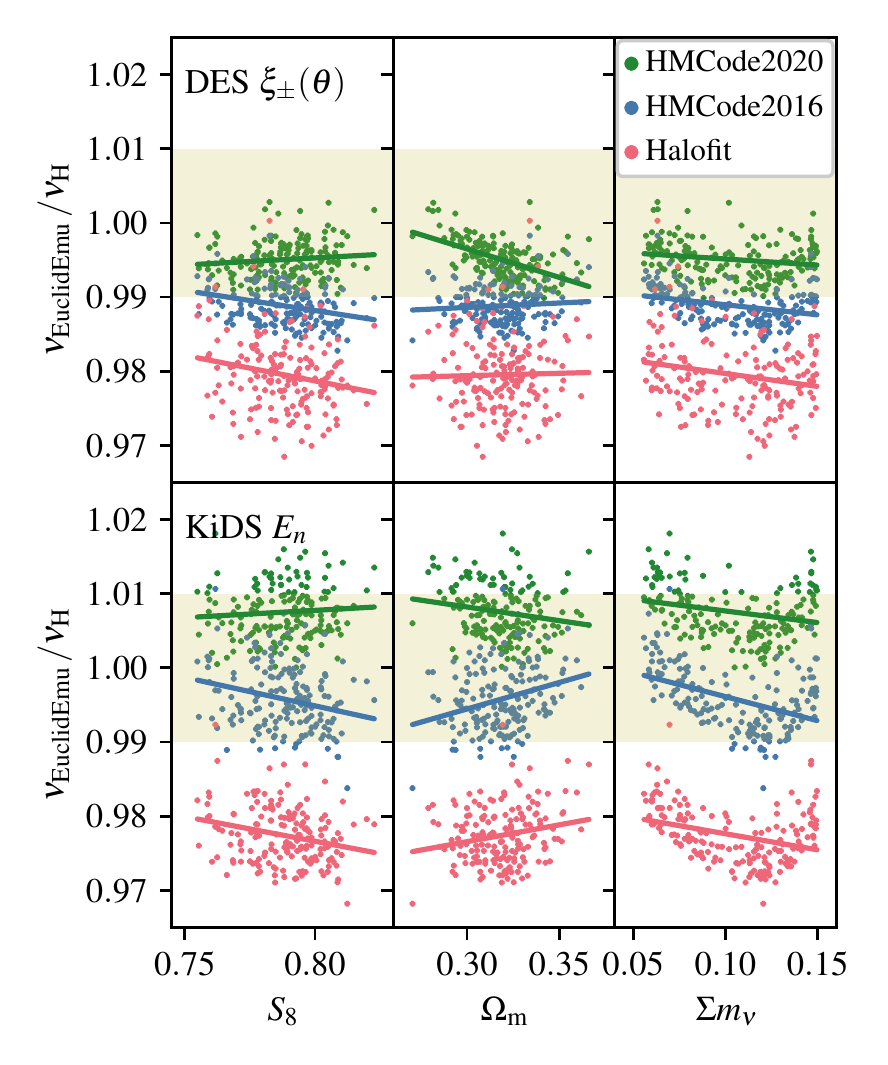}
\caption{Comparison of the DES $\xi_\pm(\theta)$ (upper) and KiDS $E_n(\theta_{\rm min}=$0\decimalarcmin5$)$ (lower) cosmic shear predictions from {\sc HMCode2020} (green), {\sc HMCode2016} (blue) and {\sc Halofit} (pink), as a ratio with the prediction from the {\sc EuclidEmulatorv2} (see Equation~\ref{eqn:snr}) for a range of cosmological parameters.   Each data point represents one sample from the DES Y3 $3\times2$pt posterior that lies within the allowed parameter range of the {\sc EuclidEmulatorv2}. The solid lines show an unweighted linear best fit to the data points as a function of input $S_8$ (left), $\Omega_{\rm m}$ (middle) and $\Sigma m_\nu$ (right).  Only weak trends are seen with all the parameters including those not shown here.  The shaded region brackets $\pm 1\%$ accuracy.}
\label{fig:2ptstatcomp}
\end{figure}   

\begin{table}
\centering                                      
\begin{tabular}{lll}         
\toprule
Parameter & Mock Input & {\sc EuclidEmulator2} range \\    
\midrule
$\Omega_{\rm b}$ & $0.041$ & $\bb{0.04,0.06}$\\
$\Omega_{\rm m}$ & $0.246$ & $\bb{0.24,0.40}$ \\
$\ns$ & $0.99$ & $\bb{0.92,\,1.00}$ \\
$h$ & $0.7251$ & $\bb{0.61,0.73}$ \\
$w_0$& $-1.0$ & $\bb{-1.30,-0.70}$\\
$w_{\rm a}$& $0.0$ & $\bb{-1.7,0.5}$\\
$\Sigma m_\nu$ & $0.06$ eV & $\bb{0.0,0.15}{\rm eV}$\\
$A_{\rm s}$ & $2.445\times10^{-9}$ & $\bb{1.7,2.5} \times 10^{-9}$ \\
\end{tabular}
\caption{Cosmological parameters adopted for the {\sc EuclidEmulator2}-based mock joint-survey analysis which correspond to an input $S_8=0.759$.  The chosen parameters lie within the range where the emulator has been shown to be accurate to $<1\%$ for $k \lesssim 10 h {\rm Mpc}^{-1}$ and $z \lesssim 3$ \citep{EuclidEmv2/etal:2021}.}
\label{tab:Euclid_input}         
\end{table}

In order to determine whether the accuracy of each non-linear model changes across a wide range of cosmological parameters, in Figure~\ref{fig:2ptstatcomp} we compress the theoretical $\xi_\pm(\theta)$ or $E_n$ tomographic data vector, $\bf{d}$, into a single signal-to-noise estimate, $\nu$, as
\be
\nu_{\rm H} = \sqrt{ \bf{d}_{\rm H}^{\rm T}\,  \bf{C}^{-1} \,  \bf{d}_{\rm H}} \, .
\label{eqn:snr}
\ee
Here $\bf{C}$ is the corresponding DES or KiDS covariance matrix and ${\rm H}$ labels the adopted non-linear model {\sc Halofit}, {\sc HMCode2016} or {\sc HMCode2020}.  Figure~\ref{fig:2ptstatcomp} shows the ratio $\nu_{\rm{EuclidEmulator}}/\nu_{\rm H}$ for a range of different cosmologies that fall within the region of parameter space covered by the {\sc EuclidEmulatorv2} (see Table~\ref{tab:Euclid_input}).   In order to test a realistic combination of cosmological parameters, each data point represents a sample from the DES Y3 $3\times2$pt {\sc Polychord} chain.     We find only weak trends with each cosmological parameter, with the {\sc HMCode2016} and {\sc HMCode2020} predictions agreeing with the {\sc EuclidEmulatorv2} model at the percent level for both statistics. {\sc Halofit} provides a poorer fit, but is nevertheless accurate at the $\sim 2\%$ level overall.  
 
Table~\ref{tab:EE2mocks} and Figure~\ref{fig:Euclidtest} present the $S_8$ constraints from the {\sc EuclidEmulatorv2}-based mock.  With the impact of the IA model choice assessed in Appendix~\ref{app:HMCodeTATTtest}, we adopt the NLA (no-z) model as an input for this analysis.  For the DES-like analysis we find an underestimate of the true $S_8$ at the level of $\sim 1.2\sigma$.  This offset is likely caused from the mismatch between the {\sc Halofit} non-linear power spectrum and {\sc EuclidEmulatorv2} shown in Figures~\ref{fig:2ptOWLS} and~\ref{fig:2ptstatcomp}, combined with the use of the TATT model which can introduce a bias in the projected marginal $S_8$ posterior\footnote{When analysing an NLA-input mock with TATT and {\sc HMCode2020} we find a $\sim 0.5\sigma (0.8\sigma)$ underestimate of the input $S_8$ when using the maximum (mean) marginalised posterior estimate (see Test G in Table~\ref{tab:EE2mocks}).}.  For the KiDS-like analysis, we find an overestimate of the true $S_8$ at the level of $\sim 0.9\sigma$, which can be understood when considering the $T_{\rm AGN}$ baryon feedback marginalisation within the KiDS pipeline.  As shown in Figure~\ref{fig:2ptOWLS}, the KiDS {\sc BAHAMAS}-range prior on $\log_{10}(T_{\rm AGN}/{\rm K})$, with $\bb{7.6,8.0}$, does not allow for a dark-matter-only cosmology.  It only includes feedback models which reduce power on small physical scales, such that baryon feedback marginalisation, in a dark-matter-only Universe, leads to an overestimate of the marginal $S_8$.   In a KiDS-like analysis modified to use a dark-matter-only matter correction for the non-linear matter power spectrum, we recover an unbiased value for $S_8$ with $S_8=0.760^{+0.013}_{-0.016}$.  These results lead to the same conclusion as Appendix~\ref{app:HMCodeTATTtest}: the KiDS-like and DES-like analysis choices can lead to $\sim 2\sigma$ $S_8$ offsets using identical data vectors, this time when considering the scenario of an $S_8=0.759$ {\sc EuclidEmulatorv2}-based dark-matter-only Universe.  

\renewcommand{\thesubsection}{\Alph{section}.\arabic{subsection}}
\subsection{A Hybrid pipeline analysis of {\sc EuclidEmulatorv2} mock data}
\label{app:euclidem}

\begin{table*}   
\centering         
\input{tabletexfiles/EE2_table_mock_analysis.tex} 
\caption{Joint survey $S_8$ constraints from mock data created using the {\sc EuclidEmulatorv2} including an NLA IA model.   The fiducial KiDS-like and DES-like analyses can be compared to our Hybrid pipeline.  A series of tests compare the baryon feedback mitigation strategies of adopting scale cuts and/or marginalising over a free nuisance parameter $T_{\rm AGN}$ using KiDS-like priors with the addition of the NLA-z $\eta$ parameter.  Where relevant, the baryon feedback is modelled using {\sc OWLS}-AGN.  We report the $S_8$ constraints and 68\% credible interval using both the maximum-marginal and mean-marginal approach, with $\Delta S_8$ quantifying the offset from the true $S_8 = 0.759$, as a fraction of the $1\sigma$ error. Some of these data are displayed in Figure~\ref{fig:Euclidtest}.}
\label{tab:EE2mocks}
\end{table*}

In defining a unified Hybrid pipeline it was clear that we would adopt the {\sc PolyChord} sampler (Appendix~\ref{app:samplers}) and {\sc HMCode2020} for the non-linear power spectrum model as this outperforms both {\sc Halofit} and {\sc HMCode2016} for the DES data vector in Figure~\ref{fig:2ptstatcomp}.  For the IA model choice we lack high signal-to-noise IA observations with which to make an informed decision between TATT and NLA.  After discussion, the NLA-z model was selected as sufficiently complex for the signal-to-noise of our current surveys \citep[see for example][]{fortuna/etal:2021}.  The KiDS cosmological parameter priors were then selected as these were found to minimise projection effects with the NLA model in Appendix~\ref{app:projection}, with the addition of the DES-like prior on the NLA-z $\eta_{\rm IA}$ parameter.  With these choices finalised, Tests A-F in Table~\ref{tab:EE2mocks} were designed to compare the baryon feedback mitigation strategies of adopting scale cuts and/or marginalising over a free nuisance parameter $T_{\rm AGN}$.

Starting with a mock survey that includes no baryon feedback, in Test A we recover the input $S_8$ within $0.1\sigma$ using only scale cuts.  Test A therefore confirms the analysis in Appendix~\ref{app:NLmodels} that {\sc HMCode2020} provides a sufficiently accurate description of the {\sc EuclidEmulatorv2} non-linear matter power spectrum, and the analysis in Appendix~\ref{app:projection} that the prior choice has minimised projection effects.  In Test B and C, we quantify the impact of including scale cuts for DES and KiDS by including baryon feedback modelled through the {\sc OWLS}-AGN hydrodynamical simulation \citep{vandaalen/etal:2011} which introduces a significant reduction of power on small physical scales (see Figure~\ref{fig:2ptOWLS}).  Adopting scale cuts for only the DES part of the joint-survey data vector (Test B) we underestimate the value of $S_8$ by $\sim 1.1\sigma$.  Including the COSEBIs scale cuts for KiDS (Test C) reduces this bias, as designed, but we are still left with an underestimate of the value of $S_8$ by $\sim 0.5\sigma$ with the maximum marginal, and $\sim 0.75\sigma$ with the mean marginal.  These tests may seem to contradict the analysis in Appendix~\ref{app:scalecuts} where we reported an offset of $\sim 0.14\sigma_{\rm 2D}$ in the $S_8 - \Omega_{\rm m}$ plane for this {\sc OWLS} AGN-contaminated data vector.  These results can however be understood by considering the weak constraints from cosmic shear alone on $\Omega_{\rm m}$.  We find that the relative error on $\Omega_{\rm m}$ is an order of magnitude larger than the relative error on $S_8$, such that a small bias within the $S_8 - \Omega_{\rm m}$ plane can translate into a large relative bias in $S_8$. Note that the \citet{krause/etal:2021} scale cut methodology, that we have adopted, was designed for the DES $3\times2$pt analysis, where $\Omega_{\rm m}$ is more tightly constrained.

\begin{figure}
\centering 
\includegraphics[width=\linewidth]{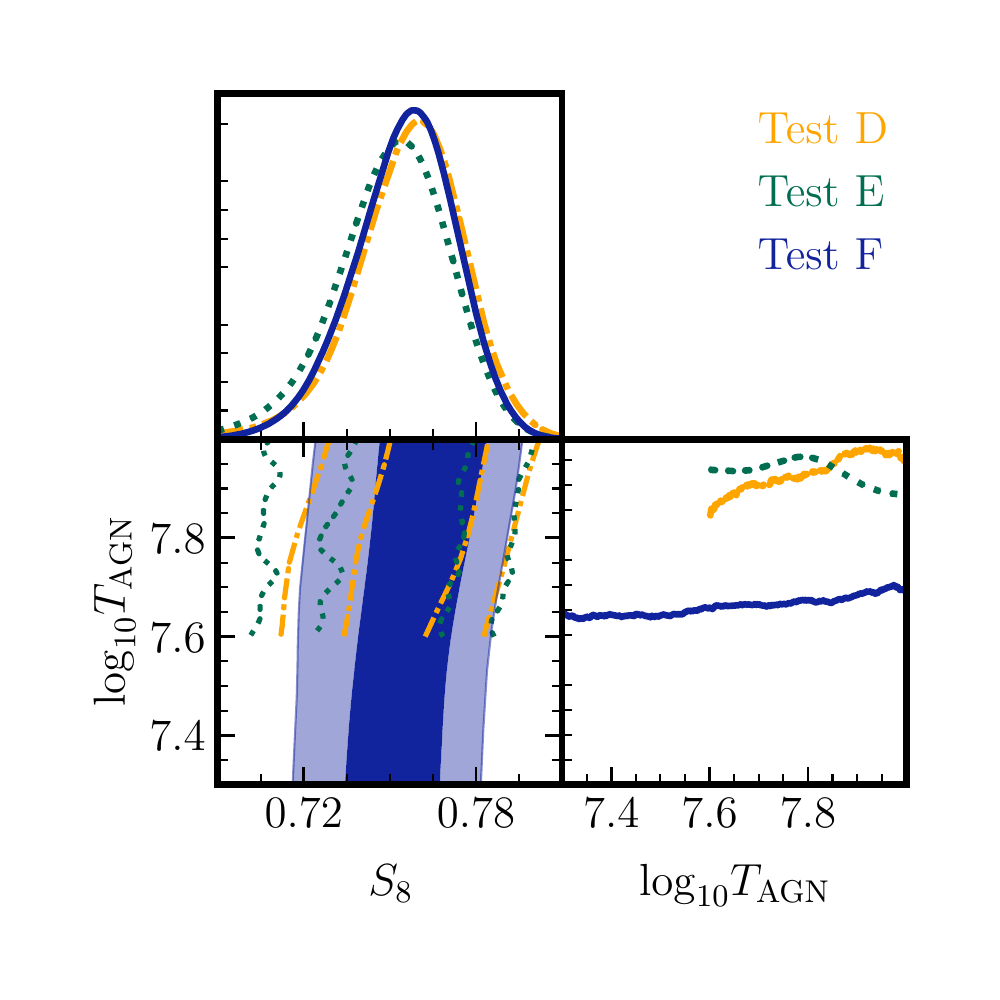}
\caption{Comparing the degeneracy in the $S_8 - \log_{10}(T_{\rm AGN}/{\rm K})$ plane for an analysis of noise-free {\sc EuclidEmulatorv2}+{\sc OWLS}-AGN mocks using different scale cuts and baryon feedback priors (see Table~\ref{tab:EE2mocks}).  Test D (orange, dot-dashed) applies no additional scale cuts to the KiDS mock data and shows the strongest degeneracy. Test E (green, dotted) shows that including scale cuts reduces the degeneracy.  Test F (blue, solid) extends the lower bound of the $\log_{10}(T_{\rm AGN}/{\rm K})$ prior to include dark-matter only scenarios. The inner 68\% and outer 95\% credible intervals are shown contoured, accurately recovering the input cosmology where $S_8 =0.759$.}
\label{fig:TAGNvsS8}
\end{figure}   

In Tests D, E, and F we analyse an {\sc OWLS} AGN-contaminated data vector, marginalising over the $T_{\rm AGN}$ baryon feedback parameter in {\sc HMCode2020}.  In all tests we recover the input $S_8$ within $0.2\sigma$ using the mean marginal value and within $\sim 0.1\sigma$ using the maximum marginal value.  These results are expected from Figure~\ref{fig:2ptOWLS} where the $T_{\rm AGN}$ prior range encompasses the {\sc OWLS}-AGN model.   Comparing Test D and Test E we quantify the impact of including the Appendix~\ref{app:scalecuts}-defined scale cuts to the KiDS data vector.  In common with the analysis in Table~\ref{tab:pipelinemods}, we find that the addition of scale cuts to KiDS does not significantly impact the $S_8$ constraining power, suggesting that the $T_{\rm AGN}$ marginalisation to mitigate baryon feedback is already removing the majority of information from these scales\footnote{\citet{longley/etal:2022} find that the addition of scale cuts to a joint cosmic shear analysis of DES, HSC and KiDS increases the error on $S_8$ by $15-30\%$, relative to an all-scale analysis with baryon feedback marginalisation.  We can understand the difference with our findings by noting that in their analysis, scale cuts are applied to DES, HSC and KiDS, in contrast to this analysis where we only investigate changes to scale cuts in KiDS.  In addition, the \citet{longley/etal:2022} $\Delta \chi^2$ scale cut methodology leads to more stringent limits on the {\sc OWLS}-AGN induced parameter bias with $\Delta S_8 < 0.2 \sigma$.  This can be compared to the \citet{krause/etal:2021} methodology, implemented in Appendix~\ref{app:scalecuts}, with a bias threshold of $< 0.3 \sigma_{\rm 2D}$.}.  In Figure~\ref{fig:TAGNvsS8} we show that the degeneracy between the baryon feedback nuisance parameter $T_{\rm AGN}$ and $S_8$ is, however, reduced when scale cuts are included.  This is important as it implies that the scale cut analysis is less sensitive to the $T_{\rm AGN}$ prior which in this case is informative, based on our current best knowledge from hydrodynamical simulations.  When expanding the informative baryon feedback prior to $\log_{10}(T_{\rm AGN}/{\rm K}):\bb{7.0,10.0}$, \citet{amon/efstathiou:2022} find a $0.55\sigma$ increase in the KiDS-1000 mean marginal value\footnote{The projection effects for the \citet{amon/efstathiou:2022} prior set have not been quantified.} for $S_8$ and a 40\% increase in the uncertainty on that value.  This study analysed the two-point shear correlation function with angular limits 0\decimalarcmin5$< \theta <$ 300\decimalarcmin0 for $\xi_+(\theta)$ and 4\decimalarcmin0$<\theta <$ 300\decimalarcmin0 for $\xi_-(\theta)$.  Adopting an uninformative prior on $T_{\rm AGN}$ would further increase the uncertainty on $S_8$, with a stronger impact expected for statistics where the high-$k$-scale information has been removed\footnote{\citet{amon/efstathiou:2022} find the constraints from a COSEBIs $E_n(\theta_{\rm min})=$ 0\decimalarcmin5 analysis with wide, but still informative, priors on a phenomenological power spectrum suppression parameter almost doubles the uncertainty on $S_8$ compared to the $\xi_\pm(\theta)$ analysis adopting the same priors.}, reducing the ability of the data to self-calibrate the baryon feedback model.

Amongst the authors there is a broad range of views on the preferred approach to mitigate baryon feedback.  Those with a high degree of confidence in the scale and redshift dependence of the {\sc BAHAMAS} model for baryon feedback and the {\sc HMCode2020} model for the non-linear power spectrum recommend a self-calibrating all-scale cosmic shear analysis with uninformative $T_{\rm AGN}$ priors.  Those with a high degree of confidence in the {\sc BAHAMAS}-defined maximum range of baryon feedback suppression, but less confidence in the finer high-$k$ details of the non-linear modelling recommend a limited scale cosmic shear analysis with informative $T_{\rm AGN}$ priors. Those with a low degree of confidence in the results from hydrodynamical simulations recommend the use of more conservative scale cuts, similar to the approach taken in Appendix~\ref{app:scalecuts} but adopting a more extreme worst case scenario for baryon feedback such as {\sc Cosmo-OWLS}-AGN:8.7.  Other opinions include the use of highly flexible baryon feedback models, with or without the combination of scale cuts.   As we have seen in Appendix~\ref{app:projection}, any change in prior and/or astrophysical model leads to different projection effects, some of which may be unexpected \citep[see for example the strong degeneracy between the {\sc HMCode2016} baryon feedback parameter $A_{\rm bary}$ and $n_{\rm s}$ when using uninformative priors in][]{yoon/jee:2021}.  A comparison of the results from the authorship's range of preferred baryon feedback mitigation strategies would therefore require both the analysis of mocks and data which we defer to a future study.  We also note that data comparisons using different angular ranges will also be subject to random noise fluctuations (see Appendix~\ref{app:KiDSCOSEBIs}).

We conclude our baryon feedback study with Test F, where we expand the lower limit of the $T_{\rm AGN}$ prior to include low feedback models that are equivalent to a dark matter only model (see Figure~\ref{fig:2ptOWLS}).  The decision to allow the non-linear model of the matter power spectrum to include no baryon feedback followed the preferred approach of \citet{amon/etal:2022}; \citet*{secco/etal:2022} and also \citet{asgari/etal:2021} who allow for a dark matter only model within their baryon feedback prior.  We find that extending the informative $T_{\rm AGN}$ prior to include no feedback, in combination with scale cuts, does not change the recovered $S_8$ value or constraining power.  This can be understood from Figure~\ref{fig:TAGNvsS8} where we see little degeneracy between $T_{\rm AGN}$ and $S_8$ for $\log_{10}(T_{\rm AGN}/{\rm K}) \lesssim 7.5$.  For our Hybrid analysis we therefore choose to combine the mitigation strategies from both surveys using scale cuts and a $T_{\rm AGN}$ prior that spans a dark matter-only Universe through to the strongest {\sc BAHAMAS} AGN feedback model. 

The `Hybrid' row in Table~\ref{tab:EE2mocks} repeats Test F with the Hybrid prior set, extending the KiDS prior set with the DES prior on neutrino mass.   This introduces an underestimate of $S_8$ by $0.46 \sigma$ in our recovered mean marginal $S_8$ value, arising from projection effects.   Here our mock input neutrino mass, $\Sigma m_\nu = 0.06$ eV, is at the lower edge of the allowed tophat prior range which extends to $\Sigma m_\nu < 0.6$ eV.  The weak correlation between $\Sigma m_\nu$ and $S_8$ then leads to a bias in the mean or maximum of the projected marginal $S_8$ distribution.   This issue should be largely circumvented, however, by reporting the MAP in contrast to the marginal values.  

In one final Test G, we replace the NLA-z model with a TATT model in our analysis to quantify the $\sim 0.8 \sigma$ level of bias from the truth when adopting this model to analyse an NLA-z mock Universe.  As shown in Appendix~\ref{app:projection}, however, this bias decreases to zero when adopting the DES priors and analysing a mock Universe that includes a TATT IA signal.

We review the different constraining power of each setup in Table~\ref{tab:EE2mocks}, focusing on the mean-marginal constraints.  Comparing the KiDS-like analysis with Test D, we find a $\sim 33\%$ increase in the error on $S_8$ resulting from the change in both IA model and sampler, which is consistent with expectations \citep[see Appendix~\ref{app:samplers}; \citealt*{secco/etal:2022};][]{asgari/etal:2021}.   We find a modest improvement of $\sim 10\%$ in constraints between the DES-like analysis, and Tests A, B and C, demonstrating that the combination of changing from TATT to NLA-z and from DES to KiDS+ priors does not make a significant difference when the input IA model is NLA, consistent with the results in Appendix~\ref{app:HMCodeTATTtest}.   Comparing Test C and E, we find a $\sim 6\%$ increase in the error on $S_8$ when including baryon feedback marginalisation.  The inclusion of a free neutrino mass parameter raises the error on $S_8$ by a further $\sim 6\%$.    For this {\sc EuclidEmulatorv2} mock we find that our Hybrid analysis choices lead to an increase in the error on $S_8$ by $\sim 43\%$ relative to the KiDS-like fiducial analysis, with similar constraining power relative to the DES-like fiducial analysis.  The difference we see between the constraints from this suite of mock analyses is largely replicated in our data analysis in Section~\ref{sec:results}.

\renewcommand{\thesection}{\Alph{section}}
\section{Comparing KiDS-1000 COSEBIs with and without scale cuts}
\label{app:KiDSCOSEBIs}
{\bf Summary}: In this appendix we measure a $0.7-0.8\sigma$ increase in the KiDS-1000 $S_8$ constraint when including scale cuts in a KiDS-like COSEBIs analysis.  We find this result to be consistent with random noise fluctuations, where a $\ge 0.7\sigma (\ge 0.8\sigma)$ offset between a $\theta_{\rm min}=$ 0\decimalarcmin5 and $\theta_{\rm min}=$ 2\decimalarcmin0 COSEBIs analysis occurs $23\% (14$\%) of the time in noisy mock KiDS-1000 simulations.  To address whether this offset could alternatively/additionally indicate an issue with unmodeled baryon feedback, we test a mock using the most extreme hydrodynamic simulation {\sc Cosmo-OWLS}:8.7.  In this case, we estimate a $0.4\sigma$ offset in $S_8$ between a $\theta_{\rm min}=$ 0\decimalarcmin5 and a $\theta_{\rm min}=$ 2\decimalarcmin0 noise-free COSEBIs analysis that adopts the {\sc HMCode2020} model, fixing $\log_{10}(T_{\rm AGN}/{\rm K})=7.8$.\\
\begin{table}  
\centering         
\input{tabletexfiles/table_all_KiDS_results.tex} 
\caption{KiDS-1000 maximum marginal constraints on $S_8$, and the intrinsic alignment parameter, $A_{\rm IA}$, using a KiDS-like pipeline for a range of different two-point statistics (2pt).  We compare COSEBIs, $E_n$, with the two-point shear correlation function, $\xi_{\pm}$, for two different scale cuts $\theta_{\rm min}=$ 0\decimalarcmin5 (the fiducial KiDS setting) and $\theta_{\rm min}=$ 2\decimalarcmin0 (the Appendix~\ref{app:scalecuts} DES-like scale cut setting).  We also compare the COSEBIs results for two versions of {\sc HMCode} (HM).  With {\sc HMCode2016} we use KiDS-like priors on the baryon feedback parameter $A_{\rm bary} = \bb{2,3.13}$.  With {\sc HMCode2020} we use the {\sc BAHAMAS}-defined baryon feedback range $\log_{10}(T_{\rm AGN}/{\rm K}):\bb{7.6,8.0}$.
$\Delta S_8$ and $\Delta A_{\rm IA}$ compares the fiducial \citet{asgari/etal:2021} $E_n$ result with the different variations, in units of $\sigma$, the 68\% credible interval for each parameter.  We also compare the $S_8$ constraining power for each variation through the ratio $\sigma/\sigma_{\rm fid}$.}
\label{tab:allkids}
\end{table}

In Table~\ref{tab:allkids} we report the maximum marginal constraints on $S_8$ and $A_{\rm IA}$ for a series of KiDS-1000 cosmic shear analyses using: COSEBIs, $E_n$, and the two-point correlation function\footnote{We refer the reader to section 2.4 and figure 1 of \citet{asgari/etal:2021} for a discussion on the advantages and disadvantages for each cosmic shear statistic.  Of note is the finding that the $\xi_\pm(\theta)$-optimised definition $S_8 = \sigma_8(\Omega_{\rm m}/0.3)^{\alpha}$, where $\alpha=0.5$, does not provide an optimal description of the COSEBIs $\sigma_8-\Omega_{\rm m}$ degeneracy where the best fit $\alpha=0.54$.}, $\xi_\pm(\theta)$; two different minimum angular scales, $\theta_{\rm min}=$ 0\decimalarcmin5 and $\theta_{\rm min}=$ 2\decimalarcmin0; and two versions of {\sc HMCode}.   The COSEBIs analysis with $\theta_{\rm min}=$ 0\decimalarcmin5 and {\sc HMCode2016} is the fiducial result from \citet{asgari/etal:2021}.   The COSEBIs analysis with $\theta_{\rm min}=$ 2\decimalarcmin0 and {\sc HMCode2020} adopts the scale cuts and updated non-linear power spectrum model used in the Hybrid analysis presented in Section~\ref{sec:hybrid}.  We find a $1.01\sigma$ increase in $S_8$ between these two analyses.  An increase of $\sim 0.2-0.3\sigma$ is found when changing from {\sc HMCode2016} to {\sc HMCode2020}, which can be understood from the percent-level change in the amplitude of the cosmic shear predictions from these two models (see Figure~\ref{fig:2ptstatcomp}), and the use of different baryon feedback priors\footnote{With {\sc HMCode2016} the most extreme baryon feedback simulation considered at the time was {\sc OWLS}-AGN which corresponds to $\log_{10}(T_{\rm AGN}/{\rm K}) \sim 7.8$.  With {\sc HMCode2020}, more extreme models were permitted by the {\sc BAHAMAS} simulation, raising the prior limit to $\log_{10}(T_{\rm AGN}/{\rm K})=8.0$.  Owing to the degeneracy between $S_8$ and $T_{\rm AGN}$, stronger feedback allows for higher $S_8$ values.}.  An increase of $\sim 0.7-0.8\sigma$ is found when changing from $\theta_{\rm min}=$ 0\decimalarcmin5 to $\theta_{\rm min}=$ 2\decimalarcmin0 for the COSEBIs $E_n$ statistic\footnote{Defining $\sigma=(\sigma_{68}^{\rm upper}+\sigma_{68}^{\rm lower})/2$, using constraints from the $E_n(\theta=$0\decimalarcmin5$)$ analysis, we find $\Delta S_8 = 0.73\sigma$ when comparing the two {\sc HMCode2016} analyses and $\Delta S_8 =0.71\sigma$ when comparing the two {\sc HMCode2020} analysis.   We note that for the {\sc HMCode2020} analysis, the $S_8$ error from the $E_n(\theta=$0\decimalarcmin5$)$ data vector is $\sim 15\%$ higher than the $S_8$ error from $E_n(\theta=2.^{'}0)$ data vector.  This may be  caused by chain-to-chain variance noise.  Defining the $S_8$ offset for the two {\sc HMCode2020} analyses in terms of the error on $E_n(\theta=2.^{'}0)$, we find $\Delta S_8 =0.81\sigma$.}.  In this appendix we focus on understanding the likely origin of a $0.7-0.8\sigma$ offset associated with scale cuts.

\begin{figure}
\centering 
\includegraphics[width=\linewidth]{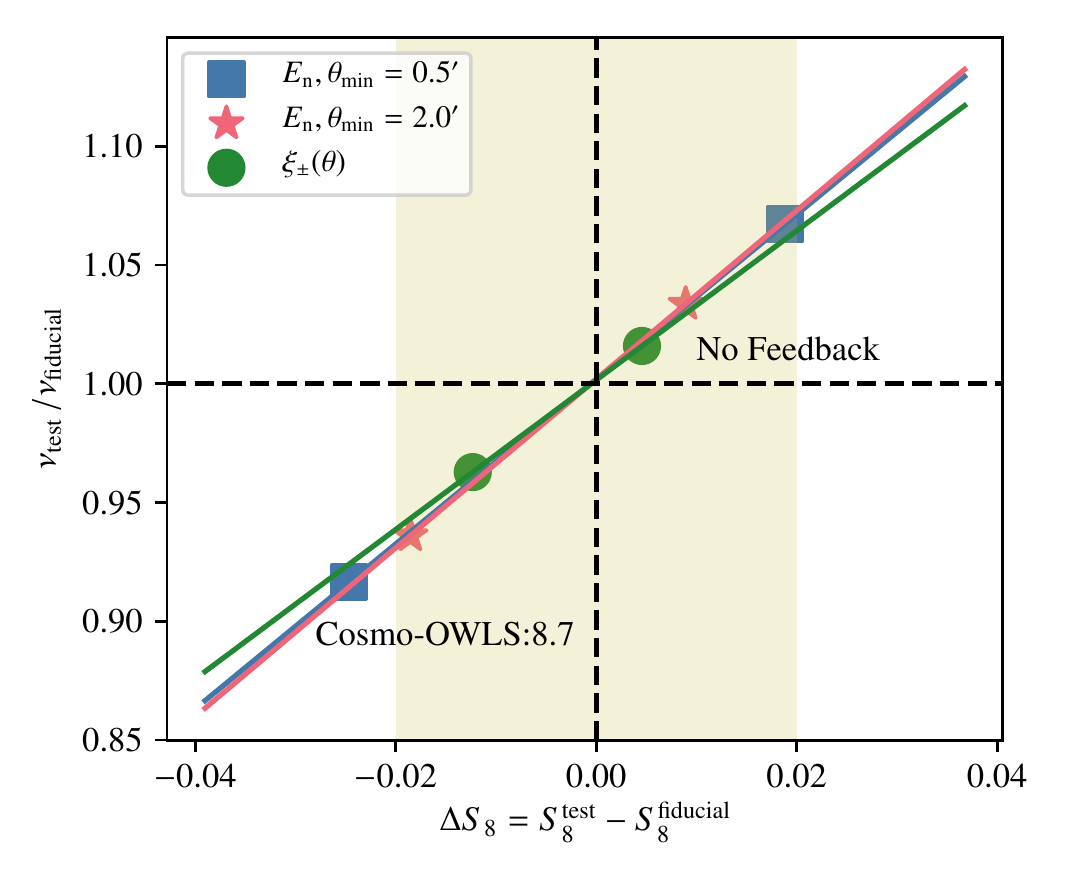}
\caption{Estimating the expected bias $\Delta S_8$ when analysing a Universe with no baryon feedback (upper right quadrant) or a Universe with strong AGN feedback, modelled with Cosmo-{\sc OWLS}:8.7 (lower left quadrant), with a fiducial theoretical model that adopts {\sc HMCode2020} with $\log_{10}(T_{\rm AGN}/{\rm K}) = 7.8$.  We determine the signal-to-noise $\nu$ (Equation~\ref{eqn:snr}) for three mock cosmic shear observables: DES Y3 $\xi_\pm(\theta)$ including scale cuts (green), KiDS-1000 COSEBIs $E_{n}$ with $\theta_{\rm min}=$ 2\decimalarcmin0 (pink) and KiDS-1000 $E_{n}$ with $\theta_{\rm min}=$ 0\decimalarcmin5 (blue). This is calculated for the fiducial model over a range of $S_8$ values (solid lines), and also for the trial cases of no feedback, and strong feedback (data points).   As expected the two scale cut statistics ($\xi_\pm(\theta)$ - circle, $E_{n}, \theta_{\rm min}=$ 2\decimalarcmin0 - star) produce the least biased estimate when the underlying feedback differs from the model.  The offset between the two COSEBIs statistics is within the $1\sigma$ error for a KiDS-like analysis (shaded).  }
\label{fig:S8BFcomp}
\end{figure}

In the interest of conducting a fast approximate analysis, we choose to compress the tomographic cosmic shear data vector into a single signal-to-noise estimate $\nu$ (Equation~\ref{eqn:snr}), following Appendix~\ref{app:NLmodels}.  By noting that the amplitude of the cosmic shear signal scales roughly as $\nu \propto S_8^2$ \citep{jain/seljak:1997}, we can write that the ratio $\nu_{\rm A}/\nu_{\rm B} \propto \Delta S_8$, for the same noise realisation of cosmology A and B whose $S_8$ values differ by a small quantity $\Delta S_8 = S_8^{\rm A} - S_8^{\rm B}$.  

In Figure~\ref{fig:S8BFcomp} we calibrate this relationship by calculating the ratio of $\nu_{\rm test}/\nu_{\rm fiducial}$ for a series of mock KiDS-1000 and DES Y3 data vectors, using test cosmologies that differ from our fiducial cosmological parameter set in Table~\ref{tab:mock_input} with $S_8^{\rm test} = S_8^{\rm fiducial} + \Delta S_8$.  We find a slightly different relationship for the DES Y3-like $\xi_\pm(\theta)$ data vector (green) and the KiDS-1000-like $E_n$ data vectors (pink, blue), but a similar response for the two COSEBIs statistic versions with $\theta_{\rm min}=$ 0\decimalarcmin5 (blue) and $\theta_{\rm min}=$ 2\decimalarcmin0 (pink).

To address the question, what offset in $S_8$ would we expect to measure from the different cosmic shear statistics if our baryon feedback model was incorrect, we calculate $\nu_{\rm test}/\nu_{\rm fiducial}$ for two further tests.  In both tests the cosmological model is fixed to the fiducial parameter set and the baryon feedback model is altered.  First we adopt a dark matter only matter power spectrum with no baryon feedback suppression.  The fiducial mock simulates baryon feedback with $\log_{10}(T_{\rm AGN}/{\rm K}) = 7.8$, reducing power relative to the dark matter only case by a factor of $\sim 5\%$ at $k=1 h {\rm Mpc}^{-1}$.  In this case $\nu_{\rm test}/\nu_{\rm fiducial} > 1$ (see the upper right quadrant in Figure~\ref{fig:S8BFcomp}).  For the second test we adopt a model from the most extreme baryon feedback simulation to date {\sc Cosmo-OWLS:8.7} \citep{lebrun/etal:2014}, which reduces power relative to the dark matter only case by a factor of $\sim 20\%$ at $k=1 h {\rm Mpc}^{-1}$, roughly a factor of four times the level of suppression relative to the fiducial $\log_{10}(T_{\rm AGN}/{\rm K}) = 7.8$ case.  In this test $\nu_{\rm test}/\nu_{\rm fiducial} < 1$ (see the lower left quadrant in Figure~\ref{fig:S8BFcomp}).  Using the $\nu_{\rm A}/\nu_{\rm B} \propto \Delta S_8$ relationship calibrated for each statistic in Figure~\ref{fig:S8BFcomp}, we can read off an estimate for $\Delta S_8$ for these two test cases.  As expected we find the COSEBIs $\theta_{\rm min}=$ 0\decimalarcmin5 statistic (blue) to be the most sensitive to baryon feedback (the largest $\Delta S_8$ values), followed by COSEBIs $\theta_{\rm min}=$ 2\decimalarcmin0 (pink) and then $\xi_\pm(\theta)$ with DES Y3 scale cuts applied.  The difference seen between the COSEBIs $\theta_{\rm min}=$ 2\decimalarcmin0 and $\xi_\pm(\theta)$ scale cut analyses demonstrates that the COSEBIs-defined scale cut from Appendix~\ref{app:scalecuts} renders the cosmic shear analysis more sensitive to the $k$-scales impacted by baryon feedback, compared to the \citet{krause/etal:2021} $\xi_\pm(\theta)$-defined scale cuts.  It may be beneficial in the future to develop the COSEBIs software to allow for redshift-dependent $\theta_{\rm min}$-limits in order to further reduce the impact of baryon feedback without incurring a significant loss of constraining power.

What is of interest for this appendix is the $S_8$ offset found between the different statistics, rather than the offset relative to the fiducial input value.    If the underlying truth was a Universe with no baryon feedback, in a noise-free analysis we would expect our best fit value for $S_8$ to decrease by $\Delta S_8 = 0.012$ when changing from $\theta_{\rm min}=$ 0\decimalarcmin5 to $\theta_{\rm min}=$ 2\decimalarcmin0.  This corresponds to a $\sim 0.6 \sigma$ decrease between the different KiDS-1000 constraints. In the data we find a higher $S_8$ value when scale cuts are included, which is more similar to the case where the underlying truth is {\sc Cosmo-OWLS:8.7}.  Here we would expect to measure an increase of $\Delta S_8 = 0.008$ between the two noise-free COSEBIs analyses, corresponding to a $0.4 \sigma$ offset between the different KiDS-1000 constraints.   

It is important to recognise that these estimates are only approximate as they compress the tomographic data vector into a single noise-averaged value, erasing second-order information on scale and redshift dependence. In one respect, however, they are conservative, as they do not account for the marginalisation step over a wide baryon feedback prior in the KiDS-like analysis that allows for models as strong as $\log_{10}(T_{\rm AGN}/{\rm K}) = 8.0$.  From this analysis we draw the conclusion that the $S_8$ offset we measure between the KiDS-1000 COSEBIs $\theta_{\rm min}=$ 0\decimalarcmin5 and $\theta_{\rm min}=$ 2\decimalarcmin0 $E_n$ constraints could be attributed to unmodeled extreme baryon feedback.  This hypothesis only accounts for roughly half of the observed difference, however, when considering {\sc Cosmo-OWLS:8.7} as the `worst-case scenario' for baryon feedback.

\citet{asgari/etal:2021} report cosmological parameter constraints from a range of two-point statistics measured from the same data set: $E_n(\theta_{\rm min}=$0\decimalarcmin5$)$, $\xi_\pm(\theta)$ and $C^{\rm EE}_{\epsilon \epsilon}(\ell)$.  Using a series of 5000 noisy mock simulations of KiDS-1000 created with {\sc SALMO}\footnote{{\sc SALMO}: \url{https://github.com/Linc-tw/salmo}.  For information see section 4 of \citet{joachimi/etal:2021}.}, they quantify the cross-correlation between the different statistics.  The range of positive, zero and negative cross-correlations for different components result from the changing sensitivity of each statistic to a range of noisy $\ell$-scales.   \citet{asgari/etal:2021} find that they expect to measure a $>1\sigma$ offset in $S_8$ constraints between the different KiDS-1000 two-point statistics, 15\% of the time \citep[see similar findings for analyses of HSC-Y1 mocks in][]{hamana/etal:2022}.   The $0.3\sigma$ offset between the fiducial $\xi_\pm(\theta)$ and $E_n$ constraints seen in Table~\ref{tab:allkids} is therefore expected.

We revisit the \citet{asgari/etal:2021} analysis to quantify the expected offset in $S_8$ between two COSEBIs analyses of the same data set using $\theta_{\rm min}=$ 0\decimalarcmin5 and $\theta_{\rm min}=$ 2\decimalarcmin0.   Using the original set of {\sc SALMO} mocks, we were able to determine the cross-correlation between the two COSEBIs versions at essentially no CPU cost.  We then took a short-cut compared to the original analysis, using the $\nu_{\rm A}/\nu_{\rm B} \propto \Delta S_8$ relationship derived in Figure~\ref{fig:S8BFcomp} to determine the expected offset in $S_8$ between pairs of $E_n$ statistics measured with and without scale cuts in each {\sc SALMO} mock.   This differs from the CPU-intensive approach of \citet{asgari/etal:2021} where full {\sc CosmoSIS} analyses were performed for each mock and each statistic.  By applying the same short-cut technique to the set of statistics tested in  \citet{asgari/etal:2021}, however, we have verified that this approximate approach results in a similar distribution in $\Delta S_8$ values.  We find that a $\ge1\sigma$ offset in $S_8$ between the two COSEBIs measurements is expected 9\% of the time.  As such the two COSEBIs versions are more correlated with each other than with $\xi_\pm(\theta)$ and $C^{\rm EE}_{\epsilon \epsilon}(\ell)$.  For the $0.7\sigma$ difference that we observe in the {\sc HMCode2016} analysis, we find that an offset of $\ge 0.7\sigma$ is expected 23\% of the time. An offset of $\ge 0.8\sigma$ is expected 14\% of the time.  As such the offsets we find between the scale cut and original version of COSEBIs with KiDS-1000 can be attributed to random noise fluctuations.

To the reader most familiar with the $\xi_\pm(\theta)$ statistic, this result may seem counter-intuitive, as the scale cut that has been applied removes only one data point from each tomographic bin.  As we can see in Table~\ref{tab:allkids}, a $\xi_\pm(\theta)$ analysis with a $\theta_{\rm min}=$ 2\decimalarcmin0 scale cut applied indeeds makes little impact, with $S_8$ changing by $\sim 0.2\sigma$.  In contrast, the COSEBIs statistic is calculated as a weighted average of finely binned noisy measurements of $\xi_\pm(\theta)$ summed over the range $\theta_{\rm min}<\theta<\theta_{\rm max}$.  The shape of the oscillatory weight functions depends on the angular range limits \citep[see equations 28 and 38 of][]{schneider/etal:2010}.  For $\theta_{\rm min} =$ 0\decimalarcmin5 and $\theta_{\rm min} = $ 2\decimalarcmin0 it is worth noting that the COSEBIs weight functions differ significantly for $\theta \lesssim$ 40\decimalarcmin0.

\section{Constraints on all parameters, additional tables and figures}
\label{app:extras}
{\bf Summary}: In this appendix we present additional results from the DES Y3 + KiDS-1000 cosmic shear analysis to complement the primary results in Section~\ref{sec:results}.  \\

\begin{table}  
\centering         
\input{tabletexfiles/Goodness_of_fit.tex} 
\caption{Goodness of fit statistics listing the $\chi^2_{\rm min}$, the effective number of parameters, $N_{\Theta}$, and the goodness of fit statistic $p(\chi^2 > \chi^2_{\rm min} | \nu)$ (see Section~\ref{sec:GoF} for details).  The Hybrid pipeline results can be compared to the DES-like and KiDS-like analysis of DES, KiDS and the joint survey.}
\label{tab:GoF}
\end{table}

\begin{table}  
\centering         
\input{tabletexfiles/table_lcdm_fullarea.tex} 
\caption{Mean marginal constraints with 68\% credible intervals for the \lcdm parameters (upper), astrophysical nuisance parameters (middle) and data calibration parameters (lower).  Cosmic shear results are reported from the joint DES Y3 and KiDS-1000 (left), DES Y3 (middle) and KiDS-1000 (right)  analyses.  Parameters marked $^*$, are considered to be unconstrained by the data as their marginal posterior, at either prior edge, exceeds 13\% of the peak posterior probability.}
\label{tab:allparams}
\end{table}

\begin{figure*}
\centering 
\includegraphics[width=\textwidth]{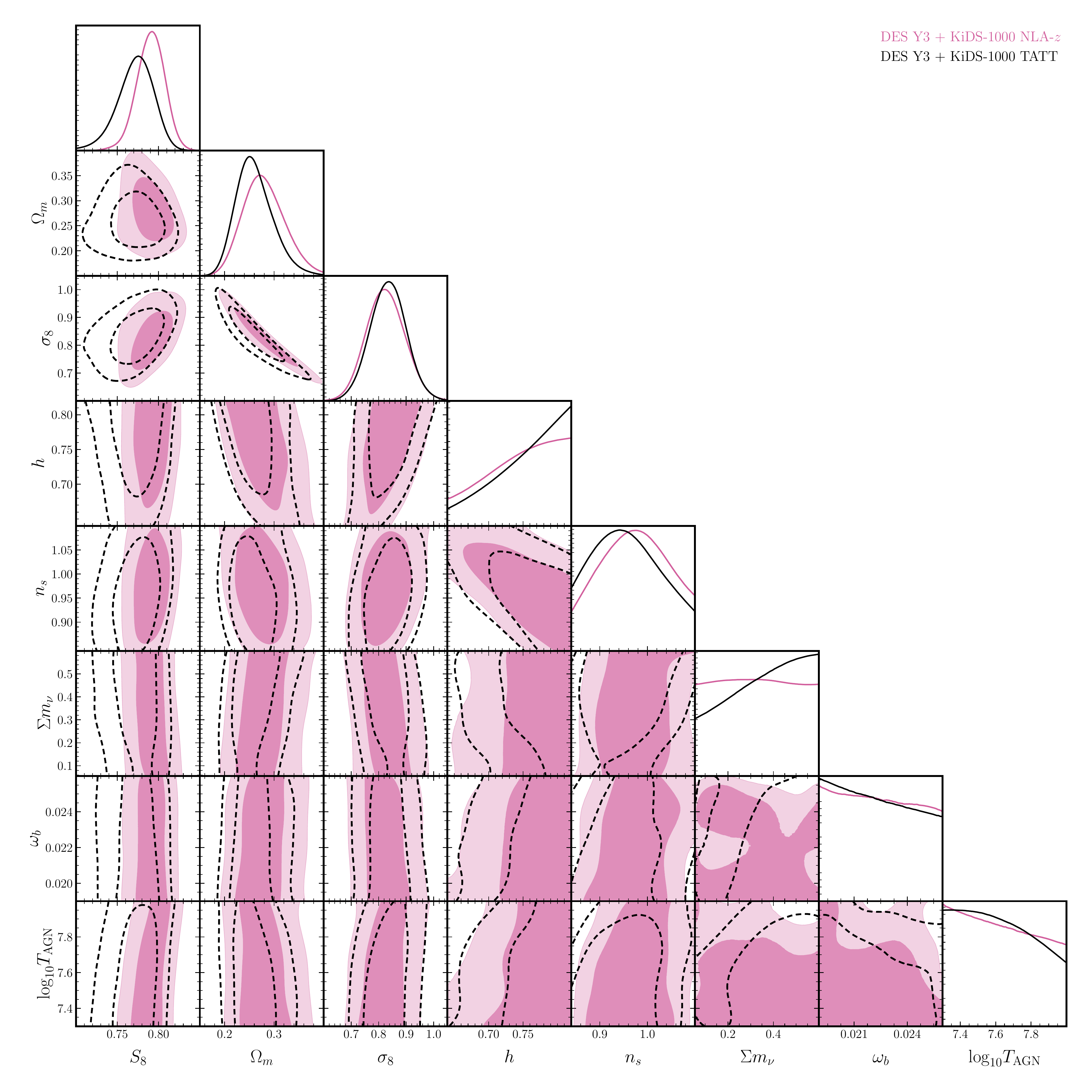}
\caption{Joint survey posteriors for the \lcdm cosmological parameters and the \textsc{HMCode2020} $T_{\rm AGN}$ parameter for baryonic feedback.  We compare the fiducial Hybrid (NLA-z) analysis in pink to a TATT analysis with black dashed lines.  The inner 68\% and outer 95\% credible intervals are shown contoured.  In the case of the parameters $h, n_{\rm s}, \Sigma m_{\nu}$, $\omega_{\rm b}$ and $T_{\rm AGN}$, the axis limits are set by the prior boundaries.}
\label{fig:allparams}
\end{figure*}

\begin{figure*}
\centering 
\includegraphics[width=\textwidth]{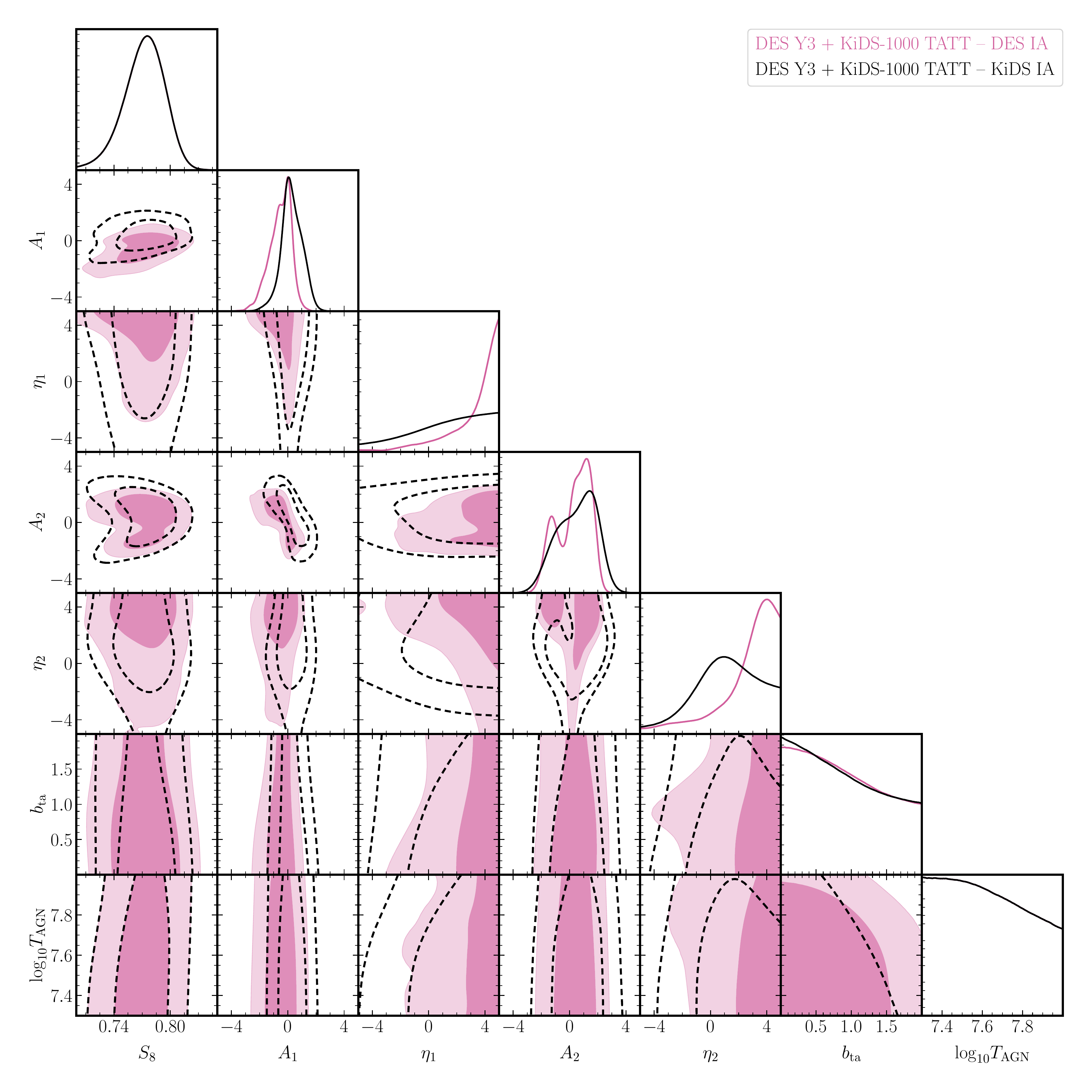}
\caption{Joint survey posteriors, comparing constraints on the TATT intrinsic alignment model (see Section~\ref{sec:IA}) from the DES Y3 (pink) and KiDS-1000 (black-dashed) surveys with the inner 68\% and outer 95\% credible intervals contoured. In all but the $S_8$ parameter, the axis limits are set by the prior boundaries. }
\label{fig:TATTparams}
\end{figure*}

\begin{list}{\labelitemi}{\leftmargin=1em}
\item Table~\ref{tab:GoF} presents the goodness of fit statistics for the DES-like and KiDS-like analyses, listing the $\chi^2_{\rm min}$, the estimated effective number of parameters, $N_{\Theta}$, and the goodness of fit statistic $p(\chi^2 > \chi^2_{\rm min} | \nu)$ for a range of different analyses of the DES, KiDS and joint-survey data vectors, with $N_{\rm data}=273, 75, 348$ respectively.   We remind the reader that the \citet{joachimi/etal:2021} approach used to estimate $N_{\Theta}$ is subject to noise at the level of $\sim 0.2 N_{\Theta}$.
\item In Figure~\ref{fig:allparams} we present the joint-survey constraints across the full cosmological parameter space, comparing the fiducial Hybrid NLA-z analysis with the alternative TATT analysis. We also include the marginal posterior for the baryon feedback parameter $T_{\rm AGN}$.  Only $S_8$, $\Omega_{\rm m}$ and $\sigma_8$ are constrained by the data.  The axis limits on $h, n_{\rm s}, \Sigma m_{\nu}$, $\omega_{\rm b}$ and $T_{\rm AGN}$ are set by the priors. As we have chosen to sample in $S_8$, the prior space is straightforward to visualise as rectangles in each 2D marginal bounded by the prior edges. Throughout the paper we use {\sc GetDist}\footnote{{\sc GetDist}: \url{https://getdist.readthedocs.io}} \citep{lewis:2019} to plot quantities of interest from our chains of Monte Carlo samples.  {\sc GetDist} applies a linear boundary kernel and multiplicative bias correction\footnote{We use the {\sc GetDist} settings: boundary\_correction\_order:0, mult\_bias\_correction\_order:1.} to remove the bias introduced when the chain-smoothing kernel passes over a prior boundary.  When the hard boundaries from the priors are aligned with the parameter coordinates, as in our case, the boundaries can be defined using the {\sc GetDist.MCSamples} `ranges' option.  With the prior ranges set, the {\sc GetDist} contoured credible intervals do not artificially close for prior-informed parameters.

\item In Figure~\ref{fig:TATTparams} we present the joint-survey constraints across the astrophysical nuisance parameter space for the alternative TATT analysis.  Only $S_8$, $a_1$ and $a_2$ are constrained by the data. We can see that the preference for a lower $S_8$ value in the TATT analysis, relative to our fiducial analysis, is primarily driven by the IA model allowing for a significant tidal torque amplitude, the $a_2$ parameter, when there is strong redshift evolution with high values for $\eta_2$.  We find a similar preference for strong redshift evolution of the IA model in the fiducial NLA-z analysis (see Figure~\ref{fig:2IAvs1IA}), but in contrast to the TATT analysis, the $\eta_{\rm IA}$ amplitude is largely uncorrelated with $S_8$.  We see similar behaviour for the IA model parameters for the DES (pink) and KiDS (black) surveys.  As with the fiducial analysis, KiDS prefers higher values for $A_1$ compared to DES.  
\item Table~\ref{tab:allparams} reports the mean 1D marginal values for each parameter for the fiducial Hybrid analyses of DES Y3 and KiDS-1000.  We use $^*$ to mark parameters that are unconstrained by the data.  In these cases the tabulated constraint comes from the prior, as the marginal posterior probability, at either prior edge, exceeds 13\% of the peak posterior value \citep[see appendix A of][]{asgari/etal:2021}.  We constrain $S_8$, $\sigma_8$, and the intrinsic alignment amplitude $A_{\rm IA}$ for each survey, and $\Omega_{\rm m}$ for the joint survey and the DES-only analysis. We have not presented figures of the marginal posteriors for the shear and redshift calibration nuisance parameters $m$ and $\Delta z$, as we recover the peak and width of the Gaussian prior for the majority of these parameters (compare the values\footnote{In this paper we define $\Delta z = \langle n^{\rm estimate}(z) \rangle - \langle n^{\rm true}(z) \rangle$ which we note has a different sign convention to the {\sc CosmoSIS} parameter DELTA\_Z\_OUT$= - \Delta z$.} in Table~\ref{tab:params} and Table~\ref{tab:allparams}).  A notable exception is the redshift calibration parameter for the second tomographic bin, $\Delta z_2$, for both the DES and KiDS data.  In the DES-only, KiDS-only and joint-survey Hybrid analysis the mean $\Delta z_2$ of each survey is $\pm 1\sigma$ offset from the Gaussian prior peak.  This is consistent with the previous findings of both survey teams \citep[see for example figure 17 of][]{amon/etal:2022}. Both teams show consistent $S_8$ results when excluding individual tomographic bins from the fiducial analysis \citep[see also figures 16 and 18 in][for the stability of the intrinsic alignment parameter constraints when removing redshift bins and varying the redshift calibration methodology]{amon/etal:2022}. Using the \citet{kohlinger/etal:2019} `three tier' set of internal consistency statistics, \citet{asgari/etal:2021} flag the second KiDS-1000 tomographic bin as an outlier (see their appendix B.2). As this bin carries insignificant levels of cosmological information, however, the results are unchanged by the inclusion or exclusion of this tomographic bin. 
\end{list}

\vspace{2cm}
\end{document}

%% file: tabletexfiles/Goodness_of_fit_Hybrid_only.tex
\begin{tabular}{lrrrc}
\midrule
Analysis & $\chi^2_{\rm min}$ & $ N_\Theta $ & $\chi^2_{\rm red}$ & $ p(\chi_{\rm min}^2 | \nu_{\rm eff} )$ \\ [0.2cm]
\midrule
DES Y3 (Full area) & 284.2 & 5.4 & 1.06 & 0.231 \\ [0.1cm]
DES Y3 (KiDS-excised) & 288.3 & 4.6 & 1.07 & 0.192 \\ [0.1cm]
KiDS-1000 & 88.3 & 7.1 & 1.30 & 0.048 \\[0.1cm]
\midrule
DES Y3+KiDS-1000: \\[0.2cm]
Fiducial & 378.0 & 9.6 & 1.12 & 0.068 \\[0.1cm]
$\Sigma m_{\nu} = 0.06 {\rm eV}$ & 376.6 & 9.7 & 1.11 & 0.074 \\[0.1cm]
Shared IA & 382.2 & 8.0 & 1.12 & 0.057 \\[0.1cm]
NLA (no z) & 379.3 & 8.8 & 1.12 & 0.065 \\[0.1cm]
TATT & 371.5 & 12.3 & 1.11 & 0.087 \\[0.1cm]
Dark Matter $P_{\delta}(k)$ & 375.5 & 10.2 & 1.11 & 0.076 \\[0.1cm]
\hline
\end{tabular}

%% file: tabletexfiles/MAP_PJHPD_table_all_Hybrid_results.tex
\begin{tabular}{ll|crc|crc|crc}
\toprule
    Survey & Analysis &
     \multicolumn{3}{c|}{Mean Marginal}
     & \multicolumn{3}{c|}{MAP+PJHPD} 
     & \multicolumn{3}{c}{Maximum Marginal} \\
\midrule 
&  & $S_8$ & $\Delta S_8$ & $\sigma/\sigma_{\rm fid}$  & $S_8$ & $\Delta S_8$ & $\sigma/\sigma_{\rm Fid}$  & $S_8$ & $\Delta S_8$ & $\sigma/\sigma_{\rm Fid}$ \\ [0.2cm]
\midrule
Joint & Fiducial  & $ 0.790^{+ 0.018}_{-0.014} $ & $ 0.00 \sigma $ & $ 1.00$ & $ 0.801^{+ 0.011}_{-0.023} $ & $ 0.62 \sigma $ & $ 1.06$ & $ 0.792^{+ 0.017}_{-0.018} $ & $ 0.11 \sigma $ & $ 1.09$  \\ [+0.1cm] 
DES Y3 & Fiducial (Full area)  & $ 0.802^{+ 0.023}_{-0.019} $ & $ 0.57 \sigma $ & $ 1.31$ & $ 0.816^{+ 0.015}_{-0.028} $ & $ 1.18 \sigma $ & $ 1.34$ & $ 0.803^{+ 0.022}_{-0.021} $ & $ 0.59 \sigma $ & $ 1.35$  \\ [+0.1cm] 
DES Y3 & Fiducial (KiDS-excised) & $ 0.806^{+ 0.021}_{-0.020} $ & $ 0.74 \sigma $ & $ 1.28$ & $ 0.803^{+ 0.024}_{-0.016} $ & $ 0.64 \sigma $ & $ 1.25$ & $ 0.807^{+ 0.021}_{-0.022} $ & $ 0.76 \sigma $ & $ 1.33$  \\ [+0.1cm] 
KiDS-1000 & Fiducial  & $ 0.763^{+ 0.031}_{-0.023} $ & $ -1.00 \sigma $ & $ 1.68$ & $ 0.776^{+ 0.029}_{-0.027} $ & $ -0.53 \sigma $ & $ 1.73$ & $ 0.770^{+ 0.026}_{-0.031} $ & $ -0.71 \sigma $ & $ 1.77$  \\ [+0.1cm] 
\midrule Joint & $\Sigma m_\nu = 0.06$eV & $ 0.797^{+ 0.017}_{-0.014} $ & $ 0.39 \sigma $ & $ 0.98$ & $0.798^{+ 0.019}_{-0.014} $ & $ 0.49 \sigma $ & $ 1.02$ & $ 0.798^{+ 0.016}_{-0.018} $ & $ 0.47 \sigma $ & $ 1.04$  \\ [+0.1cm] 
Joint & Shared IA  & $ 0.792^{+ 0.018}_{-0.013} $ & $ 0.09 \sigma $ & $ 0.98$ & $ 0.804^{+ 0.015}_{-0.020} $ & $ 0.75 \sigma $ & $ 1.09$ & $ 0.795^{+ 0.016}_{-0.018} $ & $ 0.25 \sigma $ & $ 1.06$  \\ [+0.1cm] 
Joint & NLA (no-z) & $ 0.792^{+ 0.016}_{-0.014} $ & $ 0.08 \sigma $ & $ 0.94$ & $ 0.788^{+ 0.020}_{-0.010} $ & $ -0.19 \sigma $ & $ 0.93$ & $ 0.791^{+ 0.017}_{-0.015} $ & $ 0.02 \sigma $ & $ 0.97$  \\ [+0.1cm] 
Joint & TATT  & $ 0.771^{+ 0.025}_{-0.018} $ & $ -0.88 \sigma $ & $ 1.35$ & $ 0.761^{+ 0.024}_{-0.036} $ & $ -0.98 \sigma $ & $ 1.84$ & $ 0.775^{+ 0.022}_{-0.023} $ & $ -0.66 \sigma $ & $ 1.41$  \\ [+0.1cm] 
Joint & Dark Matter $P_\delta(k)$ & $ 0.784^{+ 0.016}_{-0.015} $ & $ -0.42 \sigma $ & $ 0.95$ & $ 0.785^{+ 0.014}_{-0.016} $ & $ -0.36 \sigma $ & $ 0.93$ & $ 0.786^{+ 0.015}_{-0.016} $ & $ -0.30 \sigma $ & $ 0.97$  \\ [+0.1cm] 
\midrule {\it Planck} & Fiducial & $ 0.831^{+ 0.017}_{-0.017} $ &  & $ 1.04$ & $ 0.834^{+ 0.019}_{-0.017} $ & & $ 1.09$ & $ 0.831^{+ 0.018}_{-0.018} $ &  & $ 1.10$  \\ [+0.1cm] 
{\it Planck} & $\Sigma m_\nu = 0.06$eV & $ 0.835^{+ 0.015}_{-0.016} $ &  & $ 0.97$ & $ 0.838^{+ 0.013}_{-0.019} $ &  & $ 0.98$ & $ 0.837^{+ 0.014}_{-0.018} $ &  & $ 1.02$  \\ [+0.1cm] 

\hline
\end{tabular}

%% file: tabletexfiles/DKP_tension_table.tex
\begin{tabular}{llcc}          
\toprule
     Data &
     Analysis
     & $d_{\rm H}(S_8)$
     & $\Delta_{\rm tension}^{S_8,\Omega_{\rm m}}$ \\
\midrule
DES Y3 (Full area) & Fiducial  & $1.00\sigma$ & $0.87\sigma$ \\ [+0.1cm]
DES Y3 (KiDS-excised) & Fiducial & $0.90\sigma$ & $0.73\sigma$ \\ [+0.1cm]
KiDS-1000 & Fiducial & $2.11\sigma$ & $1.89\sigma$ \\ [+0.1cm]
DES Y3 + KiDS-1000 & Fiducial & $1.66\sigma$ & $1.67\sigma$  \\ [+0.1cm]
\midrule
DES Y3 + KiDS-1000 & $\Sigma m_\nu=0.06$eV & $1.67\sigma$ & $1.78\sigma$\\ [+0.1cm]
DES Y3 + KiDS-1000 & Shared IA & $1.57\sigma$ & $1.37\sigma$\\ [+0.1cm]
DES Y3 + KiDS-1000 & NLA (no-z) & $1.67\sigma$ & $1.62\sigma$\\ [+0.1cm]
DES Y3 + KiDS-1000 & TATT & $2.15\sigma$ & $2.33\sigma$\\ [+0.1cm]
DES Y3 + KiDS-1000 & Dark Matter $P_\delta(k)$ & $2.01\sigma$ & $1.90\sigma$\\ [+0.1cm]
\bottomrule
\end{tabular}

%% file: affiliations.tex
$^{1}$ Cerro Tololo Inter-American Observatory, NSF's National Optical-Infrared Astronomy Research Laboratory, Casilla 603, La Serena, Chile\\
$^{2}$ Laborat\'orio Interinstitucional de e-Astronomia - LIneA, Rua Gal. Jos\'e Cristino 77, Rio de Janeiro, RJ - 20921-400, Brazil\\
$^{3}$ Argonne National Laboratory, 9700 South Cass Avenue, Lemont, IL 60439, USA\\
$^{4}$ Department of Physics, University of Michigan, Ann Arbor, MI 48109, USA\\
$^{5}$ Institute of Astronomy, University of Cambridge, Madingley Road, Cambridge CB3 0HA, UK\\
$^{6}$ Kavli Institute for Cosmology, University of Cambridge, Madingley Road, Cambridge CB3 0HA, UK\\
$^{7}$ E.A Milne Centre, University of Hull, Cottingham Road, Hull, HU6 7RX, United Kingdom and Centre of Excellence for Data Science, AI, and Modelling (DAIM), University of Hull, Cottingham Road, Kingston-upon-Hull, HU6 7RX\\
$^{8}$ Institut de F\'{\i}sica d'Altes Energies (IFAE), The Barcelona Institute of Science and Technology, Campus UAB, 08193 Bellaterra (Barcelona) Spain\\
$^{9}$ Institute of Cosmology and Gravitation, University of Portsmouth, Portsmouth, PO1 3FX, UK\\
$^{10}$ Physics Department, 2320 Chamberlin Hall, University of Wisconsin-Madison, 1150 University Avenue Madison, WI  53706-1390\\
$^{11}$ Department of Physics and Astronomy, University of Pennsylvania, Philadelphia, PA 19104, USA\\
$^{12}$ CNRS, UMR 7095, Institut d'Astrophysique de Paris, F-75014, Paris, France\\
$^{13}$ Sorbonne Universit\'es, UPMC Univ Paris 06, UMR 7095, Institut d'Astrophysique de Paris, F-75014, Paris, France\\
$^{14}$ Center for Theoretical Physics, Polish Academy of Sciences, al. Lotników 32/46, 02-668 Warsaw, Poland\\
$^{15}$ Department of Physics, Northeastern University, Boston, MA 02115, USA\\
$^{16}$ University Observatory, Faculty of Physics, Ludwig-Maximilians-Universit\"at, Scheinerstr. 1, 81679 Munich, Germany\\
$^{17}$ Department of Physics \& Astronomy, University College London, Gower Street, London, WC1E 6BT, UK\\
$^{18}$ Argelander-Institut für Astronomie, Auf dem Hügel 71, D-53121 Bonn, Germany\\
$^{19}$ Kavli Institute for Particle Astrophysics \& Cosmology, P. O. Box 2450, Stanford University, Stanford, CA 94305, USA\\
$^{20}$ SLAC National Accelerator Laboratory, Menlo Park, CA 94025, USA\\
$^{21}$ Instituto de F\'{i}sica Te\'orica, Universidade Estadual Paulista, S\~ao Paulo, Brazil\\
$^{22}$ Department of Physics, Carnegie Mellon University, Pittsburgh, Pennsylvania 15312, USA\\
$^{23}$ Instituto de Astrofisica de Canarias, E-38205 La Laguna, Tenerife, Spain\\
$^{24}$ Universidad de La Laguna, Dpto. Astrofísica, E-38206 La Laguna, Tenerife, Spain\\
$^{25}$ Center for Astrophysical Surveys, National Center for Supercomputing Applications, 1205 West Clark St., Urbana, IL 61801, USA\\
$^{26}$ Department of Astronomy, University of Illinois at Urbana-Champaign, 1002 W. Green Street, Urbana, IL 61801, USA\\
$^{27}$ Institut d'Estudis Espacials de Catalunya (IEEC), 08034 Barcelona, Spain\\
$^{28}$ Institute of Space Sciences (ICE, CSIC),  Campus UAB, Carrer de Can Magrans, s/n,  08193 Barcelona, Spain\\
$^{29}$ Physics Department, William Jewell College, Liberty, MO, 64068\\
$^{30}$ Department of Astronomy and Astrophysics, University of Chicago, Chicago, IL 60637, USA\\
$^{31}$ Kavli Institute for Cosmological Physics, University of Chicago, Chicago, IL 60637, USA\\
$^{32}$ Department of Physics, Duke University Durham, NC 27708, USA\\
$^{33}$ NASA Goddard Space Flight Center, 8800 Greenbelt Rd, Greenbelt, MD 20771, USA\\
$^{34}$ Jodrell Bank Center for Astrophysics, School of Physics and Astronomy, University of Manchester, Oxford Road, Manchester, M13 9PL, UK\\
$^{35}$ University of Nottingham, School of Physics and Astronomy, Nottingham NG7 2RD, UK\\
$^{36}$ Hamburger Sternwarte, Universit\"{a}t Hamburg, Gojenbergsweg 112, 21029 Hamburg, Germany\\
$^{37}$ Department of Astrophysical Sciences, Princeton University, Princeton, NJ 08544, USA\\
$^{38}$ School of Mathematics and Physics, University of Queensland,  Brisbane, QLD 4072, Australia\\
$^{39}$ Leiden Observatory, Leiden University, P.O.Box 9513, 2300RA Leiden, The Netherlands\\
$^{40}$ Kapteyn Astronomical Institute, University of Groningen, PO Box 800, 9700 AV Groningen, The Netherlands\\
$^{41}$ Lawrence Berkeley National Laboratory, 1 Cyclotron Road, Berkeley, CA 94720, USA\\
$^{42}$ Department of Physics, IIT Hyderabad, Kandi, Telangana 502285, India\\
$^{43}$ Fermi National Accelerator Laboratory, P. O. Box 500, Batavia, IL 60510, USA\\
$^{44}$ NSF AI Planning Institute for Physics of the Future, Carnegie Mellon University, Pittsburgh, PA 15213, USA\\
$^{45}$ Universit\'e Grenoble Alpes, CNRS, LPSC-IN2P3, 38000 Grenoble, France\\
$^{46}$ Ruhr University Bochum, Faculty of Physics and Astronomy, Astronomical Institute (AIRUB), German Centre for Cosmological Lensing, 44780 Bochum, Germany\\
$^{47}$ Department of Astronomy/Steward Observatory, University of Arizona, 933 North Cherry Avenue, Tucson, AZ 85721-0065, USA\\
$^{48}$ Jet Propulsion Laboratory, California Institute of Technology, 4800 Oak Grove Dr., Pasadena, CA 91109, USA\\
$^{49}$ Department of Physics and Astronomy, University of Waterloo, 200 University Ave W, Waterloo, ON N2L 3G1, Canada\\
$^{50}$ Department of Astronomy, University of California, Berkeley,  501 Campbell Hall, Berkeley, CA 94720, USA\\
$^{51}$ Institute of Theoretical Astrophysics, University of Oslo. P.O. Box 1029 Blindern, NO-0315 Oslo, Norway\\
$^{52}$ Instituto de Fisica Teorica UAM/CSIC, Universidad Autonoma de Madrid, 28049 Madrid, Spain\\
$^{53}$ Instituto de Ciencias del Cosmos (ICC), Universidad de Barcelona, Martí i Franquès, 1, 08028 Barcelona, Spain\\
$^{54}$ Institute for Astronomy, University of Edinburgh, Royal Observatory, Blackford Hill, Edinburgh, EH9 3HJ, UK\\
$^{55}$ School of Physics and Astronomy, Cardiff University, CF24 3AA, UK\\
$^{56}$ Department of Astronomy, University of Geneva, ch. d'\'Ecogia 16, CH-1290 Versoix, Switzerland\\
$^{57}$ Santa Cruz Institute for Particle Physics, Santa Cruz, CA 95064, USA\\
$^{58}$ Center for Cosmology and Astro-Particle Physics, The Ohio State University, Columbus, OH 43210, USA\\
$^{59}$ Department of Physics, The Ohio State University, Columbus, OH 43210, USA\\
$^{60}$ Department of Physics, University of Arizona, Tucson, AZ 85721, USA\\
$^{61}$ Center for Astrophysics $\vert$ Harvard \& Smithsonian, 60 Garden Street, Cambridge, MA 02138, USA\\
$^{62}$ Department of Physics and Astronomy, University College London, Gower Street, London WC1E 6BT, UK\\
$^{63}$ Department of Astrophysical Sciences, Princeton University, 4 Ivy Lane, Princeton, NJ 08544, USA\\
$^{64}$ Australian Astronomical Optics, Macquarie University, North Ryde, NSW 2113, Australia\\
$^{65}$ Lowell Observatory, 1400 Mars Hill Rd, Flagstaff, AZ 86001, USA\\
$^{66}$ Department of Physics, Universit\'e de Montr\'eal, Montr\'eal, Canada\\
$^{67}$ Mila - Quebec Artificial Intelligence Institute, Montr\'eal, Canada\\
$^{68}$ Ciela - Montreal Institute for Astrophysical Data Analysis and Machine Learning, Montr\'eal, Canada\\
$^{69}$ Center for Computational Astrophysics, Flatiron Institute, 162 5th Avenue, 10010, New York, NY, USA\\
$^{70}$ McWilliams Center for Cosmology, Department of Physics, Carnegie Mellon University, 5000 Forbes Ave, Pittsburgh, PA 15213, USA\\
$^{71}$ Kavli Institute for the Physics and Mathematics of the Universe (WPI),The University of Tokyo Institutes for Advanced Study (UTIAS), The University of Tokyo, Chiba 277-8583, Japan\\
$^{72}$ Instituto de Astrof\'{\i}sica e Ci\^{e}ncias do Espa\c{c}o, Faculdade de Ci\^{e}ncias, Universidade de Lisboa, 1769-016 Lisboa, Portugal\\
$^{73}$ Departamento de F\'isica Matem\'atica, Instituto de F\'isica, Universidade de S\~ao Paulo, CP 66318, S\~ao Paulo, SP, 05314-970, Brazil\\
$^{74}$ WPC Systems Ltd., 55 Qingyun Road, Tucheng, New Taipei, Taiwan\\
$^{75}$ Department of Applied Mathematics and Theoretical Physics, University of Cambridge, Cambridge CB3 0WA, UK\\
$^{76}$ George P. and Cynthia Woods Mitchell Institute for Fundamental Physics and Astronomy, and Department of Physics and Astronomy, Texas A\&M University, College Station, TX 77843,  USA\\
$^{77}$ Centro de Investigaciones Energ\'eticas, Medioambientales y Tecnol\'ogicas (CIEMAT), Madrid, Spain\\
$^{78}$ Instituci\'o Catalana de Recerca i Estudis Avan\c{c}ats, E-08010 Barcelona, Spain\\
$^{79}$ Max Planck Institute for Extraterrestrial Physics, Giessenbachstrasse, 85748 Garching, Germany\\
$^{80}$ Perimeter Institute for Theoretical Physics, 31 Caroline St. North, Waterloo, ON N2L 2Y5, Canada\\
$^{81}$ Department of Physics, Stanford University, 382 Via Pueblo Mall, Stanford, CA 94305, USA\\
$^{82}$ School of Physics and Astronomy, Sun Yat-sen University, Guangzhou 519082, Zhuhai Campus, P.R. China\\
$^{83}$ Instituto de F\'isica Gleb Wataghin, Universidade Estadual de Campinas, 13083-859, Campinas, SP, Brazil\\
$^{84}$ Observat\'orio Nacional, Rua Gal. Jos\'e Cristino 77, Rio de Janeiro, RJ - 20921-400, Brazil\\
$^{85}$ Kavli Institute for the Physics and Mathematics of the Universe (WPI), UTIAS, The University of Tokyo, Kashiwa, Chiba 277-8583, Japan\\
$^{86}$ INAF - Osservatorio Astronomico di Padova, via dell'Osservatorio 5, 35122 Padova, Italy\\
$^{87}$ Department of Physics, University of Genova and INFN, Via Dodecaneso 33, 16146, Genova, Italy\\
$^{88}$ Department of Physics and Astronomy, Pevensey Building, University of Sussex, Brighton, BN1 9QH, UK\\
$^{89}$ Space Telescope Science Institute, 3700 San Martin Drive, Baltimore, MD  21218, USA\\
$^{90}$ Shanghai Astronomical Observatory (SHAO), Nandan Road 80, Shanghai 200030, China\\
$^{91}$ Key Laboratory of Radio Astronomy and Technology, Chinese Academy of Sciences, A20 Datun Road, Chaoyang District, Beijing, 100101, P. R. China\\
$^{92}$ University of Chinese Academy of Sciences, Beijing 100049, China\\
$^{93}$ Brookhaven National Laboratory, Bldg 510, Upton, NY 11973, USA\\
$^{94}$ Department of Physics and Astronomy, Stony Brook University, Stony Brook, NY 11794, USA\\
$^{95}$ Instituto de F\'isica, Pontificia Universidad Cat\'olica de Valpara\'iso, Casilla 4059, Valpara\'iso, Chile\\
$^{96}$ School of Physics and Astronomy, University of Southampton,  Southampton, SO17 1BJ, UK\\
$^{97}$ Computer Science and Mathematics Division, Oak Ridge National Laboratory, Oak Ridge, TN 37831\\
$^{98}$ Institute for Particle Physics and Astrophysics, ETH Zürich, Wolfgang-Pauli-Strasse 27, 8093 Zürich, Switzerland\\
$^{99}$ Institut de Recherche en Astrophysique et Plan\'etologie (IRAP), Universit\'e de Toulouse, CNRS, UPS, CNES, 14 Av. Edouard Belin, 31400 Toulouse, France\\
$^{100}$ Excellence Cluster Origins, Boltzmannstr.\ 2, 85748 Garching, Germany\\
$^{101}$ Universit\"ats-Sternwarte, Fakult\"at f\"ur Physik, Ludwig-Maximilians Universit\"at M\"unchen, Scheinerstr. 1, 81679 M\"unchen, Germany\\

%% file: tabletexfiles/single_survey_table_mock_analysis.tex
\begin{tabular}{lllll} 
\toprule
Analysis
     & \multicolumn{2}{c}{KiDS} 
     & \multicolumn{2}{c}{DES} \\
\midrule
Non-linear  
     & \multicolumn{2}{c}{{\sc HMCode2020}} 
     & \multicolumn{2}{c}{{\sc Halofit}} \\
IA 
     & \multicolumn{2}{c}{NLA}
     & \multicolumn{2}{c}{TATT} \\
Baryons 
     & \multicolumn{2}{c}{$T_{\rm AGN}=7.8$}
     & \multicolumn{2}{c}{Scale Cuts} \\
\midrule
Quoted $S_8$ & Max & Mean & Max & Mean \\
\midrule
 $ \SeK $ & $ 0.752^{+ 0.022}_{-0.023} $ & $ 0.747^{+ 0.026}_{-0.017} $ & $ 0.758^{+ 0.028}_{-0.033} $ & $ 0.748^{+ 0.036}_{-0.018} $  \\ [+0.1cm]   
 $ \SeD $ & $ 0.758^{+ 0.018}_{-0.017} $ & $ 0.758^{+ 0.016}_{-0.017} $ & $ 0.758^{+ 0.026}_{-0.028} $ & $ 0.753^{+ 0.030}_{-0.020} $  \\ [+0.1cm]   
 $ \SeJS $ & $ 0.758^{+ 0.012}_{-0.014} $ & $ 0.757^{+ 0.013}_{-0.012} $ & $ 0.761^{+ 0.020}_{-0.021} $ & $ 0.757^{+ 0.024}_{-0.015} $  \\ [+0.1cm]   
 $ \SeJ $ & $ 0.757^{+ 0.013}_{-0.013} $ & $ 0.757^{+ 0.011}_{-0.011} $ & $ 0.764^{+ 0.016}_{-0.019} $ & $ 0.759^{+ 0.019}_{-0.014} $  \\ [+0.1cm] \midrule 
 $ \Delta\SeK $ & $ -0.31 \sigma $ & $ -0.55 \sigma $ & $ -0.02 \sigma $ & $ -0.42 \sigma $  \\ [+0.1cm]   
 $ \Delta\SeD $ & $ -0.08 \sigma $ & $ -0.04 \sigma $ & $ -0.04 \sigma $ & $ -0.26 \sigma $  \\ [+0.1cm]   
 $ \Delta\SeJS $ & $ -0.05 \sigma $ & $ -0.15 \sigma $ & $ 0.10 \sigma $ & $ -0.10 \sigma $  \\ [+0.1cm]   
 $ \Delta\SeJ $ & $ -0.12 \sigma $ & $ -0.14 \sigma $ & $ 0.27 \sigma $ & $ 0.01 \sigma $  \\ [+0.1cm] \midrule 
 $ \sigma^{\rm KiDS}/\sigma^{\rm KiDS}_{\rm min} $ & $ 1.06 $ & $ 1.00 $ & $ 1.42 $ & $ 1.24 $  \\ [+0.1cm]   
 $ \sigma^{\rm DES}/\sigma^{\rm DES}_{\rm min} $ & $ 1.05 $ & $ 1.00 $ & $ 1.64 $ & $ 1.53 $  \\ [+0.1cm]   
 $ \sigma^{\rm J1IA}/\sigma^{\rm J1IA}_{\rm min} $ & $ 1.06 $ & $ 1.00 $ & $ 1.67 $ & $ 1.58 $  \\ [+0.1cm]   
 $ \sigma^{\rm Joint}/\sigma^{\rm Joint}_{\rm min} $ & $ 1.13 $ & $ 1.00 $ & $ 1.60 $ & $ 1.48 $  \\ [+0.1cm] \midrule 
 $ \sigma^{\rm KiDS}/\sigma^{\rm Joint} $ & $ 1.80 $ & $ 1.92 $ & $ 1.71 $ & $ 1.61 $  \\ [+0.1cm]   
 $ \sigma^{\rm DES}/\sigma^{\rm Joint} $ & $ 1.37 $ & $ 1.47 $ & $ 1.51 $ & $ 1.52 $  \\ [+0.1cm]   
 $ \sigma^{\rm J1IA}/\sigma^{\rm Joint} $ & $ 1.03 $ & $ 1.10 $ & $ 1.15 $ & $ 1.18 $  \\ [+0.1cm]   

\bottomrule
\end{tabular}

%% file: tabletexfiles/table_mock_analysis.tex
\begin{tabular}{ll|lrl|lrl}          
\toprule
	Joint Analysis
     & Modification
     & \multicolumn{3}{c|}{Maximum Marginal} 
     & \multicolumn{3}{c}{Mean Marginal} \\
\midrule    
     &
     & $S_8$
     & $\Delta S_8$
     & $\sigma / \sigma_{\rm min}$
     & $S_8$
     & $\Delta S_8$
     & $\sigma / \sigma_{\rm min}$ \\
\midrule
 {\bf KiDS-like:} & Fiducial & $ 0.758^{+ 0.012}_{-0.014} $ & $ -0.05 \sigma $ & $ 1.1$  & $ 0.757^{+ 0.013}_{-0.012} $ & $ -0.15 \sigma $ & $ 1.0$ \\ [+0.1cm] 
  & with DES Priors & $ 0.754^{+ 0.013}_{-0.013} $ & $ -0.41 \sigma $ & $ 1.1$  & $ 0.754^{+ 0.013}_{-0.012} $ & $ -0.43 \sigma $ & $ 1.0$ \\ [+0.1cm] 
  & with DES Scale cuts & $ 0.758^{+ 0.014}_{-0.013} $ & $ -0.11 \sigma $ & $ 1.1$  & $ 0.758^{+ 0.013}_{-0.012} $ & $ -0.11 \sigma $ & $ 1.0$ \\ [+0.1cm] 
 {\bf DES-like:} & Fiducial & $ 0.761^{+ 0.020}_{-0.021} $ & $ 0.10 \sigma $ & $ 1.7$  & $ 0.757^{+ 0.024}_{-0.015} $ & $ -0.10 \sigma $ & $ 1.6$ \\ [+0.1cm] 
  & with KiDS Priors & $ 0.770^{+ 0.018}_{-0.020} $ & $ 0.58 \sigma $ & $ 1.6$  & $ 0.766^{+ 0.020}_{-0.014} $ & $ 0.39 \sigma $ & $ 1.4$ \\ [+0.1cm] 
  & with NLA & $ 0.752^{+ 0.013}_{-0.014} $ & $ -0.49 \sigma $ & $ 1.1$  & $ 0.752^{+ 0.014}_{-0.013} $ & $ -0.49 \sigma $ & $ 1.1$ \\ [+0.1cm] 
  & with $\Sigma m_{\nu} = 0.06{\rm eV}$ & $ 0.772^{+ 0.017}_{-0.021} $ & $ 0.68 \sigma $ & $ 1.6$  & $ 0.766^{+ 0.022}_{-0.013} $ & $ 0.42 \sigma $ & $ 1.4$ \\ [+0.1cm] 

\bottomrule
\end{tabular}

%% file: tabletexfiles/mixed_input_and_Euclid_table_mock_analysis.tex
\begin{tabular}{ll|lr|lr|lr|lr}           
\toprule
     & 
     & \multicolumn{4}{l|}{Maximum Marginals:} 
     & \multicolumn{4}{l}{Mean Marginals:} \\
\midrule
     &
     & \multicolumn{2}{l|}{\bf KiDS-like:} 
     & \multicolumn{2}{l|}{\bf DES-like:} 
     & \multicolumn{2}{l|}{\bf KiDS-like:} 
     & \multicolumn{2}{l}{\bf DES-like:} \\
	 {\bf Non-Linear} 
     & {\bf Intrinsic Alignment}
     & $S_8$
     & $\Delta S_8$
     & $S_8$
     & $\Delta S_8$
     & $S_8$
     & $\Delta S_8$
     & $S_8$
     & $\Delta S_8$ \\
\midrule
Fiducial & Fiducial & $ 0.758^{+ 0.012}_{-0.014} $ & $ -0.05 \sigma $  & $ 0.761^{+ 0.020}_{-0.021} $ & $ 0.10 \sigma $  & $ 0.757^{+ 0.013}_{-0.012} $ & $ -0.15 \sigma $  & $ 0.757^{+ 0.024}_{-0.015} $ & $ -0.10 \sigma $  \\ [+0.1cm] 
{\sc HMCode2020}  & TATT strong & $ 0.775^{+ 0.012}_{-0.012} $ & $ 1.30 \sigma $  & $ 0.741^{+ 0.019}_{-0.022} $ & $ -0.88 \sigma $  & $ 0.775^{+ 0.012}_{-0.012} $ & $ 1.28 \sigma $  & $ 0.735^{+ 0.024}_{-0.015} $ & $ -1.28 \sigma $  \\ [+0.1cm] 
{\sc HMCode2020} & TATT weak & $ 0.763^{+ 0.012}_{-0.013} $ & $ 0.32 \sigma $  & $ 0.736^{+ 0.013}_{-0.017} $ & $ -1.55 \sigma $  & $ 0.762^{+ 0.013}_{-0.013} $ & $ 0.24 \sigma $  & $ 0.733^{+ 0.014}_{-0.014} $ & $ -1.88 \sigma $  \\ [+0.1cm] 
\bottomrule
\end{tabular}

%% file: tabletexfiles/EE2_table_mock_analysis.tex
\begin{tabular}{lccccc|lrl|lrl}          
\toprule
	\multicolumn{6}{l|}{{\sc EuclidEmulatorv2}}  
     & \multicolumn{3}{c|}{Maximum Marginal} 
     & \multicolumn{3}{c}{Mean Marginal} \\
\midrule  
    Analysis
     & IA
     & $\Sigma m_{\nu}$
     & +OWLS?
     & COSEBIs Cut?
     & $\log_{10}(T_{\rm AGN}/{\rm K})$  
     & $S_8$
     & $\Delta S_8$
     & $\sigma / \sigma_{\rm min}$
     & $S_8$
     & $\Delta S_8$
     & $\sigma / \sigma_{\rm min}$ \\
\midrule
DES-like & TATT & free & $\times$ & \checkmark & $\times$  & $ 0.738^{+ 0.017}_{-0.020} $ & $ -1.15 \sigma $ & $ 1.42$  & $ 0.734^{+ 0.020}_{-0.015} $ & $ -1.44 \sigma $ & $ 1.35$ \\ [+0.1cm] 
KiDS-like & NLA & 0.06eV & $\times$ & $\times$ & $\left[ 7.6-8.0 \right]$  & $ 0.771^{+ 0.013}_{-0.013} $ & $ 0.89 \sigma $ & $ 1.03$  & $ 0.771^{+ 0.013}_{-0.013} $ & $ 0.90 \sigma $ & $ 1.00$ \\ [+0.1cm] 
\midrule A & NLA-z & 0.06eV &$\times$ & \checkmark & $\times$  & $ 0.759^{+ 0.017}_{-0.018} $ & $ 0.01 \sigma $ & $ 1.36$  & $ 0.758^{+ 0.016}_{-0.015} $ & $ -0.09 \sigma $ & $ 1.22$ \\ [+0.1cm] 
B & " & " & \checkmark & $\times$ & $\times$ & $ 0.745^{+ 0.017}_{-0.018} $ & $ -0.83 \sigma $ & $ 1.35$  & $ 0.742^{+ 0.018}_{-0.013} $ & $ -1.10 \sigma $ & $ 1.21$ \\ [+0.1cm] 
C & " & " & \checkmark &\checkmark & $\times$  & $ 0.750^{+ 0.018}_{-0.018} $ & $ -0.52 \sigma $ & $ 1.42$  & $ 0.747^{+ 0.019}_{-0.013} $ & $ -0.75 \sigma $ & $ 1.27$ \\ [+0.1cm] 
D & " & " & \checkmark & $\times$ & $\left[ 7.6-8.0 \right]$ & $ 0.761^{+ 0.017}_{-0.022} $ & $ 0.10 \sigma $ & $ 1.53$  & $ 0.757^{+ 0.019}_{-0.015} $ & $ -0.14 \sigma $ & $ 1.33$ \\ [+0.1cm] 
E & " & " & \checkmark & \checkmark & $\left[ 7.6-8.0 \right]$ & $ 0.759^{+ 0.019}_{-0.020} $ & $ -0.00 \sigma $ & $ 1.51$  & $ 0.756^{+ 0.019}_{-0.015} $ & $ -0.16 \sigma $ & $ 1.35$ \\ [+0.1cm] 
F & " & " &\checkmark & \checkmark & $\left[ 7.3-8.0 \right]$ & $ 0.759^{+ 0.018}_{-0.020} $ & $ -0.03 \sigma $ & $ 1.48$  & $ 0.755^{+ 0.020}_{-0.015} $ & $ -0.22 \sigma $ & $ 1.39$ \\ [+0.1cm] 
\midrule Hybrid & NLA-z & free & \checkmark & \checkmark & $\left[ 7.3-8.0 \right]$ & $ 0.754^{+ 0.020}_{-0.021} $ & $ -0.25 \sigma $ & $ 1.62$  & $ 0.751^{+ 0.021}_{-0.015} $ & $ -0.46 \sigma $ & $ 1.43$ \\ [+0.1cm] 
G & TATT & free &\checkmark & \checkmark & $\left[ 7.3-8.0 \right]$ & $ 0.748^{+ 0.022}_{-0.023} $ & $ -0.49 \sigma $ & $ 1.78$  & $ 0.743^{+ 0.025}_{-0.015} $ & $ -0.81 \sigma $ & $ 1.58$ \\ [+0.1cm] 

\bottomrule
\end{tabular}

%% file: tabletexfiles/table_all_KiDS_results.tex
\begin{tabular}{lccccccc}
\hline
\hline
2pt & $\theta_{\rm min}$ &{\sc HM} & $S_8$ & $\Delta S_8$ & $\sigma/\sigma_{\rm fid}$  & $A_{\rm IA}$ & $\Delta A_{\rm IA}$ \\ [0.2cm]
\hline
$E_n$ & 0.5\' & 2016 & $0.758^{+0.017}_{-0.026}$ & $0.00 \sigma$ & 1.0 & $0.39^{+0.35}_{-0.41}$ & $0.00 \sigma$ \\ [+0.1cm] 

$\xi_{\pm}$ & 0.5\' & 2016 & $0.765^{+0.019}_{-0.017}$ & $0.33 \sigma$ & 0.83 & $0.37^{+0.36}_{-0.34}$ & $0.12 \sigma$ \\ [+0.1cm] 

\hline
$E_n$ & 2.0\' & 2016 & $ 0.774^{+ 0.020}_{-0.024} $ & $ 0.73 \sigma $ & $ 1.01$  & $ 0.62^{+ 0.35}_{-0.43} $ & $ 0.61 \sigma $ \\ [+0.1cm] 
$\xi_{\rm \pm}$ & 2.0\' & 2016 & $ 0.769^{+ 0.018}_{-0.020} $ & $ 0.52 \sigma $ & $ 0.87$  & $ 0.56^{+ 0.34}_{-0.35} $ & $ 0.45 \sigma $ \\ [+0.1cm] 
\hline
$E_n$ & 0.5\' & 2020 & $ 0.762^{+ 0.023}_{-0.028} $ & $ 0.19 \sigma $ & $ 1.19$  & $ 0.41^{+ 0.34}_{-0.46} $ & $ 0.04 \sigma $ \\ [+0.1cm] 
$E_n$ & 2.0\' & 2020 & $ 0.780^{+ 0.020}_{-0.024} $ & $ 1.01 \sigma $ & $ 1.03$  & $ 0.60^{+ 0.34}_{-0.44} $ & $ 0.56 \sigma $ \\ [+0.1cm] 

\hline
\end{tabular}

%% file: tabletexfiles/Goodness_of_fit.tex
\begin{tabular}{llrrc}
\hline
Survey & Pipeline & $\chi^2_{\rm min}$ & $ N_\Theta $ & $ p(\chi_{\rm min}^2 | \nu_{\rm eff} $) \\
\hline
DES Y3 (Full area) &  DES-like & 283.6 & 6.7 & 0.222 \\
DES Y3 (Full area) &  KiDS-like & 284.3 & 5.0 & 0.235 \\
DES Y3 (Full area) &  Hybrid & 284.2 & 5.4 & 0.231 \\
\hline
DES Y3 (KiDS-excised) &  DES-like & 285.3 & 8.0 & 0.187 \\
DES Y3 (KiDS-excised) &  KiDS-like & 286.5 & 5.1 & 0.207 \\
DES Y3 (KiDS-excised) &  Hybrid & 288.3 & 4.6 & 0.192 \\
\hline
KiDS &  DES-like & 83.7 & 8.3 & 0.078 \\
KiDS &  KiDS-like & 82.5 & 6.5 & 0.118 \\
KiDS &  Hybrid & 88.3 & 7.1 & 0.048 \\
\hline
Joint &  DES-like & 371.5 & 12.0 & 0.089 \\
Joint &  KiDS-like & 381.3 & 7.8 & 0.062 \\
Joint &  Hybrid & 378.0 & 9.6 & 0.068 \\
\hline
Joint-1IA &  DES-like & 377.2 & 10.4 & 0.068 \\
Joint-1IA &  KiDS-like & 373.9 & 8.8 & 0.094 \\
Joint-1IA &  Hybrid & 382.2 & 8.0 & 0.057 \\
\hline
\end{tabular}

%% file: tabletexfiles/table_lcdm_fullarea.tex
\begin{tabular}{lcccc}
\toprule
\ Parameter & DES Y3+KiDS-1000 & DES Y3 & KiDS-1000 \\ 
\midrule 
$S_8$ &  $ 0.790^{+0.018}_{-0.014} $  &  $ 0.802^{+0.023}_{-0.019} $  &  $ 0.763^{+0.031}_{-0.023} $  \\ [+0.1cm] 
 $\Omega_{\rm m}$ &  $ 0.280^{+0.037}_{-0.046} $  &  $ 0.297^{+0.040}_{-0.060} $  &  $ 0.270^{+0.056}_{-0.102}* $  \\ [+0.1cm] 
 $\sigma_8$ &  $ 0.825^{+0.067}_{-0.073} $  &  $ 0.816^{+0.076}_{-0.085} $  &  $ 0.833^{+0.133}_{-0.146} $  \\ [+0.1cm] 
 $h$ &  $ 0.746^{+0.069}_{-0.029}* $  &  $ 0.738^{+0.072}_{-0.040}* $  &  $ 0.749^{+0.068}_{-0.029}* $  \\ 
 $n_{\rm s}$ &  $ 0.973^{+0.074}_{-0.072}* $  &  $ 0.979^{+0.093}_{-0.062}* $  &  $ 0.984^{+0.102}_{-0.053}* $  \\ [+0.1cm] 
 $\Sigma m_{\nu}$ &  $ 0.326^{+0.110}_{-0.187}* $  &  $ 0.317^{+0.139}_{-0.223}* $  &  $ 0.327^{+0.152}_{-0.215}* $  \\ [+0.1cm] 
 $\omega_{\rm b}$ &  $ 0.022^{+0.002}_{-0.003}* $  &  $ 0.022^{+0.002}_{-0.003}* $  &  $ 0.022^{+0.002}_{-0.003}* $  \\[+0.1cm] 
 \hline
 $\mathrm{log}_{10} (T_{\mathrm{AGN}}/{\rm K)}$ &  $ 7.627^{+0.152}_{-0.299}* $  &  $ 7.632^{+0.170}_{-0.289}* $  &  $ 7.623^{+0.153}_{-0.293}* $  \\ [+0.1cm] 
 $A_{\rm IA}^{\mathrm{DES}}$ &  $ -0.025^{+0.576}_{-0.285} $  &  $ 0.323^{+0.432}_{-0.369} $  &  -  \\ [+0.1cm] 
 $\eta^{\mathrm{DES}}$ &  $ 2.047^{+2.953}_{-0.832}* $  &  $ 1.774^{+3.166}_{-0.940}* $  &  -  \\ [+0.1cm] 
 $A_{\rm IA}^{\mathrm{KiDS}}$ &  $ 1.038^{+0.543}_{-0.516} $  &  -  &  $ 0.650^{+0.884}_{-0.539} $  \\ [+0.1cm] 
 $\eta^{\mathrm{KiDS}}$ &  $ 2.211^{+2.588}_{-0.958}* $  &  -  &  $ 2.022^{+2.937}_{-0.868}* $  \\[+0.1cm] 
 \hline
 $\Delta z_{1}^{\mathrm{DES}}$ &  $ 0.001^{+0.015}_{-0.015}* $  &  $ -0.004^{+0.016}_{-0.015}* $  &  -  \\[+0.1cm] 
 $\Delta z_{2}^{\mathrm{DES}}$ &  $ 0.014^{+0.012}_{-0.011}* $  &  $ 0.011^{+0.011}_{-0.011}* $  &  -  \\ [+0.1cm] 
 $\Delta z_{3}^{\mathrm{DES}}$ &  $ -0.008^{+0.010}_{-0.010}* $  &  $ -0.004^{+0.010}_{-0.010}* $  &  -  \\ [+0.1cm] 
 $\Delta z_{4}^{\mathrm{DES}}$ &  $ -0.006^{+0.015}_{-0.014}* $  &  $ 0.000^{+0.014}_{-0.014}* $  &  -  \\ [+0.1cm] 
 $\Delta z_{1}^{\mathrm{KiDS}}$ &  $ -0.001^{+0.011}_{-0.010}* $  &  -  &  $ 0.000^{+0.010}_{-0.010}* $  \\ [+0.1cm] 
 $\Delta z_{2}^{\mathrm{KiDS}}$ &  $ -0.010^{+0.010}_{-0.011}* $  &  -  &  $ -0.008^{+0.011}_{-0.012}* $  \\ [+0.1cm] 
 $\Delta z_{3}^{\mathrm{KiDS}}$ &  $ 0.016^{+0.010}_{-0.011}* $  &  -  &  $ 0.016^{+0.011}_{-0.010}* $  \\ [+0.1cm] 
 $\Delta z_{4}^{\mathrm{KiDS}}$ &  $ 0.017^{+0.008}_{-0.009}* $  &  -  &  $ 0.015^{+0.008}_{-0.008}* $  \\ [+0.1cm] 
 $\Delta z_{5}^{\mathrm{KiDS}}$ &  $ -0.006^{+0.009}_{-0.010}* $  &  -  &  $ -0.007^{+0.009}_{-0.010}* $  \\ [+0.1cm] 
 $m_1^{\mathrm{DES}}$ &  $ -0.007^{+0.009}_{-0.009}* $  &  $ -0.007^{+0.009}_{-0.009}* $  &  -  \\ [+0.1cm] 
 $m_2^{\mathrm{DES}}$ &  $ -0.021^{+0.008}_{-0.007}* $  &  $ -0.020^{+0.008}_{-0.008}* $  &  -  \\ [+0.1cm] 
 $m_3^{\mathrm{DES}}$ &  $ -0.022^{+0.007}_{-0.007}* $  &  $ -0.024^{+0.007}_{-0.008}* $  &  -  \\ [+0.1cm] 
 $m_4^{\mathrm{DES}}$ &  $ -0.035^{+0.007}_{-0.007}* $  &  $ -0.037^{+0.007}_{-0.007}* $  &  -  \\ [+0.1cm] 
\hline
\end{tabular}